\documentclass[prb]{revtex4}

\usepackage{amsmath,bm}
\usepackage{graphicx,pdflscape}
\usepackage{color}
\usepackage[toc,page]{appendix}

\usepackage{hyperref}
\usepackage{color}
\newcommand{\beq}{\begin{equation}}
\newcommand{\eeq}{\end{equation}}
\newcommand{\barray}{\begin{eqnarray}}
\newcommand{\earray}{\end{eqnarray}}
\newcommand{\disp}[1]{Eq.~(\ref{#1})}

\newcommand{\refdisp}[1]{Ref.~(\onlinecite{#1})}
\newcommand{\figdisp}[1]{Fig.~(\ref{#1})}

\newcommand{\lsim}{\raisebox{-4pt}{$\,\stackrel{\textstyle <}{\sim}\,$}}
\newcommand{\gsim}{\raisebox{-4pt}{$\,\stackrel{\textstyle >}{\sim}\,$}}

\newcommand{\si}{\sigma}

\newcommand{\tJ}{\ $t$-$J$ \ }
\newcommand{\nn}{\nonumber}

\renewcommand{\Re}{\mathrm{Re}}
\renewcommand{\Im}{\mathrm{Im}}

\renewcommand{\emph}{\textit}

\newcommand{\G}{\mathcal{G}}

\newcommand{\GH}{{\bf g}}
\newcommand{\GHI}{\GH^{-1}}

\newcommand{\eb}{\varepsilon}

\newcommand{\chem}{\bm {\mu}}

\newcommand{\fd}{f}
\newcommand{\fdb}{\bar{f}}

\newcommand{\imag}{\Im }
\newcommand{\real}{\Re }

\newcommand{\rophi}{\rho_{\overline{\Phi}}}
\newcommand{\ropsi}{\rho_{\Psi}}
\newcommand{\rephi}{\real \ \overline{\Phi}}

\begin{document}

\title{Extremely Correlated Fermi Liquids:\\ Self consistent solution of the second order theory}
\author{ Daniel Hansen and B. Sriram Shastry}
\affiliation{ Physics Department, University of California, Santa Cruz, CA 95064, USA}
\date{January 7, 2013}
\begin{abstract}
We present detailed results from a recent microscopic  theory of extremely correlated Fermi liquids, applied to the \tJ model in two dimensions. We use  typical sets of  band parameters relevant to the cuprate superconductors. The second order theory in the parameter $\lambda$ is argued to be quantitatively  valid in the overdoped regime
for  $0 \leq n \lsim 0.75$  ($n$ is the particle density). The calculation  involves the self consistent solution of equations for an auxiliary Fermi liquid type  Greens function and an adaptive spectral weight, or caparison factor, described in recent papers by Shastry (Refs. (1) and (5)). We present the numerical results at low as well as high $T$  at various low to intermediate densities in the normal phase with emphasis placed on features that are  experimentally accessible. We display the momentum space occupation  function $m_k$, various energy dispersions locating  the peaks of spectral functions, the optical conductivity, relaxation rates for quasiparticles, and the electronic spectral functions along various directions in the Brillouin zone,  and with typical additional elastic scattering. The lineshapes have  an asymmetric shape and a broad background that is seen in experiments near and beyond  optimal hole doping, and   validate   approximate  recent recent versions of the theory. The results display  features such as the high energy kink, and provide an in depth understanding of its origin and   dependence on band parameters.
\end{abstract}
\maketitle


\section{Introduction }

The \tJ model describes the problem of  very strongly interacting   electrons, made especially difficult by  the requirement of (at most) single occupancy of the lattice sites. It is the subject of many recent works in the context of the cuprate superconductors and other correlated systems. This problem is very hard since it is precludes the application of standard  perturbative methods. This conundrum has  motivated  a new strong coupling  approach   resulting in the theory  of {\em extremely correlated Fermi liquids} (ECFL)\cite{ECFL}. Previous applications of the methodology of \refdisp{ECFL}  to the  cuprates has given encouraging results.
These  include  spectral functions that compare very well with the experimental angle resolved photoemission spectroscopy (ARPES) data \cite{Gweon,Asymmetry} including natural explanations of phenomena  such as 
``high energy kink'' also known as  the ``waterfall'' effect, and also the more subtle ``low energy kink'' seen in experiments. The theory also has led to  interesting predictions for the asymmetry of lineshapes\cite{Asymmetry}.

The formalism initiated in \refdisp{ECFL}  charted out an approach to the problem of the \tJ model using  basic insights from Schwinger's powerful approach to field theory, using source fields to write down exact functional differential equations for the propagator. Once this exact equation of motion is written down, in a crucial next step, it was recognized that complexity arising from  the non canonical nature of the (projected) electrons can be circumvented by decomposing  the propagator  as the   space time convolution of a {\em canonical electron propagator} and an adaptive spectral weight factor termed {\em the caparison factor} ; these  satisfy coupled equations of motion.  A recent work \cite{ecfl-form}  develops this idea in a systematic fashion, emphasizing the role of expanding in a parameter $\lambda$  ($0 \leq \lambda \leq 1$),   related  to the particle density, or more closely  to  $\lambda \sim (1-\frac{4}{n^2}  d)$, where $d$  is the  double occupancy  ($0 \leq d \leq \frac{n^2}{4}$). It further explores the implications of a novel set of identities for the \tJ model, termed the  {\em shift identities}, which provide an important constraint on the $\lambda$ expansion.  A method for generating a  systematic set of equations for the propagator to any orders in $\lambda$ is given, along with  explicit equations to second order  that manifestly obey  the shift identity constraints. We will refer to this theory  as (I) here and prefix equations of that paper with (I). The solution of this ECFL propagator to second order in $\lambda$ is the main focus of this work.  Broadly speaking, these equations resemble the fully self consistent skeleton diagram expansion of the standard Fermi liquid (FL) theory, as described in standard texts\cite{Nozieres,AGD,mahan}, but generalize to the case of extreme correlations.
Summarizing the arguments in \refdisp{ecfl-form} and \refdisp{ECFL}, a low order theory in $\lambda$ is already expected to capture features of extreme correlations. This, perhaps initially surprising  expectation,  arises  in view of the novel structure of the Greens functions with two self energies $\Phi$ and $\Psi$  (see in \disp{gphys} and \disp{g-def}) within this formalism. The self energy  $\Psi$ sitting in the numerator of the Greens function as in  \disp{g-def},  plays the role of an adaptive spectral weight
that balances the somewhat opposing  requirements of the  ``high energy'' weight $1- \frac{n}{2}$ and the low energy Luttinger theorem. The latter  requires a greater magnitude of the numerator  than $1- \frac{n}{2}$,   to accommodate the  particles into a Fermi surface  with the same volume as in  the Fermi gas.   A further tactical advantage of  this method  is   due to  the finite range  of variation of   $\lambda$, namely $0 \leq \lambda \leq 1$ that suffices to interpolate between the Fermi gas and the extreme correlation limit. This is in   contrast to controlling the double occupancy $d$ using a repulsive energy $U$ with its usual infinitely large range of values $0 \leq U \leq \infty$; experience shows that $U$ needs to become  very large $U \gg |t|$ in order to achieve the same end, thereby invalidating low order expansions in $U$. In summary, within the present formalism, a low order theory in $\lambda$  seems  well worth examining in detail, this is our task here.
  
 We note that apart from a few exact solutions in 1-dimension, and some calculations for finite sized systems (see below), there are {no systematic analytical  calculations for the dynamics of the \tJ model in higher dimensions}, working directly in the thermodynamic limit. 
 Therefore while the importance of the \tJ model was understood many years ago, there  has been little detailed comparison with the ARPES experiments until recently\cite{Gweon}. This gap is one of the main motivations for this (and our related) work.   In this paper, we present {\em the first controlled   calculations for the spectral functions of the  \tJ model}, by solving the above equations to evaluate thermodynamical variables, the spectral functions, ARPES lineshapes and optical conductivity of the \tJ model. The ECFL formalism and the $\lambda$ expansion method provides an inbuilt  criterion to judge the validity of the expansion at any order. Using this criterion we argue that
  our present $O(\lambda^2)$
  calculations are valid in the high hole doping limit, known as the overdoped regime which corresponds to low and intermediate {\em electron density}. Future work will be aimed at higher order calculations in $\lambda$ which would presumably enable us to address densities close to optimal doping ($n\sim.85$) with greater accuracy. The results are compared with other approximations as well as a few experiments. Needless to say, even in such an overdoped regime,  experimental evidence points to   the important role of strong correlations \cite{ali,basov}. Further, our work provides qualitative  support to the approximate and phenomenological developments from ECFL used at higher particle densities\cite{Gweon,Anatomy,kazue}.

While analytical methods beyond crude mean field theories have been in short supply, there is a valuable body of numerical results for the \tJ model from exact diagonalization\cite{StephanHorsch}, high temperature series expansions\cite{Singh}, variational wave functions \cite{Gros1,Gros2,mohit}, and finite temperature Lanczos methods \cite{prelov1,prelov2,prelov3,prelovJaklic}. Noteworthy are the results of \refdisp{prelov2} from Prelovsek and co workers, who handle the series expansion in inverse temperature in a stochastic fashion, thereby obtaining results down to fairly low temperatures. Owing to finite size effects and the inherent nature of the high $T$ expansion, the results from this theory, although broadly comparable to ours, seem more grainy.

The Hubbard model for  large on-site coupling U tends to the \tJ model (apart from   $O(t^2/U)$ correction terms),  so the large U studies of this model are of interest. Quantum Monte Carlo methods, despite the difficulties associated with the sign problem, yield some valuable insights into the spectral features such as kinks \cite{Moritz}. We note that the dynamical mean field theory (DMFT) for the Hubbard model \cite{DMFT1,DMFT2} gives a numerically exact solution in high enough dimensions of the Hubbard  model.  Although  the strong coupling (i.e. $U>W$) relevant to the \tJ model results is challenging, there is impressive progress overall. A recent DMFT study\cite{georges} at strong coupling obtains detailed spectral functions that are roughly comparable to what we find here for the \tJ model.

In the light of the above progress,   it may be   reasonable to  restate the rationale for  developing yet another  technique. The ECFL formalism has several advantages, since it is essentially an analytical method  with a  computational aspect that is lightweight, in comparison with other methods listed above.
 The only present  limitation is the density attainable with the second order theory.  Despite this, we are able to make useful comparisons below, with  state of the art calculations from other methods in overlapping domains ($n \lsim .75$). For example, in  comparing  our results with the already impressive results of \refdisp{prelov2}, we note that  they obtain a spectral function which is broadly comparable to ours (compare their Fig. (2d) with our Fig. ~(7,8)). Their approach starts from a finite lattice of ~20 sites and incorporates temperature dependence through stochastic means. The smooth momentum dependence is obtained by averaging  over twisted boundary  conditions of their small lattice. Our approach, on the other hand, is essentially in the  thermodynamic limit, with 3600 sites in Fig.~(7,8). The momentum grid available is therefore dense, and hence we are guaranteed to catch features that might be missed in an averaging procedure on small lattices. Taking the T=0 limit is straightforward as well studying finite T, we can also  vary the parameters of the model such as the range of  hopping in a simple fashion.   We also  note that the analytical approach available through ECFL complements methods such as DMFT \cite{georges} and Quantum Monte Carlo \cite{Moritz} where the analytical underpinning of the results is somewhat  suboptimal.

The work proceeds as follows. In Section (II), we  present a summary of the equations solved here from (I). In Section (III), we  discuss the computational strategy and  explain the  scheme, using the fast Fourier transform method (FFT),  so that  the spectral functions can be computed efficiently. Section (IV) presents the detailed  results of the calculation. Section (V) contains a summary and concluding comments.

\section{Summary of the $O(\lambda^2)$ Theory}

In the ECFL formalism developed in (I) the physical Greens function $\G$ can be factored in the momentum space as
  \beq
  \G(k) = \GH(k) \ \mu(k),  \;\;\;\;\mbox{where} \;(k) \equiv (\vec{k}, i \omega_k).\label{gphys}
  \eeq
Here $\mu(k)$, termed the caparison factor,  plays the role of an adaptive spectral weight while $\GH(k)$ is a canonical FL,  derivable from a Luttinger-Ward functional. Each factor is then written in terms of a frequency and momentum dependent ``self energy'' such that $\GH(k)$ and $\mu(k)$ may be written schematically in terms of a perturbation parameter $\lambda$ as
\barray
\mu(k)&=&1- \lambda \ \gamma(\lambda,n) + \lambda \ \Psi(k;\lambda)  \label{mu-def} \\
\GHI(k)&=& i \omega_n + \chem - c(\lambda,n) \   \varepsilon_k - \lambda \ \Phi(k;\lambda) \label{g-def}
\earray
where $\Psi$ and $\Phi$ play the role of two self energies, 
 and   the band energy  $\varepsilon_k$ is  multiplied by the  renormalization factor $c(\lambda,n)$.
Quantities such as   $\gamma(\lambda,n), c(\lambda,n)$ and the two self energies $\Psi(k;\lambda), \Phi(k;\lambda) $  are  expanded in $\lambda$.
   The self energies    vanish at $\lambda=0$,  so that $\G(\lambda\rightarrow0)$ is a non-interacting gas of canonical Fermions. The \tJ model $\G$ corresponds to $\lambda=1$. Thus both $\mu(k)$ and $\GH^{-1}(k)$ can be expanded to any order in the parameter $\lambda$, where higher order terms are determined recursively via functional derivatives of the \tJ equation of motion as in (I). The second order equations are the lowest order where  non-trivial frequency dependence arises,  and will be the focus of this work.  The strategy is to  set the parameter $\lambda \to 1$  at the end of a calculation to a given order.  As mentioned in the Introduction,  the $\lambda$ expansion is already expected to be meaningful at low order.  In confirmation of this argument, we shall see below that  the truncation of the $\lambda$ series at second order results in an interesting spectral function  exhibiting many key features of strong correlation.  Below we discuss in detail the criterion for the quantitative validity of the present second order expansion, after presenting the relevant equations. As shown in equations  (I-83, I-84, I-85), the second order equations for the ECFL Greens function are:
\barray
\mu(k) &=& 1- \lambda  \frac{n}{2}  + \lambda^2 \ \frac{n^2}{4} +\lambda^2 \Psi(k) , \label{mu} \\
\Psi(k)&=&  -  \sum_{ p,q}   \left( \varepsilon_p +\varepsilon_{k+q-p}  +  \varepsilon_k   +\varepsilon_q +  J_{k-p}-u_0\right) \ \GH(p) \ \GH(q) \ \GH( q+ k -p )  \label{psi} \\
 \GHI(k) & = & i \omega_n + \chem' - \overline{\varepsilon}_k - \lambda^2 \  \overline{\Phi}(k)  \label{ginv} \\
\overline{\varepsilon}_k &=&  \left( 1- \lambda \ n + \lambda^2 \ \frac{3n^2}{8}  \right)\varepsilon_k + \lambda \  \sum_q \frac{1}{2} J_{k-q} \ \GH(q)  \label{eband} \\
  \overline{\Phi}(k) & = & - \sum_{q,p} \GH(q) \ \GH(p) \ \GH(k+q-p) \nn  \\
&& \times \left(  \varepsilon_{k} +\varepsilon_{p}+  \varepsilon_{q} +  \varepsilon_{k+q-p} + J_{k-p} -u_0 \ \right)  \  \{ \varepsilon_{k} +\varepsilon_{p}+ \varepsilon_{q} + \varepsilon_{k+q-p} + \frac{1}{2} \left(J_{k-p} + J_{p - q} \right) - u_0 \   \}, \label{phi}
  \earray
where $\sum_k \equiv \frac{1}{\beta N_s} \sum_{\vec{k}, \omega_n}$ with $N_s$ being the number of lattice sites and $\beta$ is inverse temperature. As written here, $\mu(k)$ and $\GH(k)$ have acquired a variety of static terms as well as frequency dependent terms called $\Psi$ and $\overline{\Phi}$, respectively.  This is written with slight change of notation $[\Phi(k)]_1 \to \overline{\Phi}(k)$ from (I-85), and we have introduced the effective band energy $\overline{\varepsilon}_k$ in \disp{eband} that gets a static contribution from shrinking of the bare energies $\varepsilon_k$, as well as from the exchange energy $J$. The role of the parameter $u_0$ as a second chemical potential will be described below. All terms are understood to be correct up to $O(\lambda^2)$, and possess corrections of $O(\lambda^3)$ that are ignored here.
  
The number of the physical electrons is fixed by a sum rule, 
  \barray
  \frac{n}{2} &=& \sum_k \G(k) \ e^{i \omega_n 0^+},  \label{sumrule-1}
  \earray
while, in order that $\G$ will satisfy the Luttinger volume theorem, the auxiliary Fermions described by $\GH$ must be equal in number, and therefore satisfy a {\em second} sum rule:
  \barray
  \frac{n}{2}&=& \sum_k \GH(k) \ e^{i \omega_n 0^+}   \label{sumrule-2}.
  \earray
A conventional skeleton diagram expansion would have only one such sum rule which could be enforced by using the chemical potential as a Lagrange multiplier. Here there are two independent sum rules and  a second Lagrange multiplier arises naturally in the ECFL formalism, thanks to the role of the shift identities, as shown in (I). The two Lagrange multipliers have very different roles. The first, $\chem'$, is similar to a standard chemical potential in that it sits next to the band energies, $\overline{\varepsilon}_k$, in the denominator of $\GH$. Changes in $\chem'$ will therefore shift the entire band intact, up or down in frequency. The second Lagrange multiplier $u_0$  has a role  similar to that of the Hubbard $U$ in the effective Hamiltonian in (I). It controls the broadening of the spectral function through the magnitude of $\Phi$ and $\Psi$. Through the $u_0$ dependence of the magnitude  of $\Psi$,  it is also responsible for the creation of asymmetric spectral tails which extend to very high frequency and create dramatically skewed lineshapes similar to ARPES data from the cuprates. These two independent Lagrange multipliers $\chem'$ and $u_0$, which shift and  skew the spectrum respectively, are determined such that the two equations (\disp{sumrule-1} and \disp{sumrule-2}) are satisfied simultaneously. Furthermore they dutifully satisfy the ``shift invariances'' described in \refdisp{ecfl-form}, i.e. any uniform shift in $\varepsilon_k$ or $J_k$ will be absorbed in $\chem'$ and $u_0$ such that the spectral function is invariant. Neither of these Lagrange multipliers is the physical  thermodynamic chemical potential of the Grand Canonical Ensemble. The physical  chemical potential $\mu_{phys}$, denoted by $ \chem$, can be obtained as a function of $\chem'$ and $u_0$  as shown  in Eq. (179) of (I):
\beq
\chem=\chem'+u_0 \frac{\lambda n}{2}(1-\frac{\lambda n}{4})-\left(J_0\frac{\lambda n}{4}(1-\frac{\lambda n}{2})+2\lambda (1-\frac{\lambda n}{8})\sum_q \varepsilon_q \GH(q)\right)+O(\lambda^3) \label{physmu}.
\eeq
 The structure of \disp{physmu} is revealing; at $O(\lambda^0)$ we find the Fermi gas result $\mu_{phys}^{(0)}=\chem'$ and $\G=\GH$. The second sum rule is therefore satisfied trivially and $u_0$ is absent from all expressions.  This tells us that when $\mu(k)=1$ the Luttinger Ward volume theorem and the particle density sum rule are simultaneously determined by the one chemical potential $\chem$, as in standard weak coupling theories. However at finite $\lambda$ we find  $\mu(k) \neq 1$ from \disp{mu-def}, and hence this spectral depletion (arising from $(1- \gamma) $ in \disp{mu-def})  and the Luttinger Ward volume theorem become decoupled. Now  two parameters  $u_0$  and   $\chem'$ are  available  to satisfy the two simultaneously. As a net  consequence the spectral function shapes are influenced by $\mu(k)$ significantly, and develop a bias or a skew towards occupied energies as discussed in \refdisp{Gweon}.

We now discuss the criterion for validity of equations to a second order in $\lambda$. As stated above, dropping terms of $O(\lambda^3)$ in \disp{mu} - \disp{phi}  limits the regime of validity of these calculation to densities not too close to unity.  To see this, note from \disp{mu}  that this theory would give a high frequency behavior of $\G\sim \frac{c_0}{ i \omega}$ with $c_0 =  1 - \frac{n}{2}+ \frac{n^2}{4}$,  rather than the exact value $c_0 =  1 - \frac{n}{2}$, and thus introducing an error. This slight error in the high frequency physics is a result of keeping a few terms in  the expansion in $\lambda$. Note  however that  the low frequency physics encoded by the Luttinger Ward sum rule is untouched by this, and is exactly obeyed to each order in $\lambda$. Thus at $n\sim .78$ we have an error of $\frac{n^2}{4-2 n} \sim 25$\% in the high frequency spectral weight in this theory, a value  somewhat beyond where we can push this approximation. The $O(\lambda^3)$ terms are expected to extend the range of this approximation to higher particle densities.
  
\section{Computation of Spectral Functions}
\subsection{ Definitions}
Computationally it is expedient to employ a spectral function notation as described for example  in \refdisp{mahan}. The Matsubara frequency object $\G(k,i\omega_n)$  is analytically continued to the real axis and its spectral density written as:
\beq
\rho_{\G}(k,\omega)= - \frac{1}{\pi} \imag \ \left[ \G(k,i\omega_n \rightarrow \omega+i\eta) \right] \equiv A(k,\omega), \label{akw}
\eeq
were $A(k,\omega)$ is the  notation for the spectral function used in most experimental literature.  The real part of the analytically continued function can be obtained by a Hilbert transform
\beq
\real \ \G(k,\omega) = \text{P.V.} \int_{-\infty}^{\infty} \frac{\rho_{\G}(k,\nu)}{\omega-\nu}d\nu. \label{hilberteq}
\eeq
An analogous definition is given for spectral representation $\rho_{\GH}(k,\nu)$, $\rophi(k,\nu)$, $\ropsi(k,\nu)$ used for $\GH$, $\overline{\Phi}$, $\Psi$, etc, and hence, the full set of equations above can be rewritten in terms of these spectral functions.  Since $\G$ is a product as in \disp{gphys}, we note that within the $O(\lambda^2)$ theory
\beq
\rho_{\G}(k, \omega) = \rho_{\GH}(k, \omega) \ \left( 1-   \frac{n}{2}  +  \frac{n^2}{4} +  \Re  \ \Psi(k, \omega)\right)  + \rho_{\Psi}(k, \omega) \ \Re \ \GH(k, \omega), \label{rhoproduct}
\eeq
so the two sum rules \disp{sumrule-1} and \disp{sumrule-2} can be written as
\barray
\frac{n}{2}&=&\sum_k \int{d\omega \rho_{\GH}(k,\omega)f(\omega) }\nn\\
\frac{n^2}{4}(1- \frac{n}{2})&=&-\sum_k \int {d\omega \biggl(\rho_{\GH}(k,\omega) \ \Re  \ \Psi(k,\omega)+\real \ \GH(k,\omega) \ \ropsi(k,\omega)\biggr)f(\omega) }\label{second2}
\earray
where $f(\omega)=(1+\exp(\beta \omega))^{-1}$ and $\overline{f}(\omega)=1-f(\omega)$. The  auxiliary  spectral function is in the usual Dysonian form
\beq
\rho_{\GH}(k,\omega)=\frac{\rophi(k,\omega)}{\left\{\omega+\chem'-  \overline{\eb}_k -  \rephi(k,\omega)\right\}^2+\bigl(\pi \rophi \bigr)^2}. \label{rhog}
\eeq

The spectral functions for $\Psi$ and $\overline{\Phi}$ have the form
\barray
\rophi(k,\omega)&=&\frac{1}{N_s^2}\sum_{pq}\int d\nu_1 d\nu_2 \ \rho_{\GH}(p,\nu_1) \rho_{\GH}(q,\nu_2) \rho_{\GH}(p+q-k,\nu_1+\nu_2-\omega)\times \nn\\
&      &\left\{\fd(\nu_1)\fd(\nu_2)\fdb(\nu_1+\nu_2-\omega)+\fdb(\nu_1)\fdb(\nu_2)\fd(\nu_1+\nu_2-\omega)\right\}\times \nn\\
&      &  \left( \varepsilon_p +\varepsilon_{k+q-p}  +  \varepsilon_k   +\varepsilon_q +  J_{k-p}-u_0\right) \  \{ \varepsilon_{k} +\varepsilon_{p}+ \varepsilon_{q} + \varepsilon_{k+q-p} + \frac{1}{2} \left(J_{k-p} + J_{k - q} \right) - u_0 \   \}  \label{rhophi-2}\\
\ropsi(k,\omega)&=&\frac{1}{N_s^2}\sum_{pq}\int d\nu_1 d\nu_2 \ \rho_{\GH}(p,\nu_1) \rho_{\GH}(q,\nu_2) \rho_{\GH}(p+q-k,\nu_1+\nu_2-\omega)\times \nn\\
&      &\left\{\fd(\nu_1)\fd(\nu_2)\fdb(\nu_1+\nu_2-\omega)+\fdb(\nu_1)\fdb(\nu_2)\fd(\nu_1+\nu_2-\omega)\right\}\times \nn\\
&  & \left( \varepsilon_p +\varepsilon_{k+q-p}  +  \varepsilon_k   +\varepsilon_q +  J_{k-p}-u_0\right),    \label{rhopsi-2}
\earray
which differ only through the presence of momentum dependent factors. These frequency integrals are solved by discretizing frequency over a finite window which captures the finite support of the spectral functions. In Appendices A and B we outline how this is accomplished efficiently with Fast Fourier Transforms (FFTs) and implemented in an iterative process.
leading to a  self consistent $\G$. 

\section{Results}

\subsection{ Physical Variables  \label{paras}}
The  computational program has several parameters which can be varied. These include the tight binding bandstructure (through hopping parameters t, t' etc.), the spin coupling J, density, and temperature.   For the parameters of the model, we study the following important situations:
\begin{itemize}
\item  {\bf Case (A)} the   nearest neighbor hopping model with $t\sim 3000$K and $J\sim900$K;
\item  {\bf Case (B)} with  $t'=-.4 \ t$ and the other parameters unchanged. 
\item  {\bf Case (B')} with six hopping parameters as in \refdisp{Bansil},  $t\sim 3000$K and $J\sim900$K;
\item  {\bf Case (C)} with  $t'=+.4 \ t$ and the other parameters unchanged as in \refdisp{Fulde}. 
\end{itemize} {\bf Case (A)} is the most natural one to study since it is minimal. However at the bare level, its  Fermi surface (FS) near half filling remains closed around the $\Gamma =(0,0)$ point in the Brillouin zone (BZ). This is in contrast to the ARPES reconstructed FS of, say, $BISSCO$ which opens up to a hole like surface,  i.e.  no longer closes around the $\Gamma$ point. For this reason we study the {\bf Case (B)} as well, with a FS that closes around the $\Gamma$ point for $n \lsim .55$ but becomes hole-like at higher n. The Fermi momentum along the $<11>$ direction is measured at $n \sim .85$ in experiments on $BISSCO$ as $\sim .43$\AA$^{-1}$  and provides a useful check on the parameters. Within the {\bf Case (A)} we find a considerably larger value $\sim .55$\AA$^{-1}$ while {\bf Case (B)}  gives a better fit  $\sim .45$\AA$^{-1}$.  The  {\bf Case (B)}  has the   minimal set of hoppings needed to give a hole like FS, but ends up giving a very flat dispersion near the $\Gamma$ point.
 {\bf Case (B')} is  a fine tuned tight binding fit from \refdisp{Bansil}, and captures the exact Fermi momentum as well as the curvature near $\Gamma$,  and gives the best overall representation of the dispersion.
 It is somewhat harder from the computational view point, so we use this case  only to illustrate its  distinction with   {\bf Case (B)} waterfall below in \figdisp{waterfall}.

While the Fermi volume is conserved by the theory reported here, the shape of the FS can, in principle, change since the self energy $\overline{\Phi}(k)$ strongly depends upon the wave vector $\vec{k}$, so these numbers may in principle change somewhat in the final spectrum. However, in this $O(\lambda^2)$ calculation we find that the ``bare'' FS is preserved to high accuracy. We also note that for {\bf Case (B)}, the band density of states has a van Hove singularity (vHs) close to the bare Fermi energy for density $n\sim .58$. This ends up influencing the results for this case significantly in a range of densities around $n\gsim.55$. The contrast between this and {\bf Case (A)} therefore gives an idea of the role played by the van Hove points in strongly correlated systems.

 While {\bf Cases (A)} and {\bf (B), (B')} have parameters which are most relevant for the hole doped cuprates. We will also present a third case
 {\bf Case (C)}, it  is more relevant to the electron doped side of the cuprate phase diagram \refdisp{Fulde}. This case will not be explored in as much detail, but provides an interesting situation  where the  waterfall effect is magnified further from {\bf Case (A)}.

\subsection{ Other parameters in the programs} 
The program can be implemented on lattices of various size and spatial dimension. For a given choice of these parameters an appropriate choice must be made for computational grid. This includes the lattice size as well as the discretized frequency grid. We look at converged spectral functions for a wide variety of these parameters.
 
The majority of the following results were performed on a square lattice which has dimension $ L \times L$ with L=36,  and  periodic boundary conditions are imposed.  We therefore work in a momentum representation with an $ L \times L$ sized  k-grid of points $k_{i,j}=\frac{\pi}{a L}(i,j)$ where $1\le i,j \le L$ and the lattice parameter is $a=3.82$\AA. The spectral functions have compact support which extends to $|\omega|\lsim 8 \times t$. We choose a frequency range $-\frac{1}{2} \omega_c \leq \omega \leq  \frac{1}{2} \omega_c$,   with $\omega_c = 30 \times t $, a range that  is sufficient to capture the full range of the spectral functions. We discretize this frequency range in $N_{\omega}=3000$ bins each of width $\Delta \omega=\frac{\omega_c}{N_\omega}=.01t=30K$. $\Delta \omega$ is the lowest resolvable frequency scale in the calculation so it is prudent to disallow any spectral features from becoming any sharper than this scale. Therefore, we introduce the convergence factor $\eta_{min}=\Delta\omega$ which will act as a lower limit on the width of spectral features. Thus in the Dysonian form of $\rho_{\GH}$ \disp{rhog} we set $\rophi \ \rightarrow \ \rophi+\frac{\eta}{\pi}$. It is sometimes interesting to include larger $\eta$ in the calculation to imitate the effects of impurity scattering events in ARPES experiments as detailed in \figdisp{etapanel}
\subsection{Thermodynamics}
\begin{figure}[t]
\includegraphics[width=3.5in]{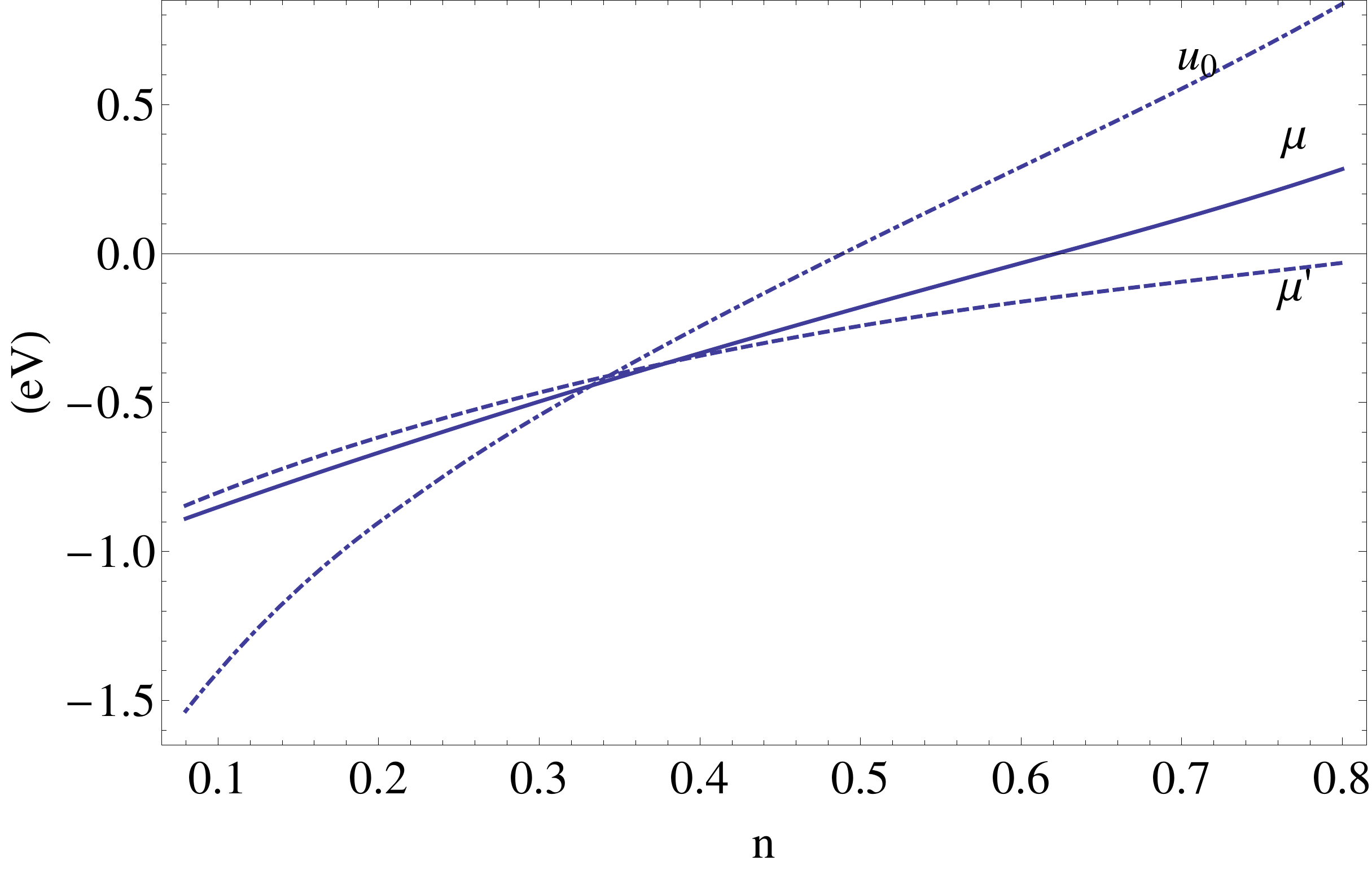}
\caption{
 $T=280K$. The density dependence of the two Lagrange multipliers, $\chem'$ and $u_0$ and the chemical potential $\chem$ for {\bf Case (A)} (dashed,dot-dashed, and solid, respectively). The auxiliary chemical potential $\chem'$ show a trend expected for a canonical FL with particle-hole symmetry approaching 0 as $n\rightarrow1$, i.e. half filling. $u_0$ rises monotonically with density indicating an increasingly asymmetric spectral function in this limit. The physical chemical potential (constructed from \disp{physmu}) goes to a positive number at high densities. This is natural for a state where spectral weight has been removed by the single occupancy constraint such that n=1 corresponds to a filled rather than half-filled band. The scale of $\chem$ and $u_0$ corresponds to the scale of the renormalized bandwidth in the high density limit.   } 
\label{MiscScales}
\end{figure}

In \figdisp{MiscScales} we display the density dependence of the two Lagrange multipliers, $\chem'$ and $u_0$,  as well as the physical chemical potential, $\chem$, (reconstructed from \disp{physmu}) for {\bf Case (A)}. The plot illustrates a few points regarding the nature of $\GH$ and $\G$, and their differences. First, we see that $\chem\approx\chem'$ in the low density limit. With increased density, $u_0$ grows monotonically causing  the development of large asymmetric tails in the spectral background. Furthermore, with the growth of both density and $u_0$, $\chem$ grows faster than $\chem'$. While $\chem'$ approaches 0, a limit expected for a half-filled FL with particle-hole symmetry, $\chem$ approaches a limit which is comparable to the renormalized bandwidth, i.e. the top of the band. This difference therefore also  signals the reduction of spectral weight in $\G$ relative to $\GH$, a symptom of the main constraint of this theory, namely the removal of states with double occupancy. 

\begin{figure}
\includegraphics[width=3.5in]{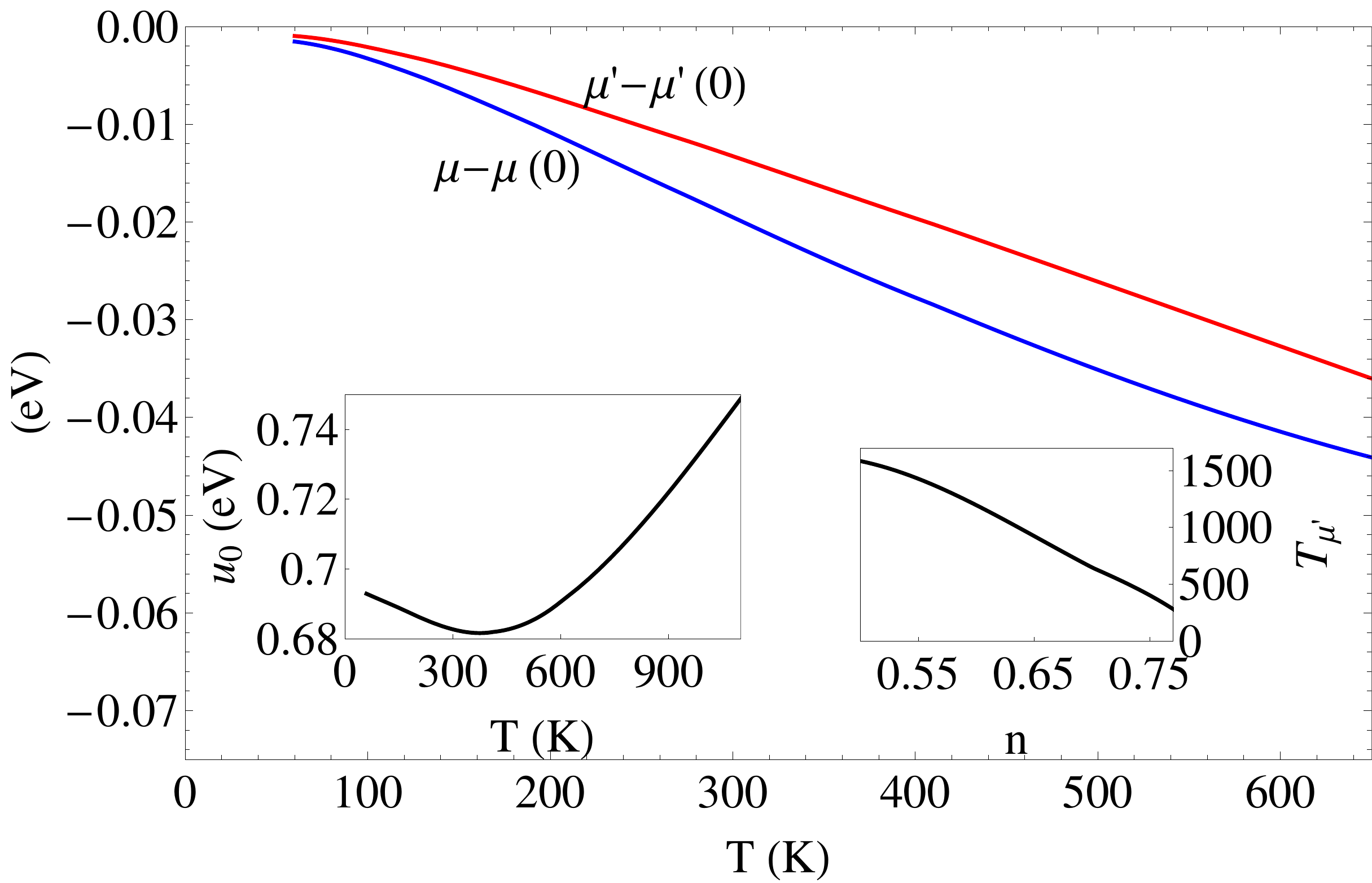}
\includegraphics[width=3.5in]{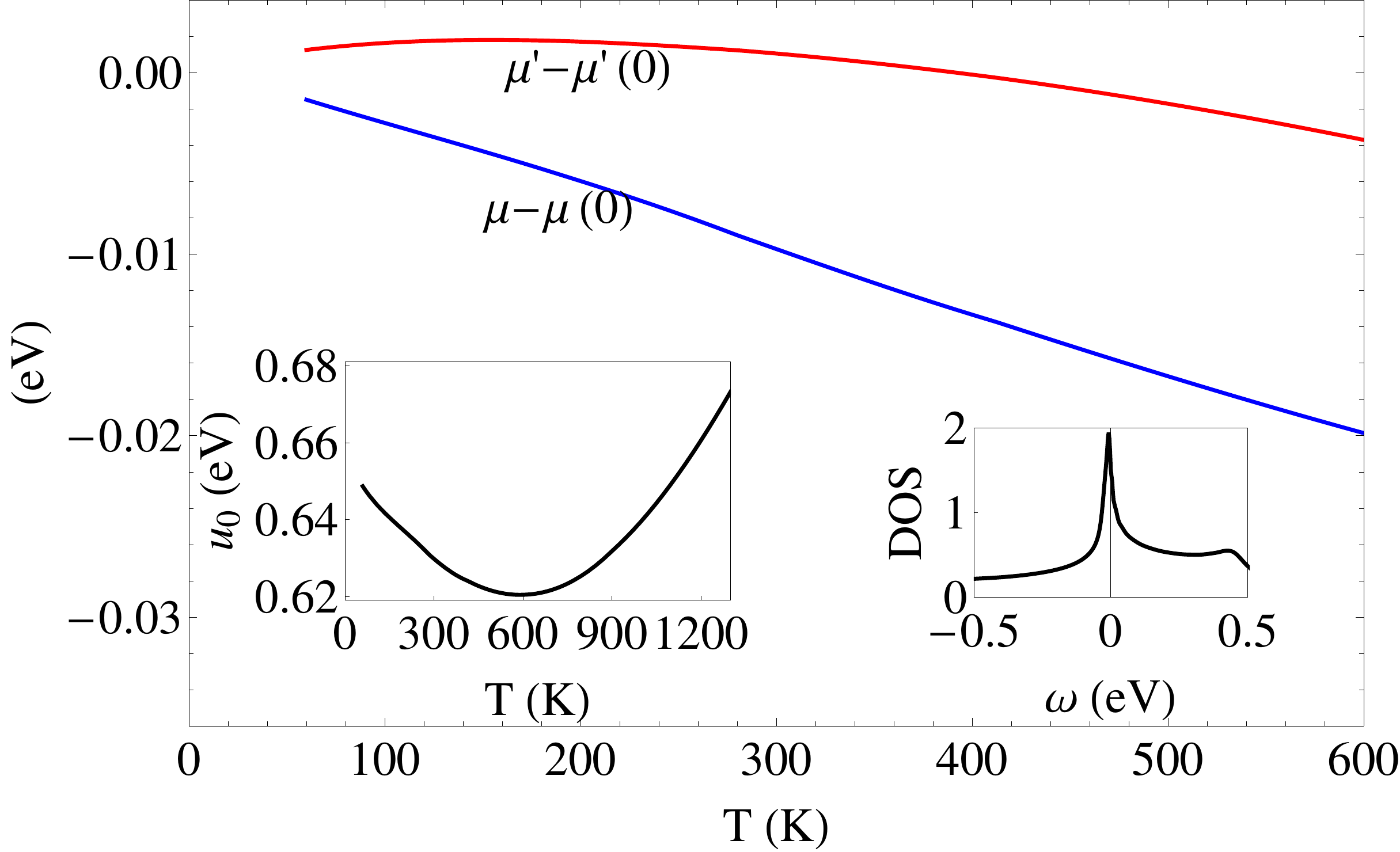}
\caption{  $\chem'$ and $\chem$  versus  T at $n=.75$ for {\bf Cases (A)} and {\bf (B)} (left and right respectively). An effective Fermi temperature $T_{\mu'}$, found from \disp{tchem},  is  shown  in {\bf Case (A)} as a function of the density. In {\bf Case (B)} the quadratic dependence is disrupted by the presence of a vHs at low frequency (the inset of DOS illustrates the vHs).  Note that the scale of variation of the $\chem'$, $\Delta \chem' \sim 10$ meV in heating from $T=100$K to $T=300$K, is quite large  and is potentially observable in ARPES. A shrunken overall energy scale, seen clearly in \figdisp{DispersionPanel}, is ultimately responsible for this sensitivity.  The temperature dependence of $u_0$ is shown in each case to be non-monotonic with a minimum at finite temperature. At high temperature $u_0$ rises linearly with T, similarly to  $\chem'$.  } 
\label{Kelvin}
\end{figure}

The temperature dependence of $(\chem',\chem,u_0)$ is shown below in \figdisp{Kelvin} revealing the temperature scale for the thermodynamics of the ECFL for {\bf Cases (A)} and {\bf (B)} at a single density (n=.75). At the lowest temperatures we expect that the chemical potential goes as $\chem'(T)=\chem'(0) - b T^2$ (to focus on the quadratic term we subtract off the $T=0$ intercepts in \figdisp{Kelvin}). We can define a temperature scale $T_{\chem'}$ which sets the strength of the $O(T^2)$ term according to 
\beq
T_{\chem'}=\sqrt{\left|\frac{\chem'(0)}{b}\right|}. \label{tchem}
\eeq
where we have used the auxiliary ``chemical potential'' $\chem'$ because it is most closely related to the FL aspects of the spectral function. {\bf Case (A)} exhibits the expected quadratic dependence at low T for all densities allowed by the $O(\lambda^2)$ approximation. $T_{\chem'}$ has been extracted and plotted in the inset as a function of the density n. However, in {\bf Case (B)}, $\chem'$ sits very close to a van Hove singularity (vHs) in the very overdoped regime. Consequently, the quadratic form of $\chem'(T)$ is disturbed and we do not attempt to extract a $T_{\chem'}$ for this case. The density of states of {\bf Case (B)} is shown in the inset to illustrate the presence of the vHs at low frequency. Note the reduction of the scale of $T_{\chem'}$ in {\bf Case (A)}  as we approach half filling, i.e. as $\chem'$ approaches a vHs. At high T ($>600K$), $\chem'$ rises linearly with T in typical FL fashion for both cases. 

We also observe in \figdisp{Kelvin} that    the temperature dependence of $\chem$ comes from $\chem'$ as well as that of $u_0$ and hence, unlike in a simple FL, is substantially driven  by many body effects  even at low T.  Furthermore, the scale of the temperature variation is larger than would be expected for weakly coupled systems. We note the large variation of the chemical potential with temperature,  $\Delta\chem\sim20meV$ on heating from 100K to 300K, and about half this number  for the change for $\Delta \chem'$. This large variation should be  be readily   measurable in ARPES, and appears to have been overlooked in most studies so far.

Unlike $\chem'$, $u_0$ does not follow the standard behavior of a FL chemical potential. Whereas the low T temperature dependence of $\chem'$ arises from small changes in occupation at low frequency only, $u_0$ feels {\em all frequencies} due to its explicit appearance in $\overline{\Phi}$ and $\Psi$. The scaling with T is therefore difficult to predict by a Sommerfeld type argument. Numerically, we observe linear-T behavior at low T and at high T separated by a minima at an  intermediate T. The scale of the minimum is greater in {\bf Case (B)}; we see that  $u_0$  asymptotes to a  T linear behavior for $T\gsim 650K$.

\subsection{Frequency independent Variables}

\begin{figure}[h]
\includegraphics[width=3.5in]{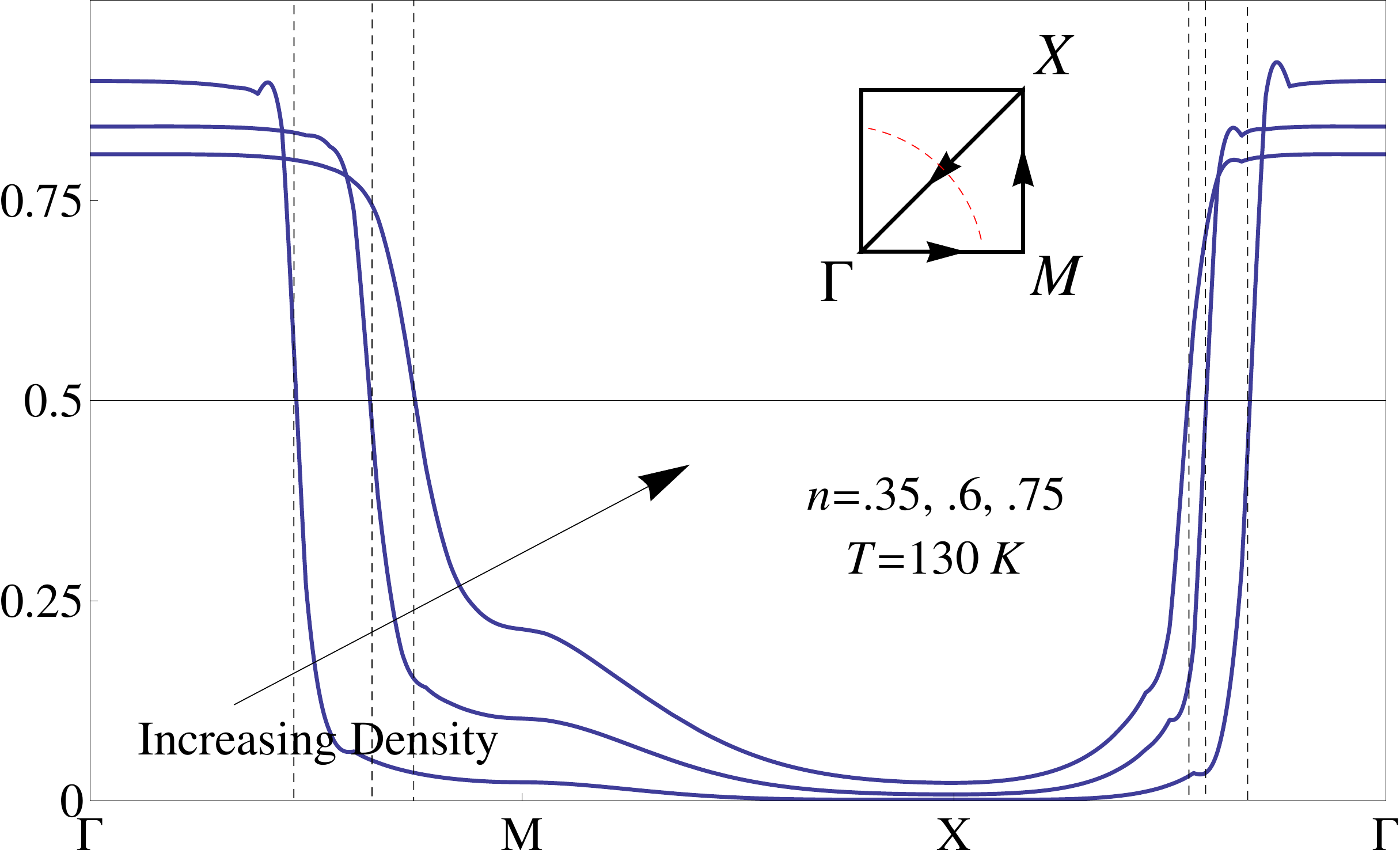}
\includegraphics[width=3.5in]{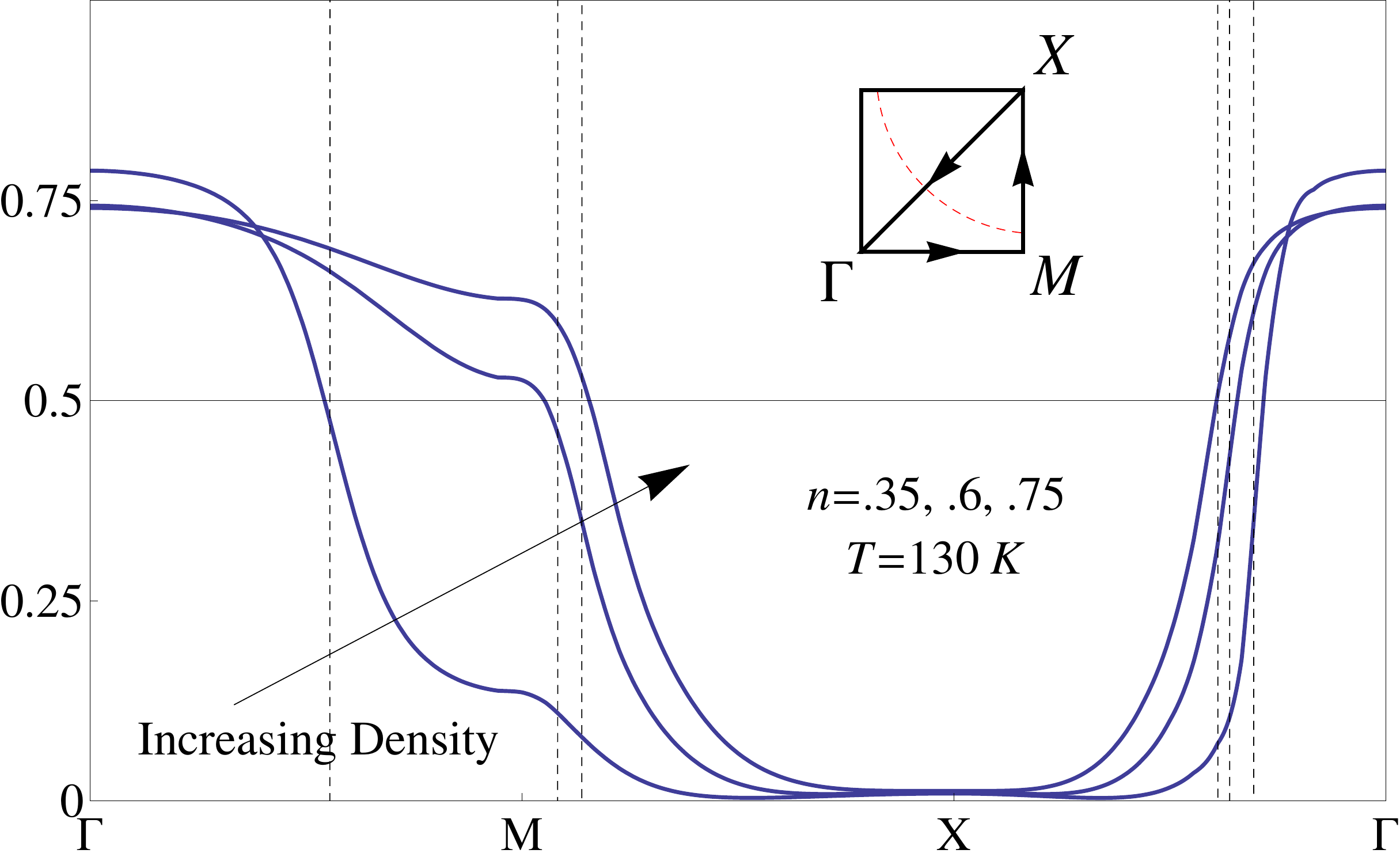}
\caption{The momentum distribution function $m_k$ is plotted along three principle lines of the BZ with $T=130K$. The panel on the left shows {\bf Case (A)} with  a FS that  is closed around $k=(0,0)$. {\bf Case (B)} is shown in the right panel and has an open FS due to the presence of the finite t'. In each case the FS is the same as in the non-interacting problem. The Luttinger Ward crossing ${\Re \ \G^{-1}(\vec{k},0)}=0$ is indicated for each density by the vertical dashed lines. For each density and each bandstructure the Luttinger Ward crossings correspond well with the condition $m_k=\frac{1}{2}$. {\bf Case (A)} shows  less variation as a function of density, because the gradient of $\varepsilon_k$ is relatively small in the vicinity of the M-point.}
\label{mklow}
\end{figure}

\begin{figure}[h]
\includegraphics[width=3.5in]{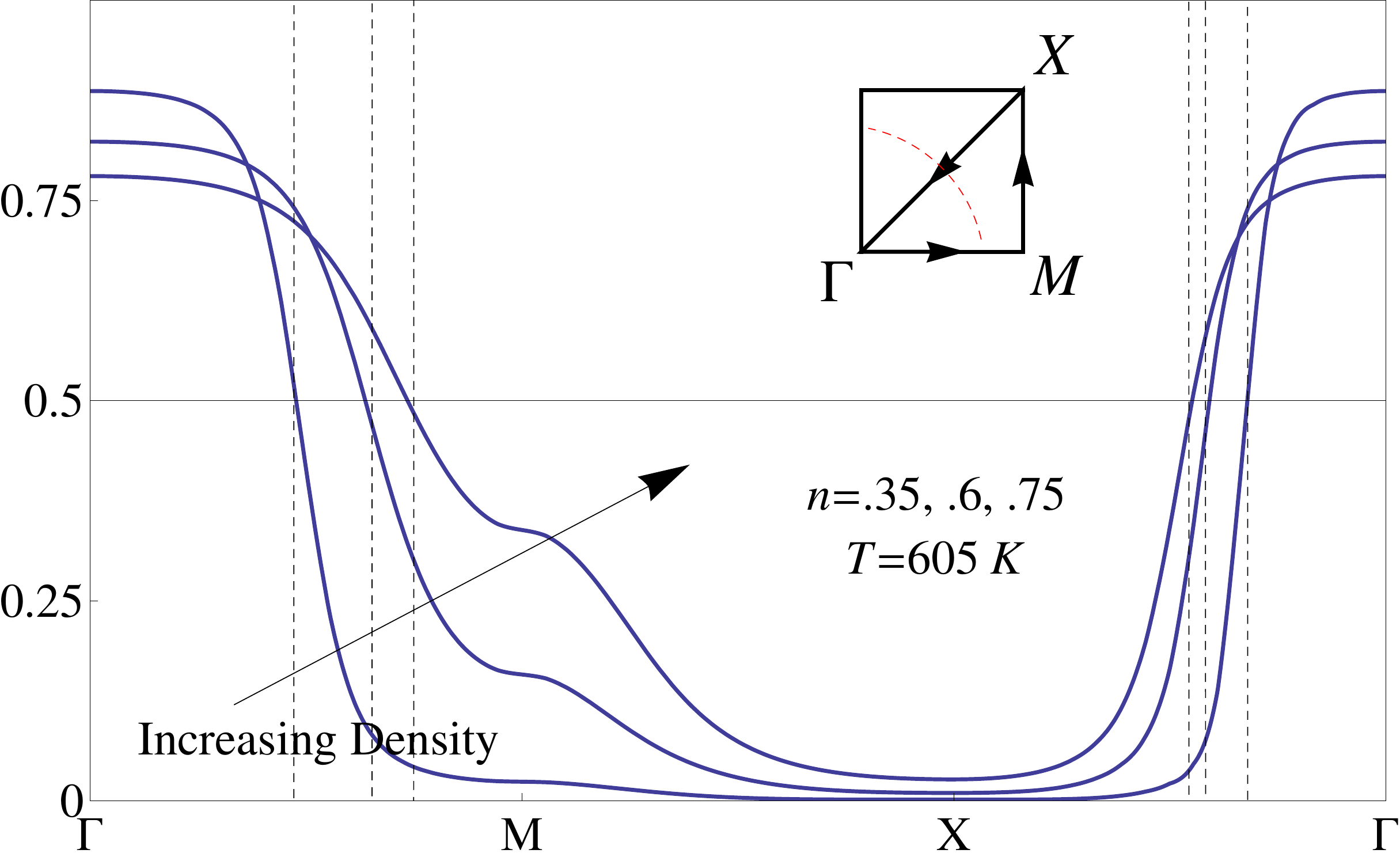}
\includegraphics[width=3.5in]{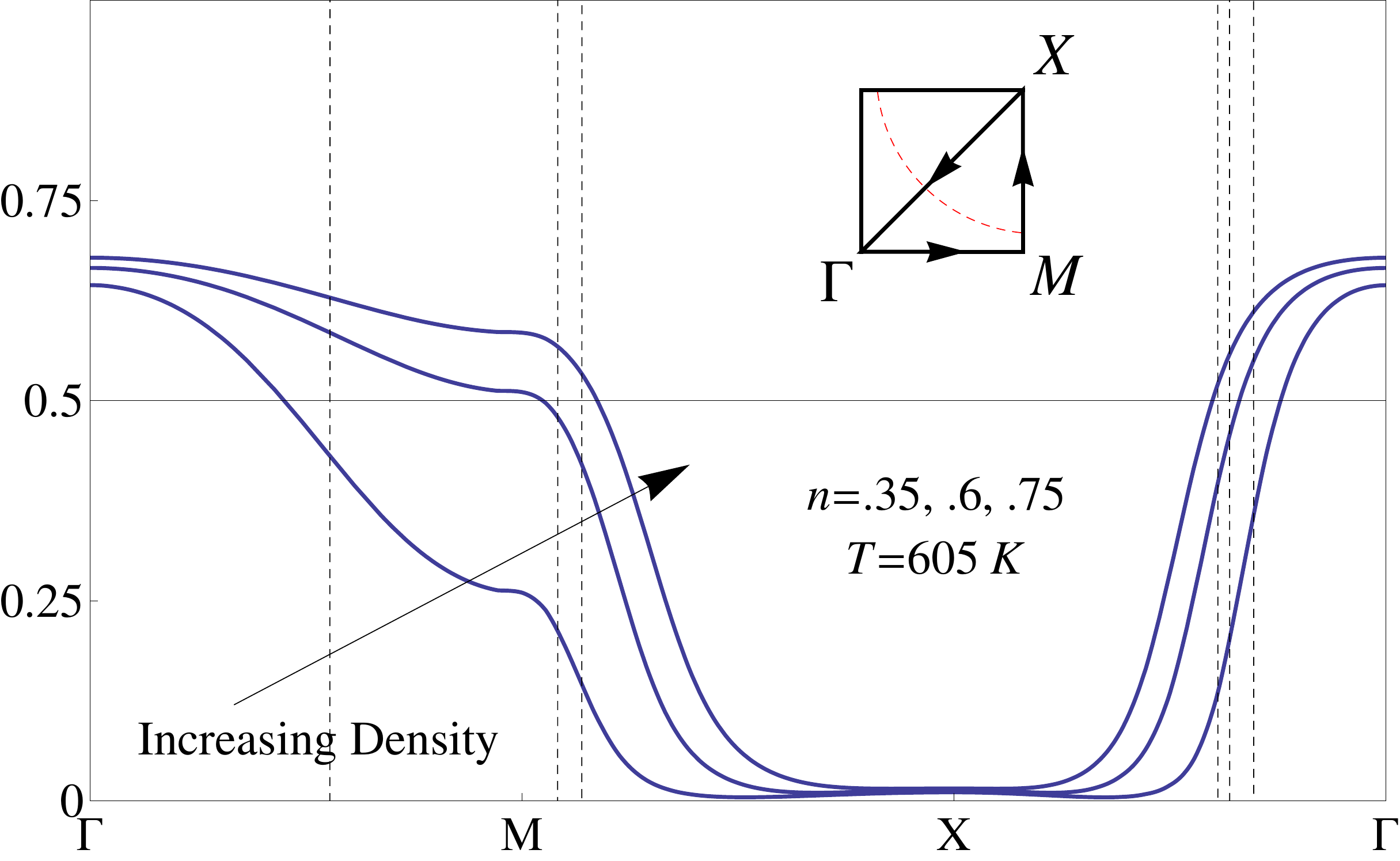}
\caption{The momentum distribution function $m_k$ is plotted along principle lines of the BZ for $T=605K$. {\bf Case (A)} ({\bf Case (B)}  ) is to the left (right). Unlike  the low  T data seen in \figdisp{mklow}, the condition $m_{k_F}=\frac{1}{2}$ is not satisfied closely in {\bf Case (B)}, although it {\em does work} to  reasonable precision  for {\bf Case (A)}.  }
\label{mkhigh}
\end{figure}

We now proceed to study the FS in this theory, starting with the momentum occupation function
$m_k$ of the Gutzwiller projected Fermions:
\beq
m_k \equiv \langle \hat{C}^\dagger_{k \si} \hat{C}_{k \si}  \rangle =\int_{-\infty}^{\infty} \rho_{\G}(\vec{k},\omega) f(\omega)d\omega. \label{mk}
\eeq
A sharp drop in this function  helps to locate the FS at low $T$. This can be compared with the Luttinger - Ward surface  defined by a sign change in $\Re \  \G(k,0)$, also  given in terms of the spectral function by
\beq
\Re \ \G(\vec{k},0)=\int_{-\infty}^{\infty}\frac{\rho_{\G}(\vec{k},\omega)d\omega}{\omega}
\eeq
At  $T=0$  the  FS in $\vec{k}$ space  is  traced out  by ${\Re \ \G^{-1}(\vec{k},0)}=0$,  as   dictated  by  the Luttinger  Ward sum rule or the volume theorem. The momentum distribution $m_k$ is plotted in \figdisp{mklow} at $T=130K$ for various densities of {\bf Cases (A)} and {\bf (B)} along three principle directions of the BZ. The Luttinger Ward zero crossings ${\Re \ \G^{-1}(\vec{k},0)}=0$ are depicted by dashed vertical lines. There is a close  correspondence between these crossings and the point where $m_k=.5$, similarly to that  noted previously by Stephan and Horsch \cite{StephanHorsch} in an exact diagonalization study. Since this correspondence is not on any rigorously firm basis, it is difficult to do more than to list the conditions for its approximate validity. Using high temperature expansions for the \tJ model Singh and Glenister  \cite{Singh} found the FS to be that of the Fermi gas  by various  criteria, and noted that the condition $m_{k_F} \sim 0.5$ is only satisfied approximately at high T.  In \figdisp{mkhigh} we   find that   at higher temperature where the QP near the FS have been significantly broadened, and thus there are deviations from $m_{k_F} \sim 0.5$ in {\bf Case (B)}, although {\bf Case(A)} has much  smaller deviations. 
In \figdisp{mklow} and \figdisp{mkhigh}, a point of considerable interest is the spillover of the occupation to the regions in k space that are unoccupied in the Fermi gas- as noted in various variational wave function studies of the \tJ model already\cite{Gros1,Gros2,mohit}.
From \disp{mk} we note that  the magnitude of $m_k$  for momenta $k > k_F$, 
   provides an estimate of the spectral weight $\rho_{\G}(k,\omega)$ {\em at occupied  energies } at low $T$. In early  analyses of ARPES data,  the significance of this piece of information was not always realized, and often  substantial spectral weight were discarded  as belonging to some unspecified background. Only recent studies such as \refdisp{Gweon} have taken note of the significance of the background.

\subsection{Quasiparticle weight and dispersion}

The spectra obtained here contain sharp peaks, and also  substantial background due to extreme correlations which can be quantified after some care is taken in defining a suitable $Z_k$.  In a conventional FL the QP weight is defined by 
$
Z_k=\frac{1}{1-\frac{\partial \Sigma}{\partial \omega}}/_{(k_F,0)}
$
where $\Sigma$ is the Dyson self energy. This definition does not immediately work for us as we do not have a conventional Dysonian form for our $\G$. To obtain an appropriate definition for $Z_k$ we note that the ECFL Greens function can be written in a Dyson-Mori form 
\beq
\G=\frac{A+\Psi}{x-\overline{\Phi}}=\frac{A}{x-\Sigma_{DM}}\label{sigma_def}
\eeq
where $A= 1- \frac{n}{2}$ exactly  and $x=i\omega+\chem-\overline{\epsilon}_k$. However  in the present approximation $A=1-\frac{n}{2}+\frac{n^2}{4}$  owing to the second order in $\lambda$ approximation.  In analogy to the standard FL we now define
\beq
Z_k=\frac{A}{1-\frac{\partial }{\partial \omega}\Sigma_{DM}}. \label{zk}
\eeq
  While it may be tempting to drop the factor  $A$  from \disp{zk}, it represents an important piece of physics in the larger context and therefore must be retained\cite{fnh}. To elaborate this, note that  
the full spectral function of canonical electrons, e.g. in a large U Hubbard model, would have features at the scale of U that correspond to the upper Hubbard band, and are thrown out in the \tJ model thereby  isolating  the lower Hubbard band.  Thus in a comprehensive  canonical theory, the (low) value of $Z_k$ representing a faint QP feature found with the present definition,    would  be compensated by a large background  piece with net  weight  $1-Z_k$ contained  partly in the lower Hubbard band, and  partly in the upper Hubbard band  that lies outside the domain of the \tJ model. It is therefore \disp{zk} that can be compared to the values found in experiments, and also in studies of the Hubbard model with large U. We will refer to $\Sigma_{DM}$ as simply $\Sigma$ later in the paper in order to simplify notation.

\begin{figure}[h]
\includegraphics[width=3.5in]{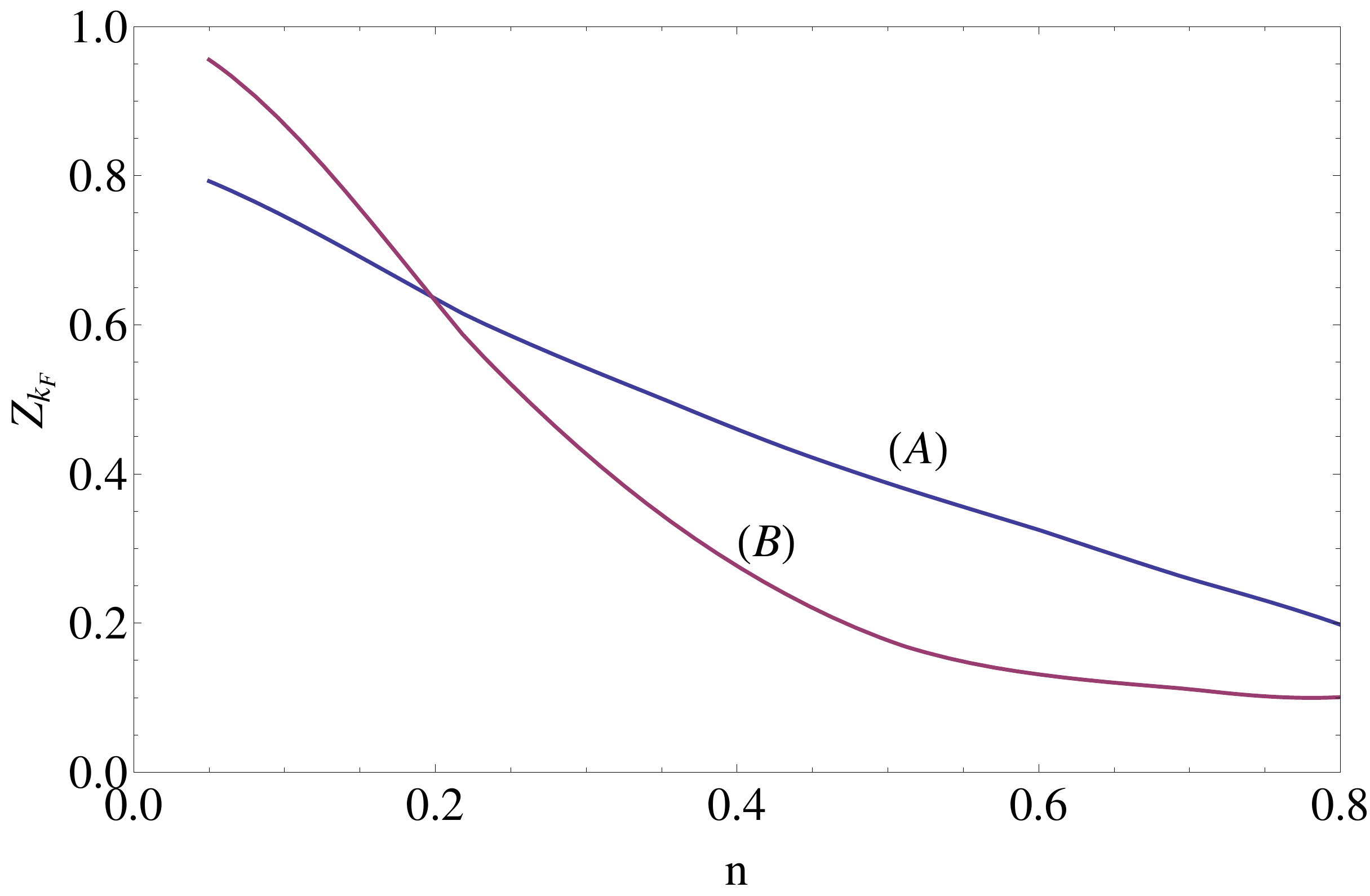}
\caption{The QP weight $Z_{k_F}$ is plotted as a function of density for {\bf Cases (A)} and {\bf (B)}. In each case, $Z_{k_F}$ decreases with increased density,  more quickly in {\bf Case (B)}, due to the vHs at $n\sim .58$. The incoherent spectral contribution is therefore already $\sim 4$ times  greater than the QP part at $n \sim .7$, and this ratio appears to increase further near half filling. }
\label{zkfig}
\end{figure}

Numerics are performed at finite temperature and the T=0 value of $\frac{\partial \Sigma}{\partial \omega}$ is obtained by extrapolating the finite T data. It is found that $Z_k$ falls quickly with density in both {\bf Cases (A)} and {\bf(B)} as plotted in \figdisp{zkfig}. At the highest densities, $Z_k$ is lower for {\bf Case (B)} but it falls more abruptly at low densities in {\bf Case (A)}. It is clear that such small values of $Z_k$ speak of enormous backgrounds in the spectral functions. Below we capture the energy dependence of these backgrounds in detail.

\subsection{Various excitation energies }
We next task is to study features in the spectral function. For this purpose    we may fruitfully study  three  functions defined in \refdisp{Anatomy},  to gauge the effect of the many body renormalizations of the dispersion relations.
\barray
\overline{\varepsilon}_k &=&  \left( 1-  \ n +  \ \frac{3n^2}{8}  \right) \ \varepsilon_k + \ \frac{1}{2} \sum_q  J_{k-q} \ m_q  \nn \\
E_k&=& \overline{\varepsilon}_k - \chem' + \Re \ \overline{\Phi}( k, E_k) \nn \\
E_k^* &=& \mbox{max}[  \rho_{\G}(k,\omega): \omega],  \label{energies}
\earray
Here $\overline{\varepsilon}_k$ defines the bare energy times  its static renormalization, while $E_k$ locates the vanishing point for the real part of the auxiliary Greens function $\GH$  defining  the Luttinger Ward surface  through  a change of sign. $E_k^*$ locates the highest peak of the physical Greens function $\G$ and hence defines QP excitations when they are sufficiently sharp.    ARPES experiments performed with constant $k$, termed the EDC's, locate $E_k^*$ as the peak locations and thus  $E_{EDC}(k)\leftrightarrow E^*_k$. Another type of experiments are commonly performed at constant $\omega$ by scanning $k$, and   are termed the momentum distribution curves (MDC's). These  motivate a fourth dispersion spectrum $E_{MDC}$ from the peak locations.  The  MDC dispersion may be defined by inverting the MDC peak locations through:
\beq
k^*({\omega})=\mbox{max}[  \rho_{\G}(k,\omega): k], \;\;  E_{MDC}(k) = \mbox{Inverse of} \ k^*({E}). \label{mdc}
\eeq
Experimentally near optimum doping $n \sim .85$, the $E_{MDC}$ and $E^*(k)$ split off from each other slightly, this is the subtle phenomenon of a low energy kink. The   fits of the  data to  phenomenological ECFL line shapes \refdisp{Gweon} capture this  phenomenon very naturally, and in this region the peaks $E(k)$ and $E^*(k)$ are also split off from each other as noted in \refdisp{Anatomy}.  In the present work, we are confined to a lower density of particles, and here this (low energy kink) anomaly is not  visible in our computed spectral functions.  However, at greater binding energies and for $k$ far from the Fermi surface, there is a second anomaly, the so called high energy kink (or waterfall) seen in 
experiments \refdisp{highkink}.   The present calculation captures it very well,
we see in this regime that  $E_{MDC}$ and  $E_k^*$, as well as  the $E_k$ and $E_k^*$ differ from each other.

\begin{figure}[b]
\includegraphics[width=3.5in]{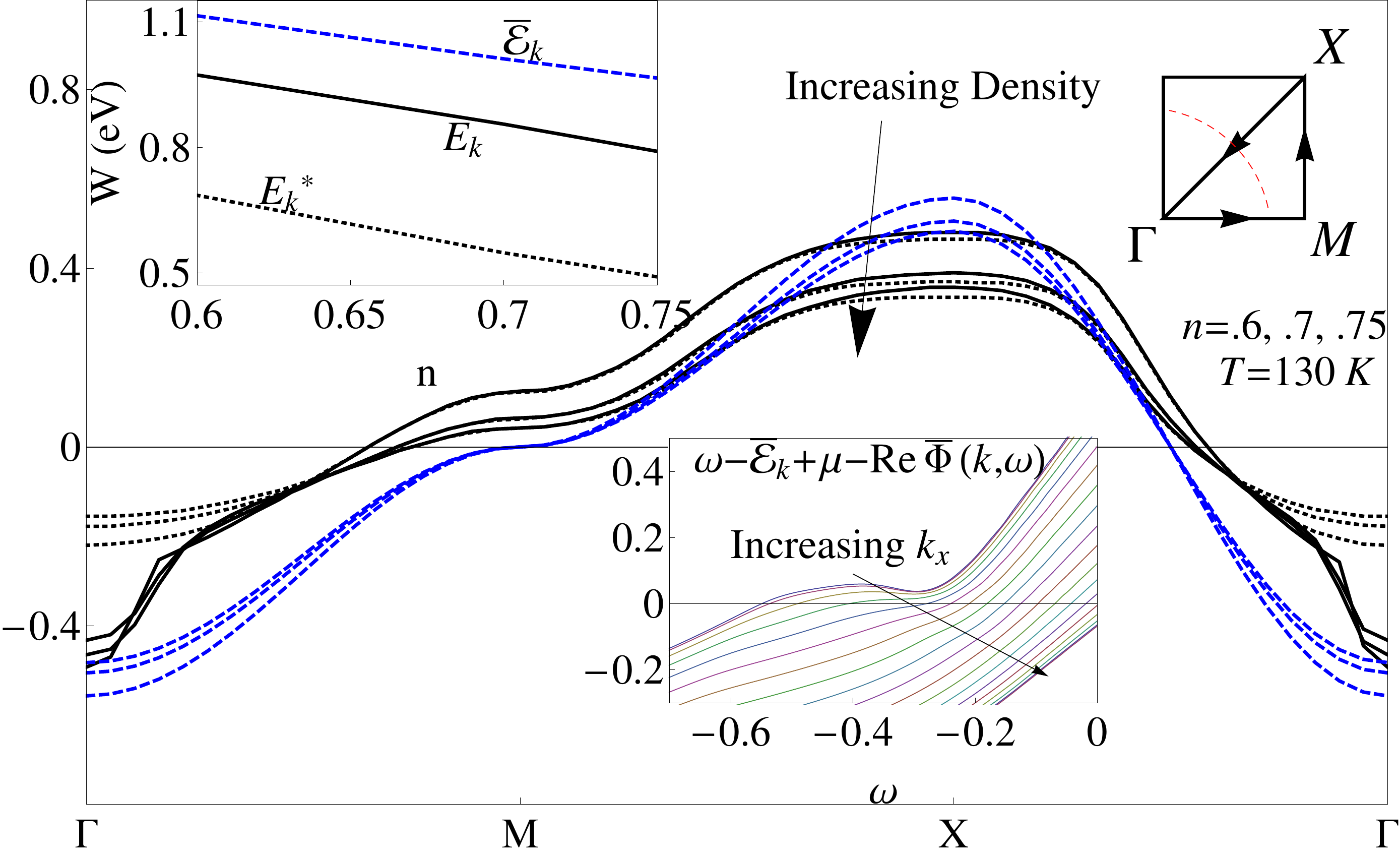}
\includegraphics[width=3.5in]{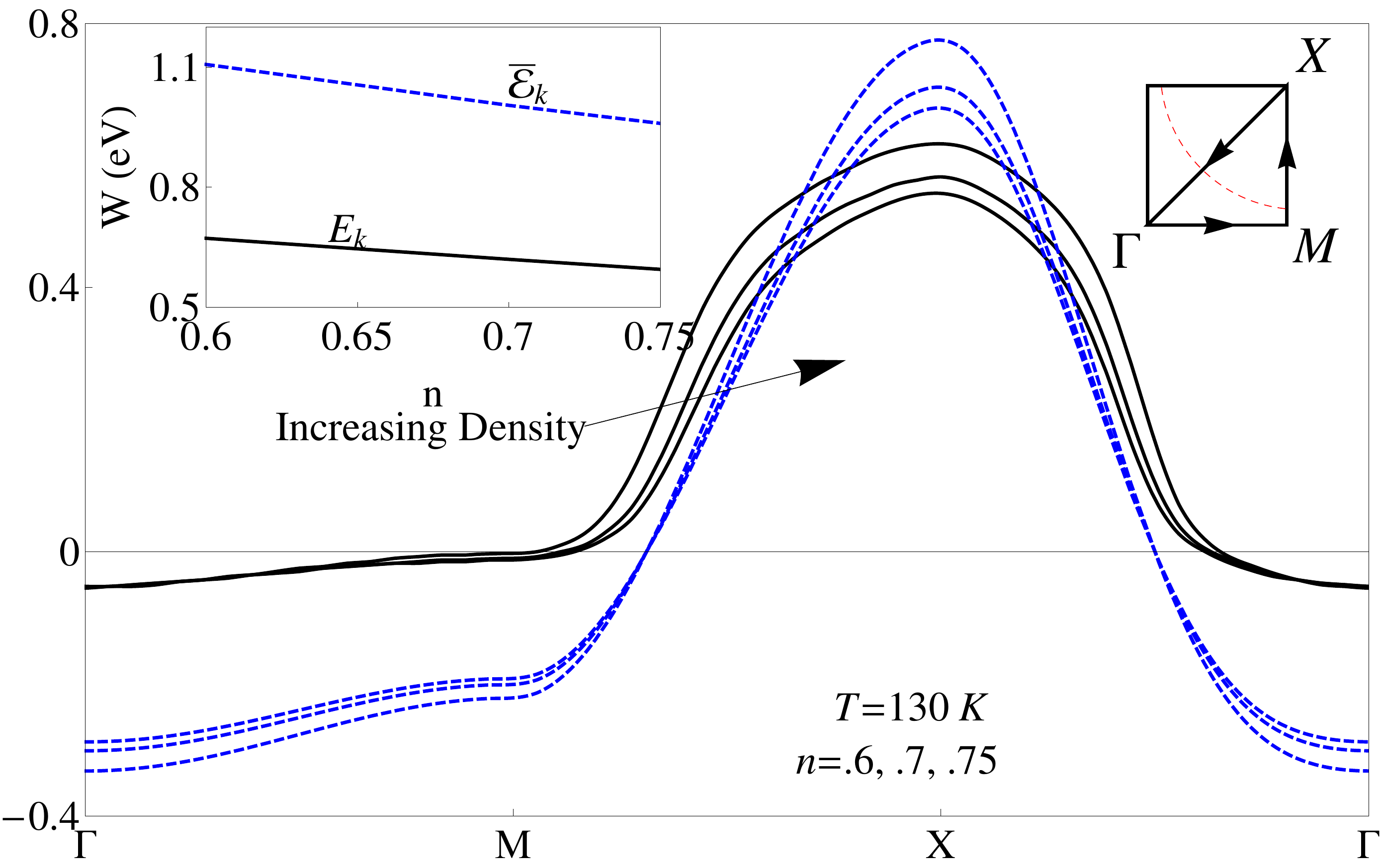}
\caption{$T=130K$.  The three dispersions defined in \disp{energies} are plotted along principle directions for three different densities. The vertical energy scale is in $eV$.  Again {\bf Case (A)} ({\bf Case (B)})  is to the left (right). The upper insets show the bandwidth of the dispersions as a function of the density. In {\bf Case (A)}, the bare bandwidth is $2eV$ but each of these dispersions shrinks compared to that scale. 
The bandwidth renormalization due to $\rephi$ in \disp{energies} is k-dependent, and  so $E_k$ has a different shape than $\epsilon_k$. Note that $E_k \sim E_k^*$ near the FS. However, in {\bf Case (A)} $E_k^*$ differs from $E_k$ near the $\Gamma$-point for each of the densities. The {\bf lower inset} on the left panel shows the evolution of real part of the denominator of $\GH(k,\omega)$ with $\omega$ to illustrate the origin of the difference between $E_k$ and $E_k^*$. In {\bf the inset} $E_k$ is determined by the zero crossings of the curves. Notice that a relatively flat feature develops at low k for which there is a shallow minimum near $\omega=-.3eV$. The minimum corresponds to the peak $E_k^*$. For increasing k, the flat feature quickly disappears and the zero crossing move quickly upward in frequency producing the observed kink in $E_k$.} 
\label{DispersionPanel}
\end{figure}

In \figdisp{DispersionPanel} we illustrate the density dependence of the  three dispersions in \disp{energies} for {\bf Cases (A)} and {\bf (B)}. The  insets shows the bandwidths, W(n), of the three dispersions as a function of the density. Note that the bare bandwidth of $\epsilon_k$ is 2eV for both cases. Near the FS we see that $E_k\approx E_k^*$ but they differ near the $\Gamma$-point where  $E_k^*$ and $E_{MDC}$ are also split off from each other, this is the operational definition of the waterfall. We now discuss the origin of these splittings.

Although $E_k$ is not experimentally relevant, it does play a significant role in the theory so we first comment on the splitting between $E_k$ and $E^*_k$ near the $\Gamma$ point, in {\bf Case (A)}.
Since $E_k$ is defined as the root of $\Re \ \GH^{-1}(k,E_k)=0$, we plot $\omega+\chem'-\overline{\varepsilon}_k-\rephi(k,\omega)$ at various $k$ as a function of $\omega$ in the inset of \figdisp{DispersionPanel}. A strong $\omega$ dependence of $\Re \ \overline{\Phi}(k, \omega)$ causes a flattening of the curves near the zero crossing between $-0.6 $ and $-.3$ eV, and this causes the $E_k$ to fall rapidly with $k$ in the main figure \figdisp{DispersionPanel} for  {\bf Case (A)}.

\begin{figure}[h]
\includegraphics[width=5.0in]{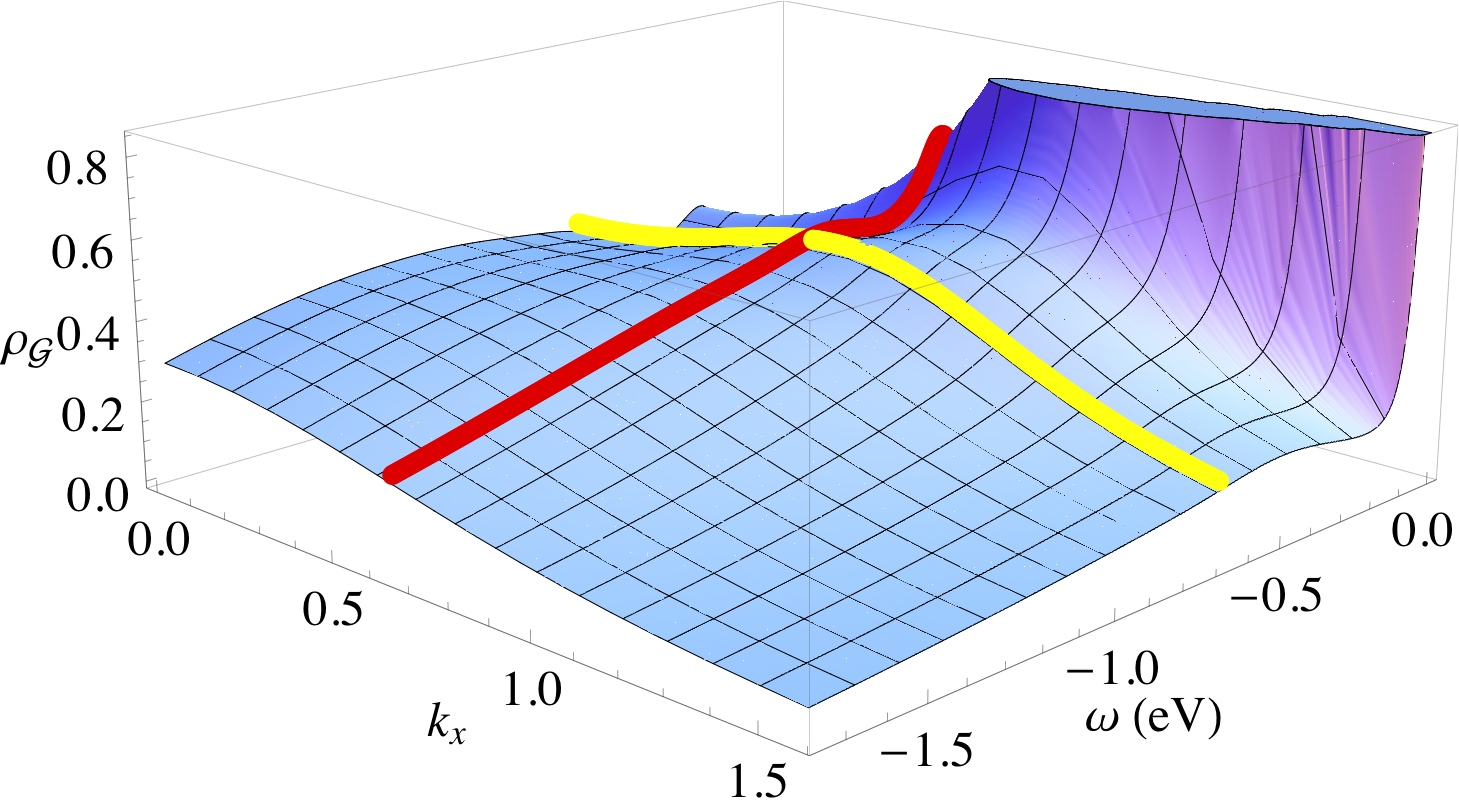}
\caption{ A typical ECFL spectral function. An MDC slice is shown in yellow and an EDC slice in red. The highlighted slices illustrate the phenomenon of the waterfall or the high energy kink. Near $(k_x,\omega)=(1.5,0)$ the QP band is strong and sharp, and the background has little weight. At lower momentum the QP peak has lost much of its weight to a broad background and a weak secondary peak, visible close to the intersection of the two cuts.  At a fixed $k_x$, the energy scans are dominated by the QP peaks near $\omega\sim 0$. The fixed energy cuts however, are sensitive to the broad secondary peaks of the background, and thereby obtain a different peak position than the EDC slice. In the region where EDC and MDC peaks disagree, the QP peaks becomes much weaker than those seen near the Fermi level. } 
\label{wfzoom}
\end{figure}

\begin{figure}[h]
\includegraphics[width=3.50in]{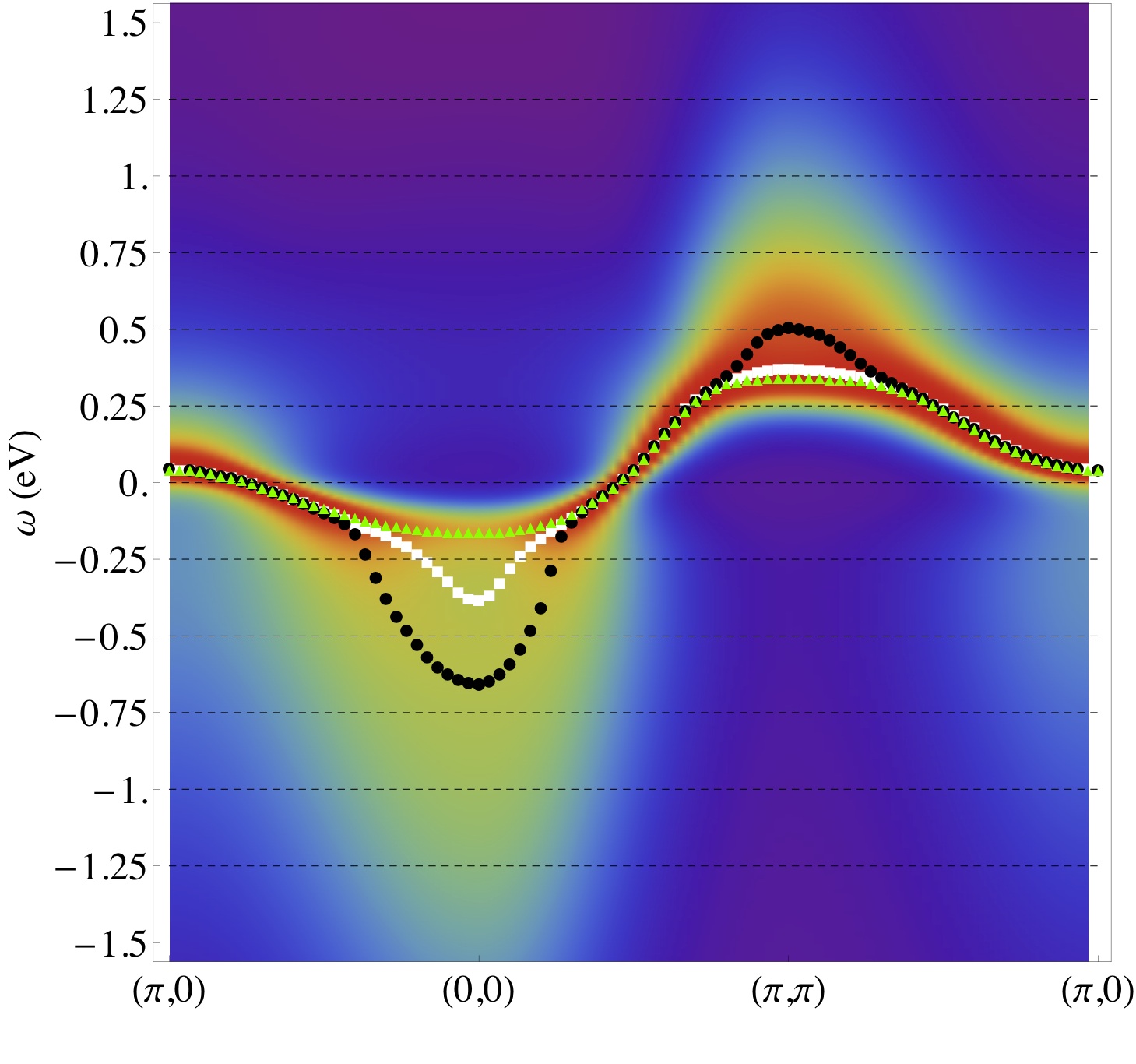}
\includegraphics[width=3.50in]{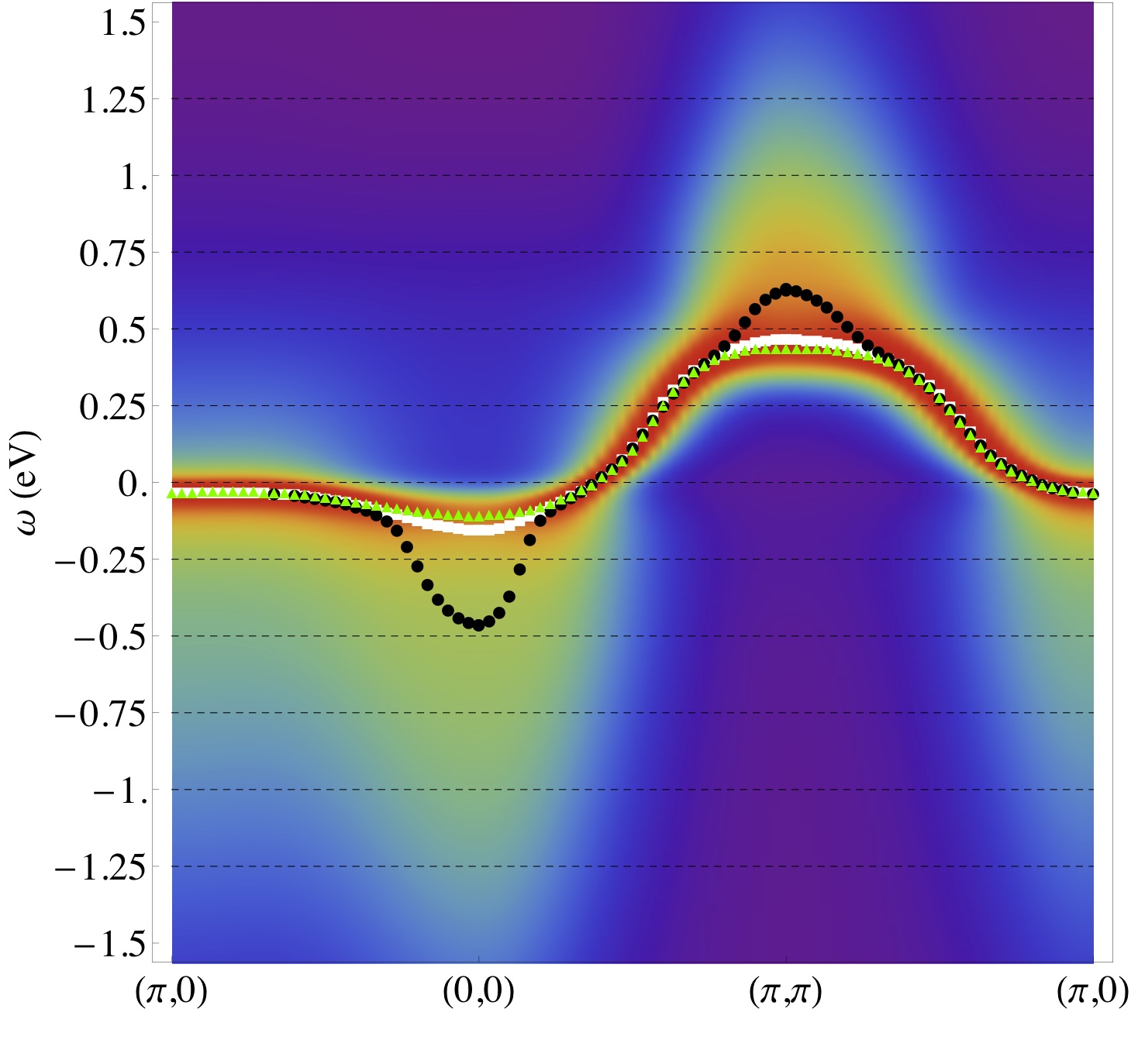}
\caption{$L=60$, $(n,T)=(.75,300K)$. The spectral function of {\bf Left:  Case (A)} and {\bf Right:  Case (B')}  is presented as a density plot to illustrate the structure of the background spectrum.  We have overlaid  the $E_k$, $E_k^*$, and $E_{MDC}(k)$ spectra  as white, green and black points, respectively. In the low frequency region near $k_F$ we see that the MDC, EDC, and $E_k$ peaks coincide. Of particular interest however, is the region near $k=(0,0)$ where the MDC peaks occur at a significantly higher energy scale than $E_k$ or $E_k^*$. Furthermore, in this region the EDC peak loses weight. Notice that the red peaks of the density plot, representing a sharp QP band, fade out to yellow near the $\Gamma$-point. This pronounced difference between the EDC and MDC peaks indicates that a waterfall-like phenomenon is present in this theory. There is also a new feature at positive frequency near $k=(\pi,\pi)$ which looks like an inverted waterfall, and is explored further below in \figdisp{wfpanel}. In the curve on the right with {\bf Case (B')},
 inclusion  all of the in-plane hopping parameters quoted for BSCCO in \refdisp{Bansil} recovers a waterfall feature for a FS which is nonetheless hole-like. On the other hand {\bf Case (B)} does not display the waterfall, as \figdisp{wfpanel} shows.  The essential difference between the full BSCCO dispersion and the  {\bf Case (B)} is the magnitude of the curvature near the $\Gamma$-point. This suggests that  the band curvature  determines  the magnitude of the waterfall. } 
\label{waterfall}
\end{figure}

To understand the splitting between the MDC and EDC dispersions,  \figdisp{wfzoom} illustrates how the waterfall phenomenon arises as a consequence of an EDC curve with {\em two maxima}, one somewhat broader and flatter  than the other. The EDC dispersion will favor the sharper peak created by the QP pole while the lesser peak in the background will be overlooked. The MDC curve has no such prejudice toward the QP.  In \figdisp{waterfall} the spectral function for {\bf Case (A)} and  {\bf Case (B')}  is depicted as a color density plot with the dispersions $(E_k,E_k^*,E_{MDC})$ overlayed. Near the $\Gamma$-point where $k=(0,0)$ the QP becomes incoherent and the bulk of its spectral weight is spread out to high negative frequencies. In this region $E_{MDC}$ differs considerably from $E_k^*$, and recovers the scale of the bare dispersion $\epsilon_k$. 


\begin{figure}
\includegraphics[width=3.5in]{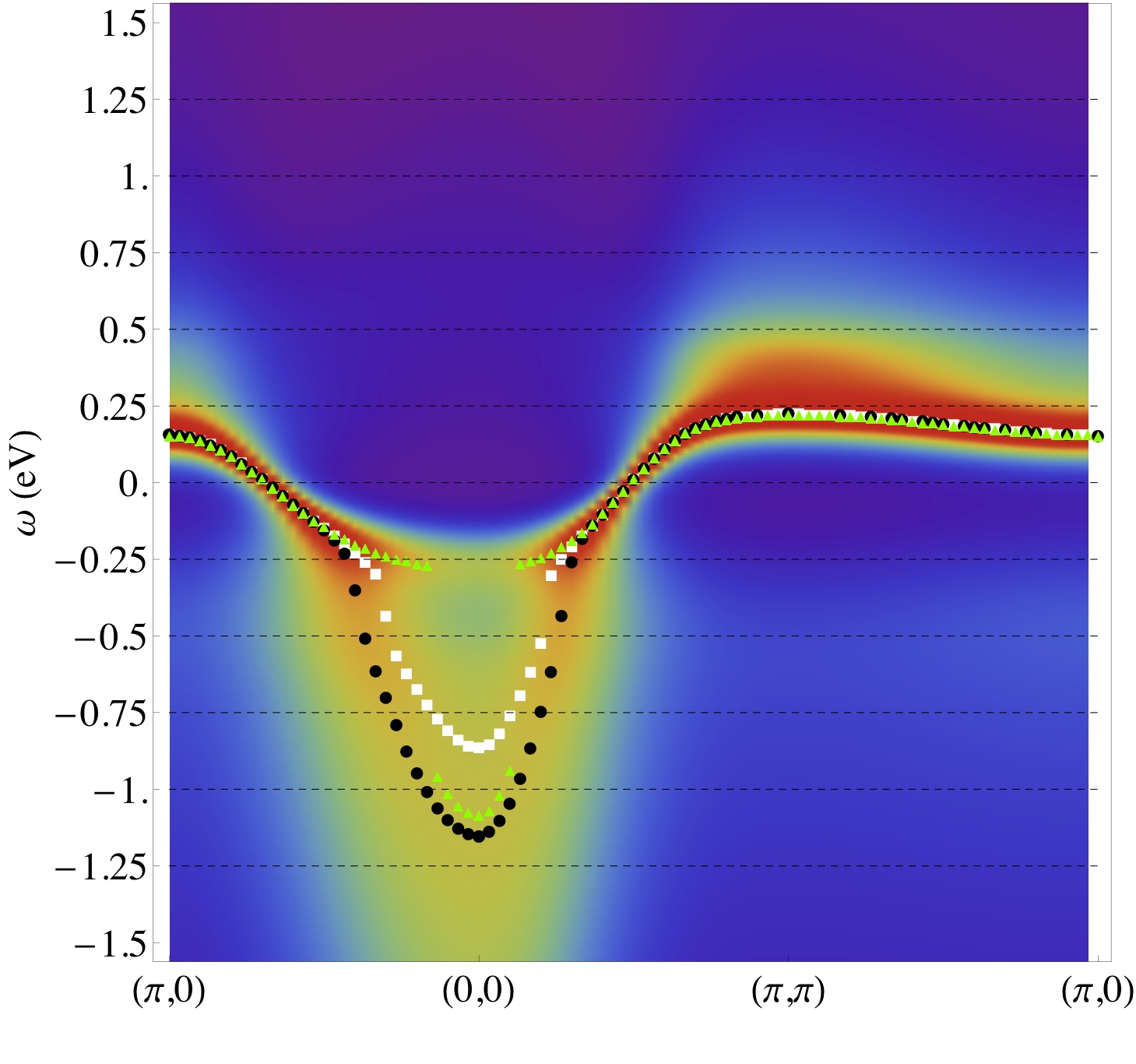}
\includegraphics[width=3.5in]{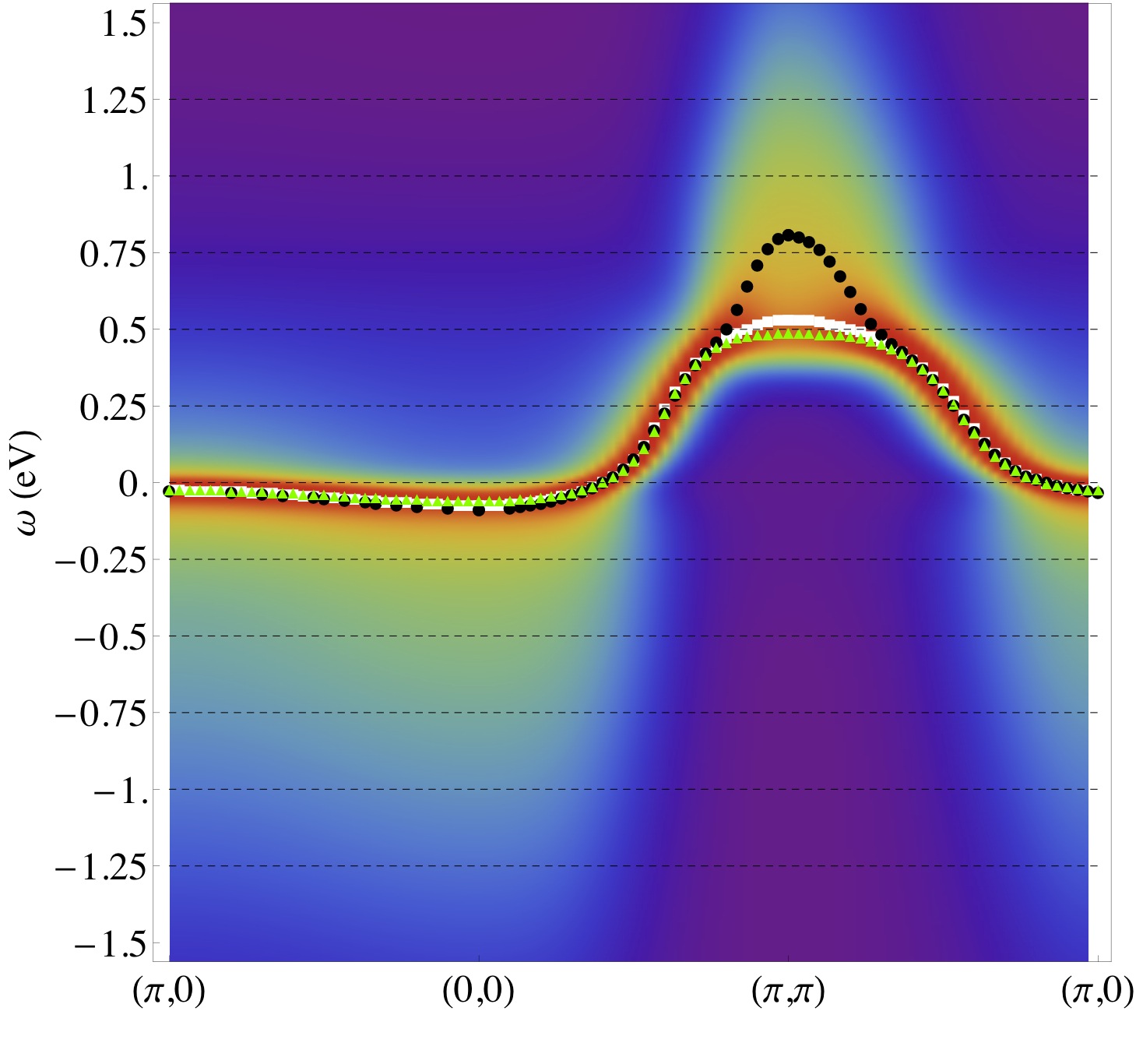}
\caption{$L=60$, $(n,T)=(.75,300K)$.  Left  {\bf Case (C)} $t'/t=.4$ is used to model electron doped High Tc superconductors, while at right  is {\bf Case (B)} $t'/t=-.4$. These demonstrate that the background and the kink features are  sensitive to the bare band parameters. In these plots we overlay three dispersions to help with the visualization of the kinks,   the  $E_{MDC}$, $E_k$, and $E_k^*$  in black, white, and green, respectively. A positive $t'/t$ ({\bf Case (C)}  results in the largest binding energy scale for the kink anomaly  near the $\Gamma$-point. This feature shrinks as $t'/t$ is decreased. Therefore this calculation suggests that the relative magnitude of the waterfall in hole and electron doped cuprates is a consequence of the sign of $t'/t$. Simultaneously, a reciprocal kink at positive frequency occurs around $k=(\pi,\pi)$ with increasing magnitude for more negative $t'/t$.} 
\label{wfpanel}
\end{figure}

Within this theory we can explore various parameter dependences of the waterfall anomaly. In \figdisp{wfpanel} we show density plots of the spectral function with negative and positive $t'/t$. The case $t'=-.4 t$ corresponds to {\bf Case (B)}, while a  third {\bf Case (C)} has  $t'/t=.4$. {\bf Case (C)} has  been identified with the phenomenology of the electron doped cuprates.  In these plots we can see two major trends in the development of the high energy features of the spectral function. Firstly, in {\bf Case (C)} both the kink frequency and the bottom of the waterfall happen at approximately twice the frequency seen in {\bf Case (A)}. This is roughly in accord with experimental data at various dopings \cite{Park}. Furthermore, while a remnant of the QP band remains intact in {\bf Case (A)}, sharing weight with features at the kink scale, the QP peaks of {\bf Case (C)} lose all of their weight to the kink scale features. On the other hand, {\bf Case (B)}, in contrast to {\bf Case (A,B')},   has no measurable waterfall near the $\Gamma$-point. The background at negative frequency is essentially featureless and the QP peaks maintain their spectral weight. However, at positive frequencies, an inverted waterfall like feature develops near $k=(\pi,\pi)$ as $t'/t$ is decreased. {\bf Case (B)} shows the best example of this inverted waterfall. 
{\bf Case (A)} is an intermediate example where both features are present.

Comparing   {\bf Cases (A,B')} in \figdisp{waterfall} showing the waterfall,  and {\bf Case (B)} in \figdisp{wfpanel} without one near the $\Gamma$ point, we deduce that  the magnitude of the waterfall is correlated with the curvature of the dispersion. Regions where the QP band is relatively flat show agreement between $E_{MDC}$ and $E_{EDC}$. At much lower densities $(n=.35)$ the waterfall is yet visible, albeit with a decreased magnitude. The experimental observation \cite{Park} that the electron-doped compounds have a  larger magnitude of the  waterfall relative to the hole-doped  cuprates, also finds  a natural  explanation in this ECFL calculation via the dependence on  band parameters in  the \tJ model, as seen in \figdisp{wfpanel}.   The curvature is increased  by having a large and positive ratio $t'/t$, resulting in 
the  greater jump seen here.  A  similar result for  the  Hubbard model using  quantum Monte Carlo methods is reported in   \refdisp{Moritz}.  On the other hand, a large and negative $t'/t$ flattens the bottom of the tight-binding band
at the $\Gamma$ point, as in  {\bf Case (B)}, while increasing it at the $X$ point with $k= (\pi,\pi)$. Thus the waterfall is removed from the former and reappears at the latter point.

\begin{figure}[h]
\includegraphics[width=2.5in]{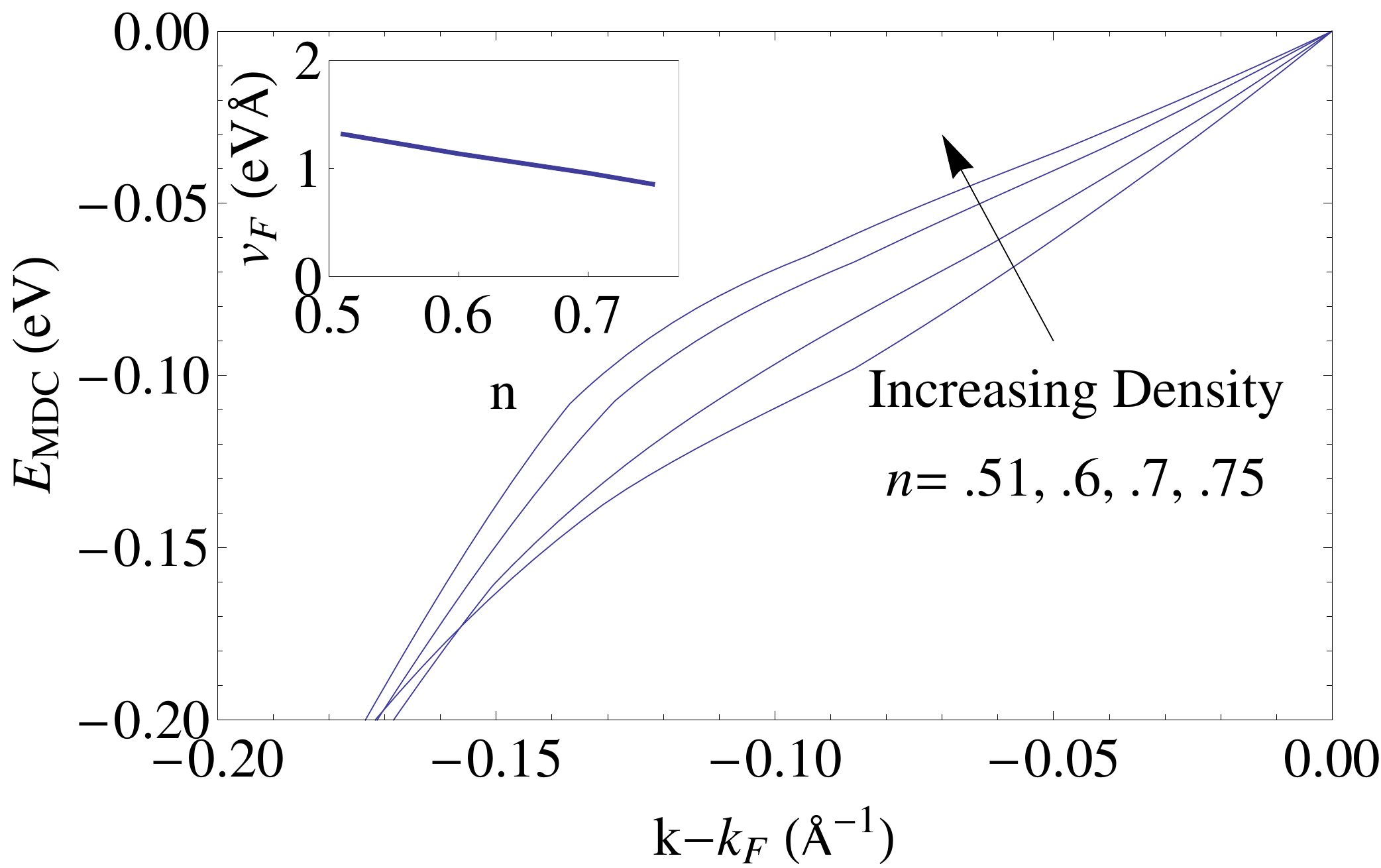}
\includegraphics[width=2.5in]{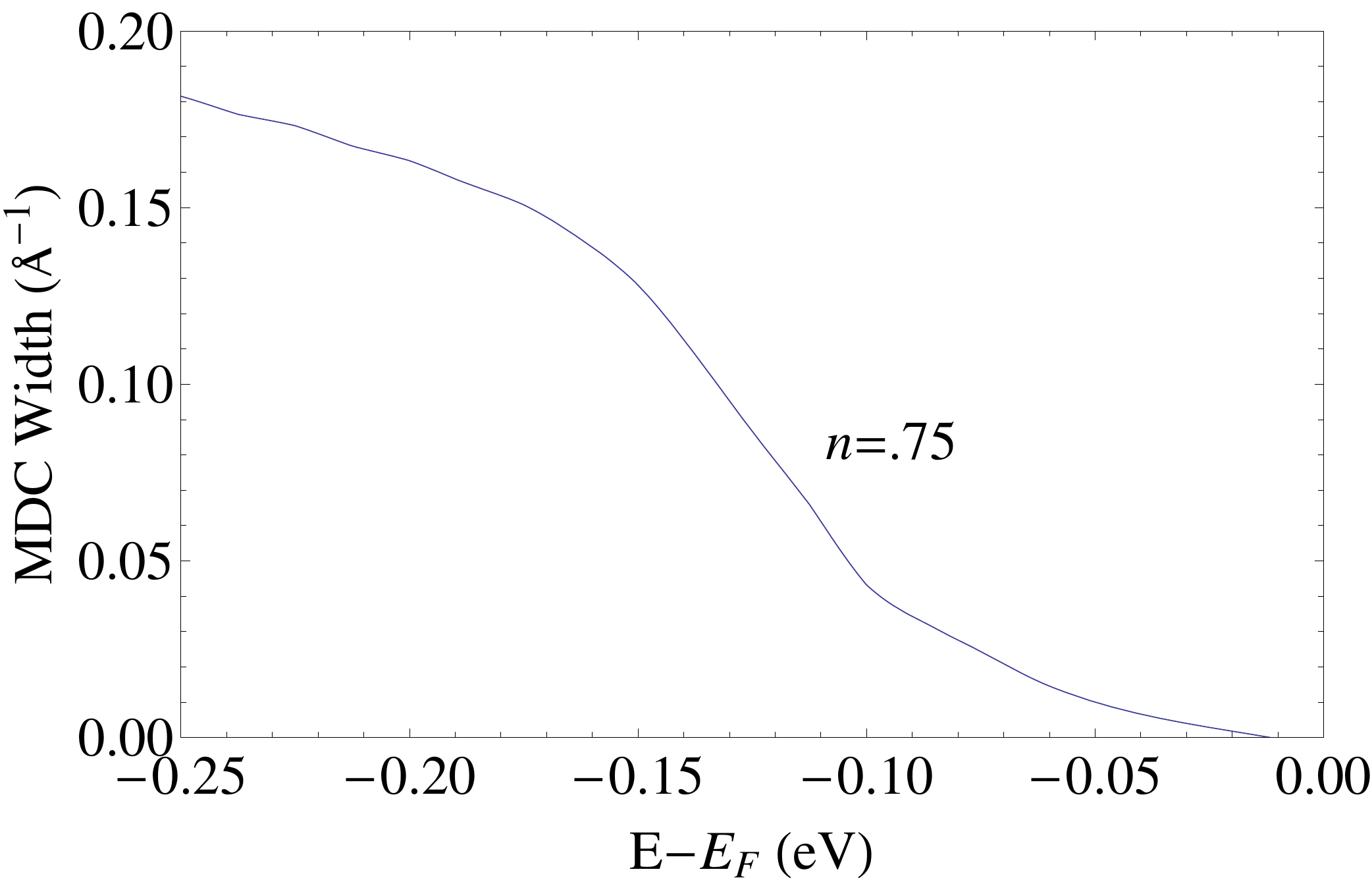}
\caption{ The computed Fermi velocity in the present scheme {\bf Case (A)} has a slight  density dependence in the range studied. Furthermore, the kink becomes more pronounced  at higher densities, a trend which is seen in  the experimental literature\cite{zhou}. The $\omega$ dependence of the MDC width is also shown for a single density. At the frequency scale associated with the kink, the MDC peaks broaden abruptly, indicating that this feature  does not have the character of single particle excitations.} 
\label{velocities}
\end{figure}

The density dependence of the kink scale and Fermi velocity along the nodal direction have been recently studied in \refdisp{zhou}. The nodal $v_F$ in the cuprate compounds is found to have a very weak density dependence for hole dopings $x\lsim.25$, leading to the suggestion of a universal Fermi velocity. Our calculated  nodal QP dispersion for {\bf Case (A)} is plotted in \figdisp{velocities} for several densities on the overdoped side (results for {\bf Case (B)} are similar although the kink is absent). The nodal dependence is presented because it is always available to experiment, even in the superconducting phase, whereas other parts of the BZ may become obscured by the superconducting gap. The universal $v_F$ seen in ARPES, $v_F=1.8\pm.2eV$\AA \cite{FermiVel} exceeds our computed value for {\bf Case (B)} for the same object by a factor of about 2 at $n=.75$. This discrepancy is similar to that in several computed scales in our calculation, and could  be adjusted by varying  the magnitude of $t$. On the overdoped side, where our calculations are performed, we find that  $v_F$ has a  slightly greater  density dependence than seen at higher densities in experiments. More encouraging is the broadening of the MDC width for frequencies greater than the kink scale. This is shown in the right panel of \figdisp{velocities} and compares well with a similar figure in \refdisp{zhou}.

\subsection{Detailed Spectral Lineshapes (EDCs)}
In this section, we present detailed lineshapes for the spectral function. In an earlier work\refdisp{Gweon}, we have compared the results of the simplified ECFL formalism,  including some phenomenological inputs, with the experimental data at somewhat higher particle densities $n \sim 0.85$, and  found remarkably good  agreement with  the  lineshapes. We are content in this work to present the results at lower particle densities, but {\em  from a microscopic calculation} of ECFL, without any phenomenological input. This is   made possible  by the solving  $O(\lambda^2)$ equations numerically. The lineshapes obtained here have a  similar  general nature as the ones in \refdisp{Gweon}, giving support to that work. However, as one expects from a lower density situation, we find somewhat less dynamical asymmetry about zero energy. More detailed comparison with data near optimal doping with the microscopic ECFL theory must await the solution of the third or higher order equations where the criterion for validity discussed above (see para following \disp{sumrule-2}) is satisfied more closely than here.  

\begin{figure}[h]
\includegraphics[width=2.5in]{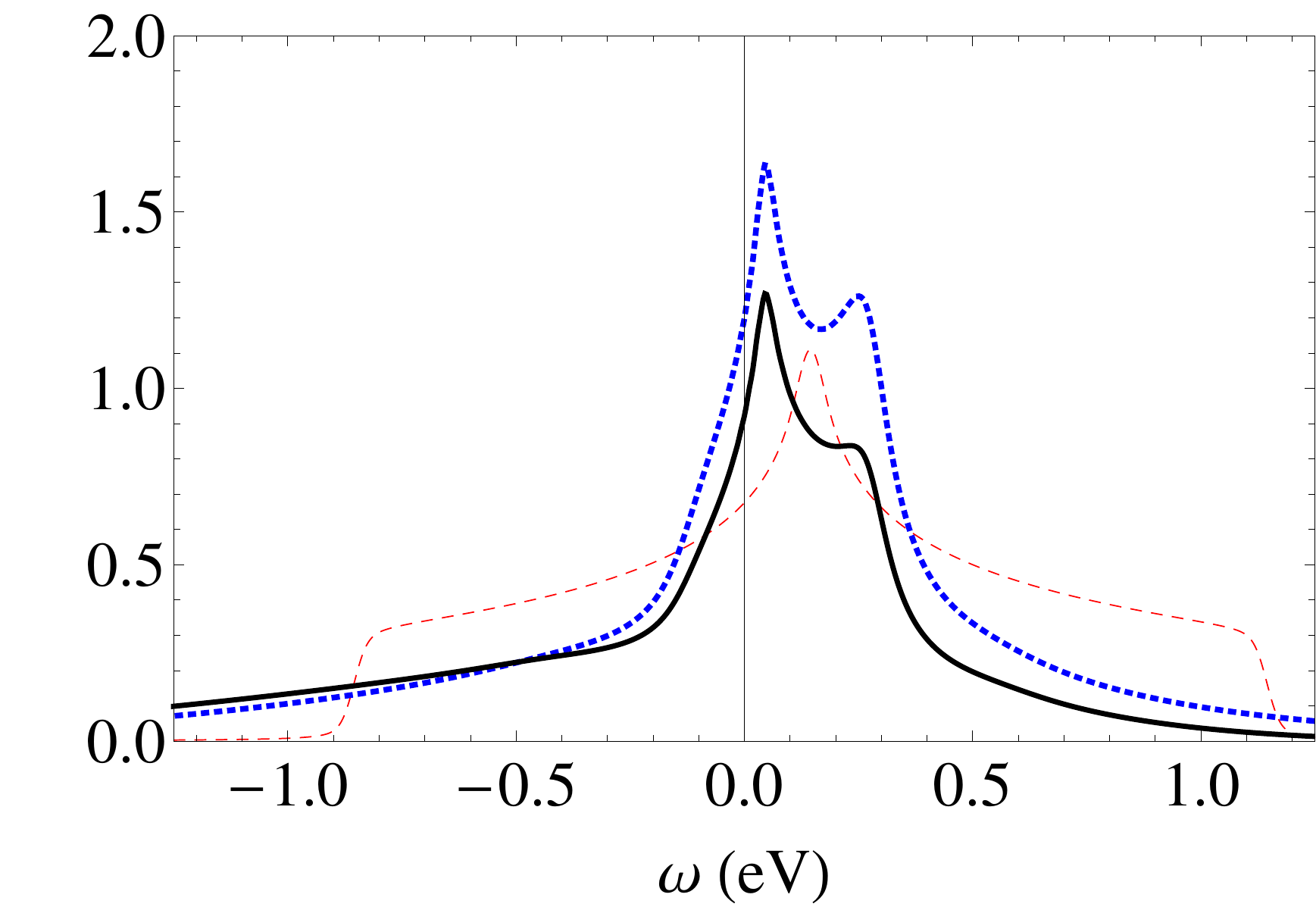}
\includegraphics[width=2.5in]{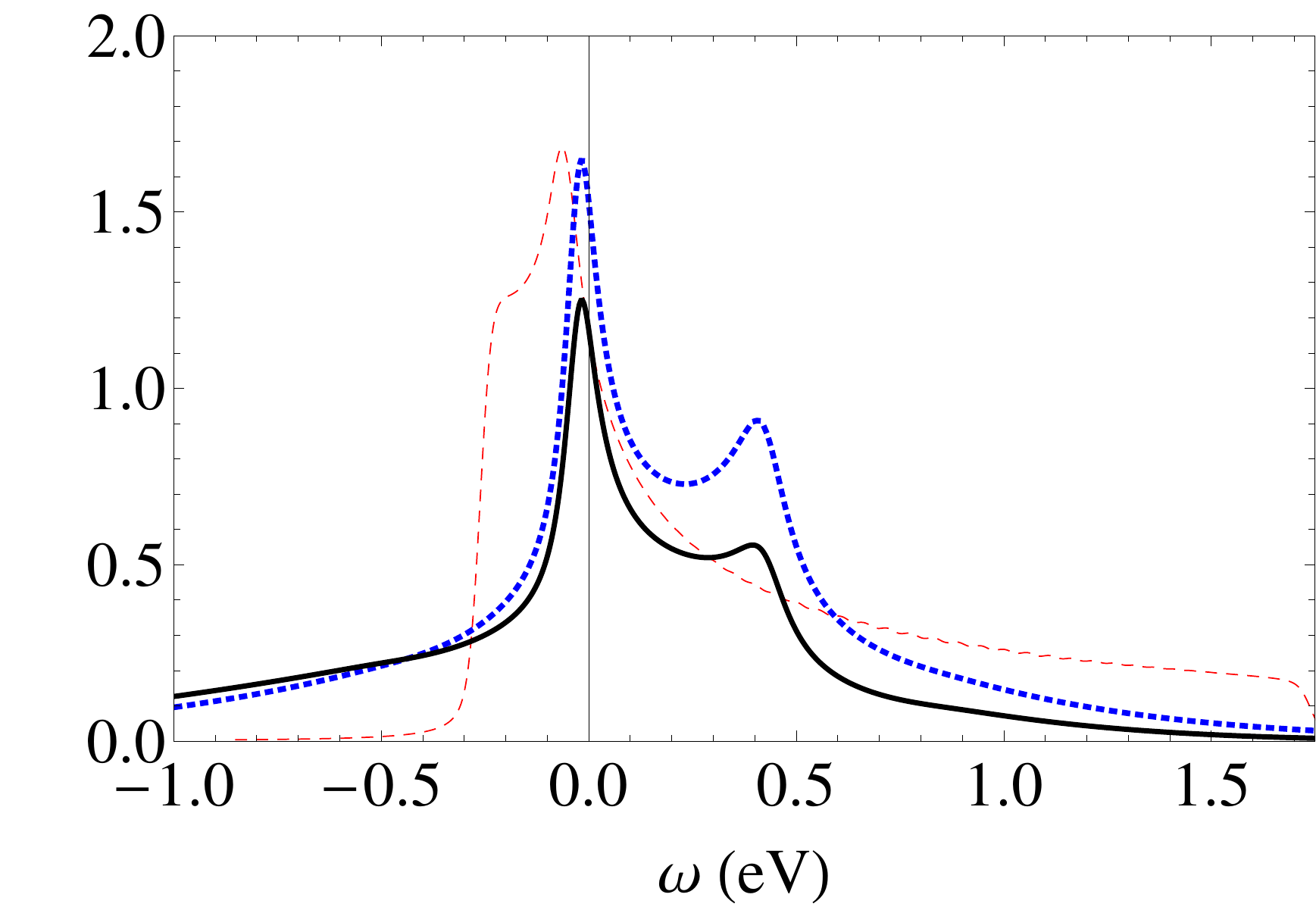}
\caption{
The  the local density of states (LDOS) for {\bf Cases (A)} and {\bf (B)} (left and right) with $n=.75$. The LDOS of the physical $\G$ is shown in black while the bare DOS is the dashed red curve. The LDOS of the auxiliary $\GH$ is shown as a dotted blue curve. The QP bandwidth is considerably  narrower,  and in each case there is a long tail extending to large negative frequency. Furthermore, it is clear that the tail of $\G$ decays more slowly (quickly) than $\GH$ for large negative (positive) frequencies. This arises from the asymmetry produced by the caparison factor. The LDOS also develops a second spectral peak which arises from a strongly k-dependent feature in the self energy.} 
\label{dos}
\end{figure}

Let us first  examine the local density of states (LDOS) at $n=0.75$ for both cases at low T in \figdisp{dos}. We note again that the Fermi energy is rather close to the peak in the LDOS for {\bf Case (B)} at these fillings, and this proximity influences the results for this case significantly. In both cases, however, there are several shared features. Note that the width of the main peak is much narrowed compared with the bare LDOS. There is furthermore a long tail which extends to frequencies much greater than those seen in the bare LDOS. Finally we note that in each case the LDOS acquires a second peak at positive frequency. This peak arises due to some k-dependent features in $\rho_{\Sigma}$ (discussed below).

\begin{figure}[h]
\includegraphics[width=2.5in]{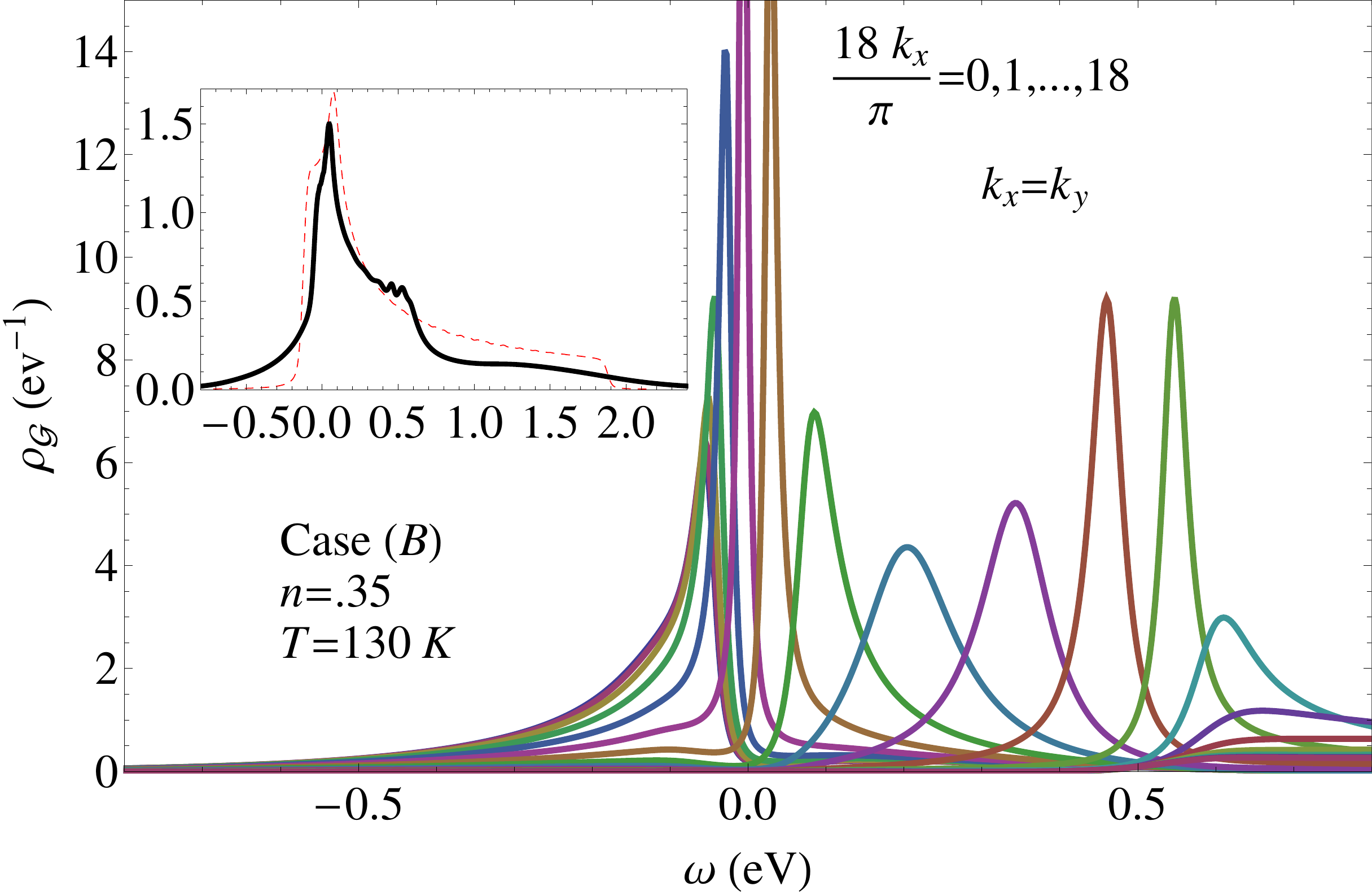}
\caption{{\bf Case (B)} at very low density with $(n,T)=(.35,130K)$. The physical electronic spectral function $\rho_{\G}$ at several k points along the $<11>$ direction is plotted, showing the evolution of lineshapes across the nodal FS. At this rather low density, the high frequency tails decay  more rapidly  than they do  at higher densities. The inset shows the ECFL density of states overlayed with the bare  LDOS for the same parameters. The difference between the Fermi gas and ECFL at this density is much less than was seen in \figdisp{dos}. Note also that the rather  sharp QP's  at $k\gg k_F$ are visible at  this low density, they persist to higher densities as seen below. Further, these sharp QP's create a distinct feature in the LDOS at $\omega > 0$.  } 
\label{lowdensity}
\end{figure}

We begin by displaying the spectral lineshapes at very low density of particles in \figdisp{lowdensity}, where the spectral lines for $k < k_F$ begin to show a characteristic tails at high binding energy relative to the lines at $k>k_F$, which are more Lorentzian. Also interesting is the sharpening of the lines at higher momenta $k\gg k_F$, indicating longer lived QPs far outside the FS. This is a consequence of the strong k-dependence of $\rho_{\Sigma}$ at positive frequencies. Furthermore, the inset of the LDOS reveals how this set of long-lived QP's results in the emergence of a secondary spectral peak in the density of states which is absent from the bare LDOS.

\begin{figure}[h]
\includegraphics[width=2.75in]{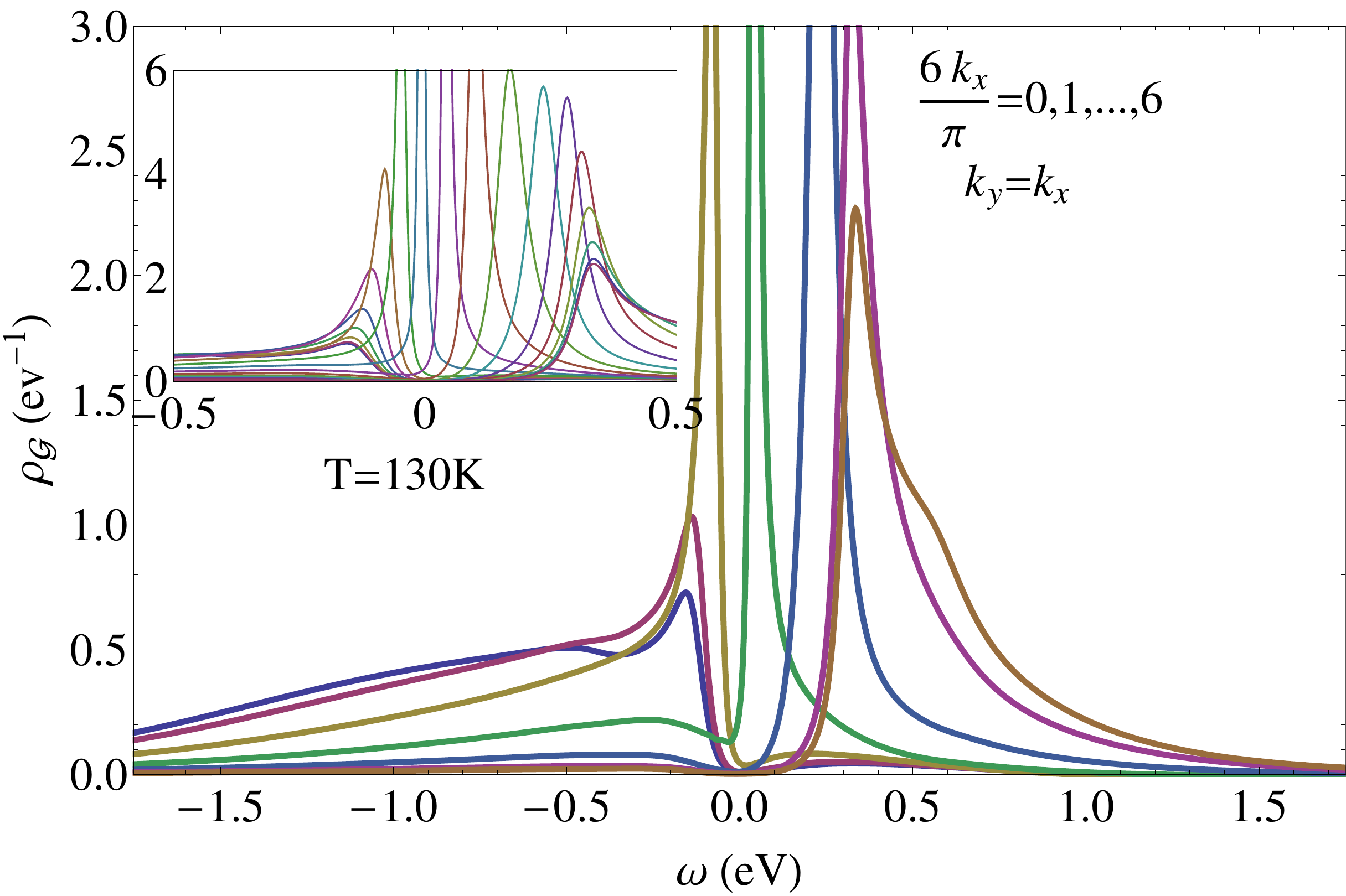}
\includegraphics[width=2.75in]{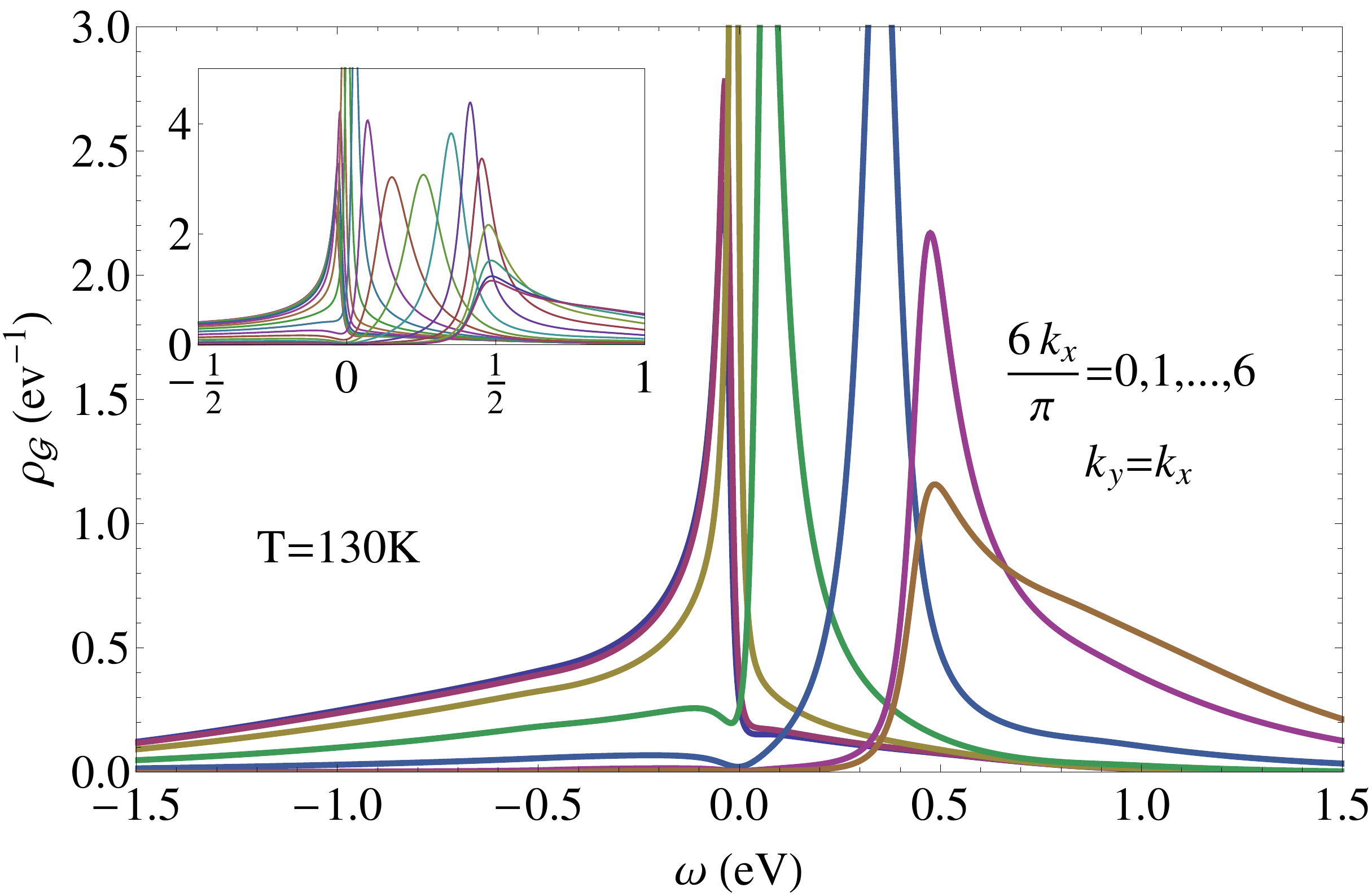}
\includegraphics[width=2.75in]{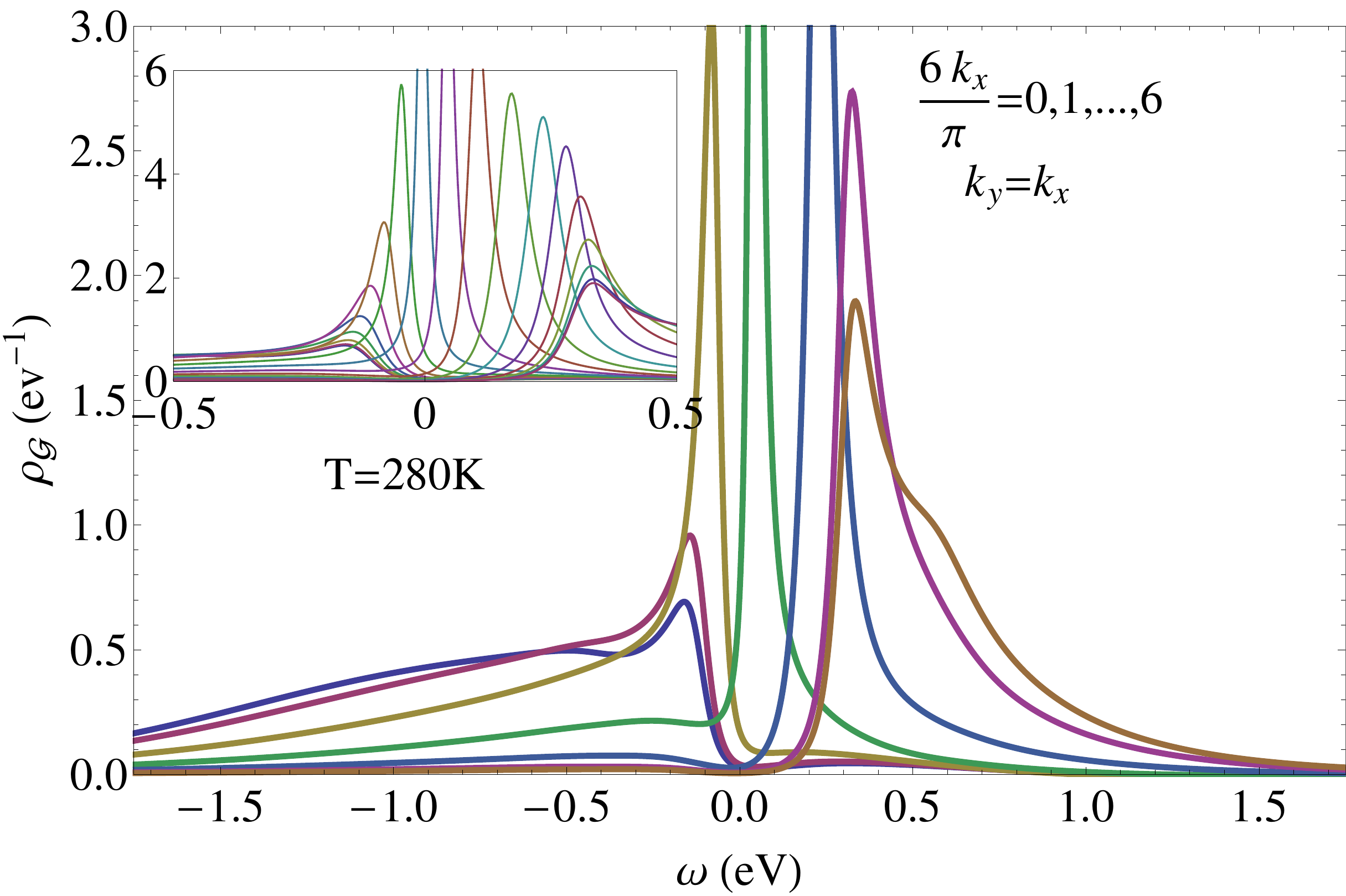}
\includegraphics[width=2.75in]{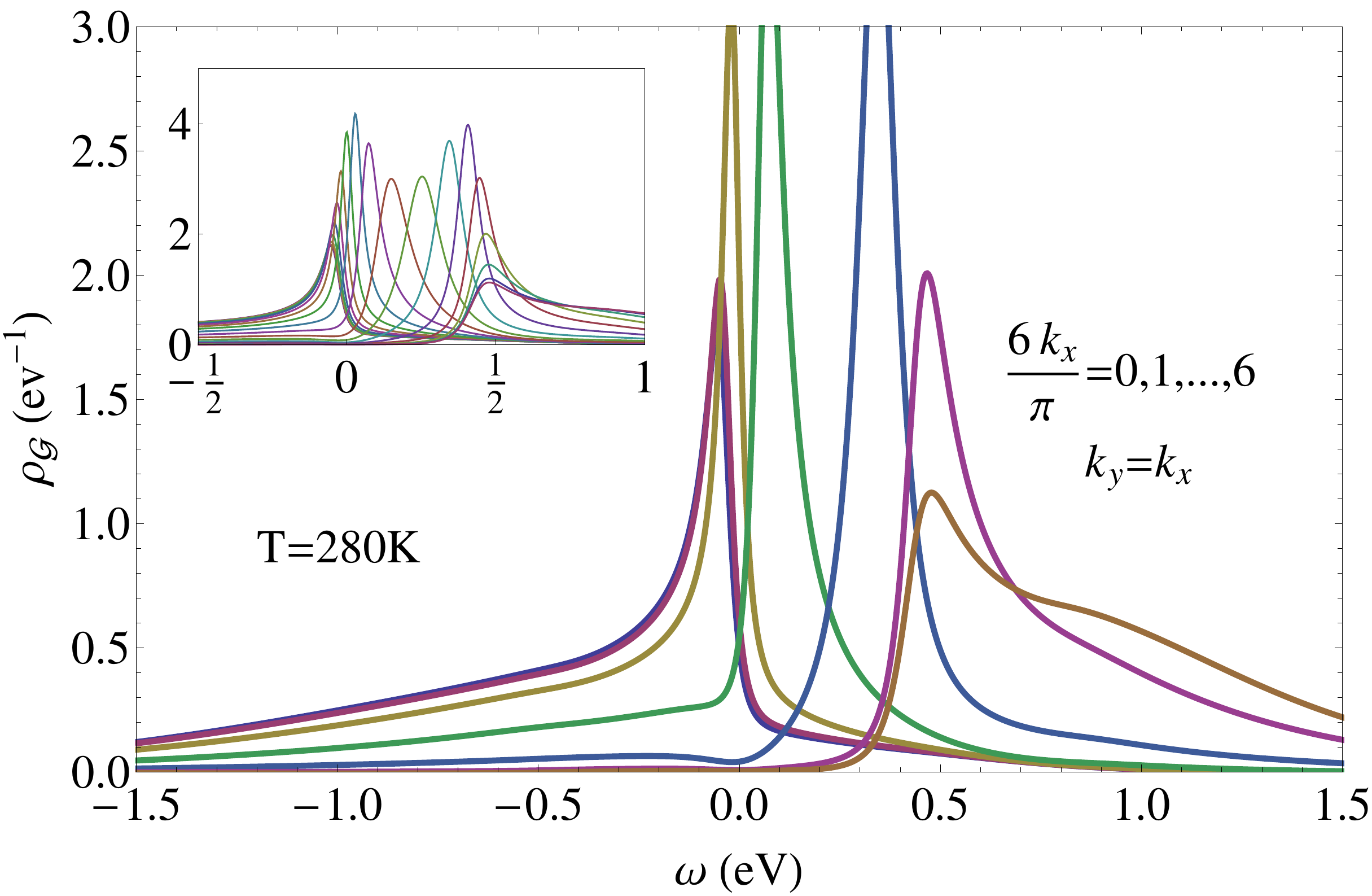}
\includegraphics[width=2.75in]{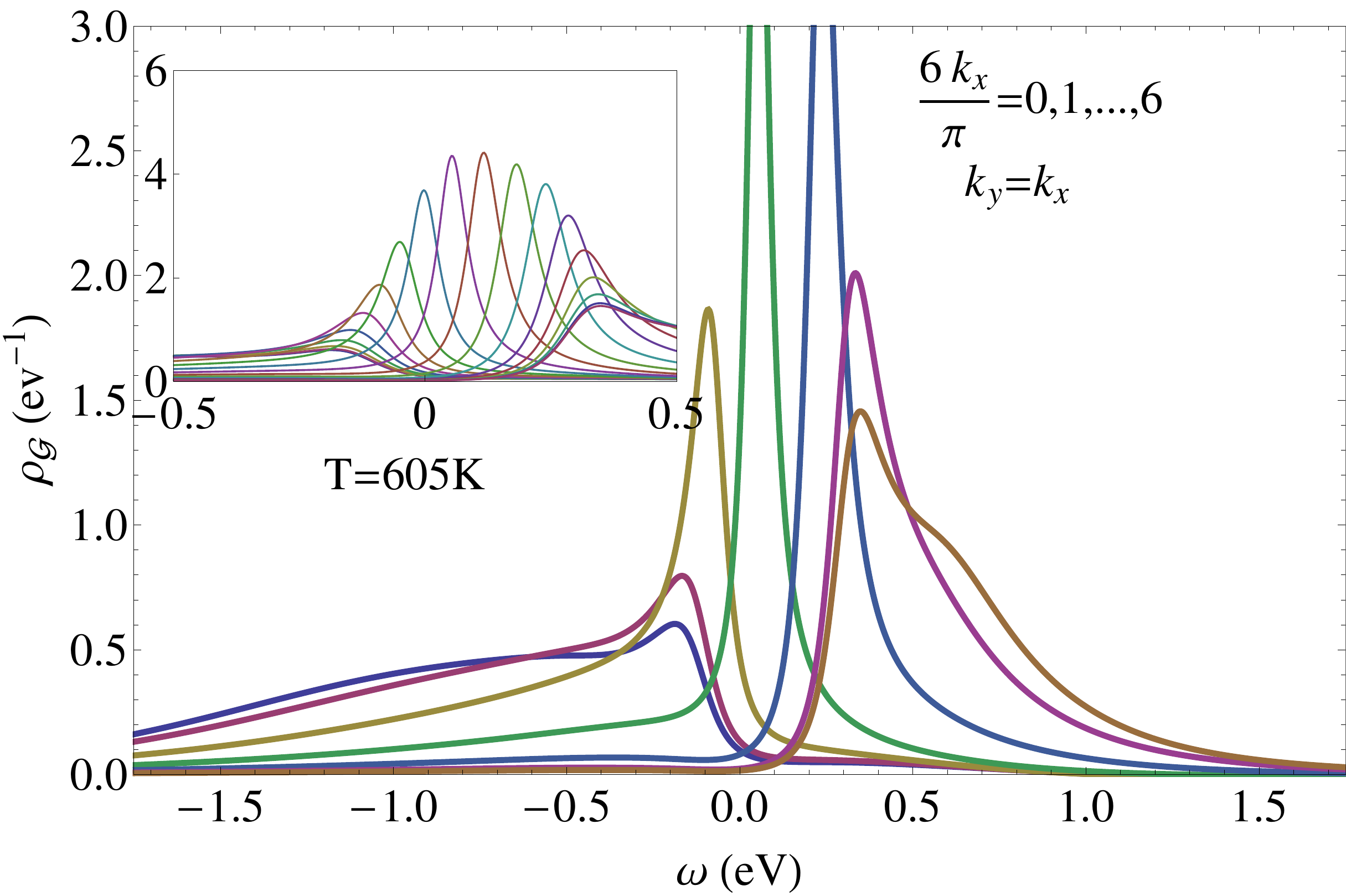}
\includegraphics[width=2.75in]{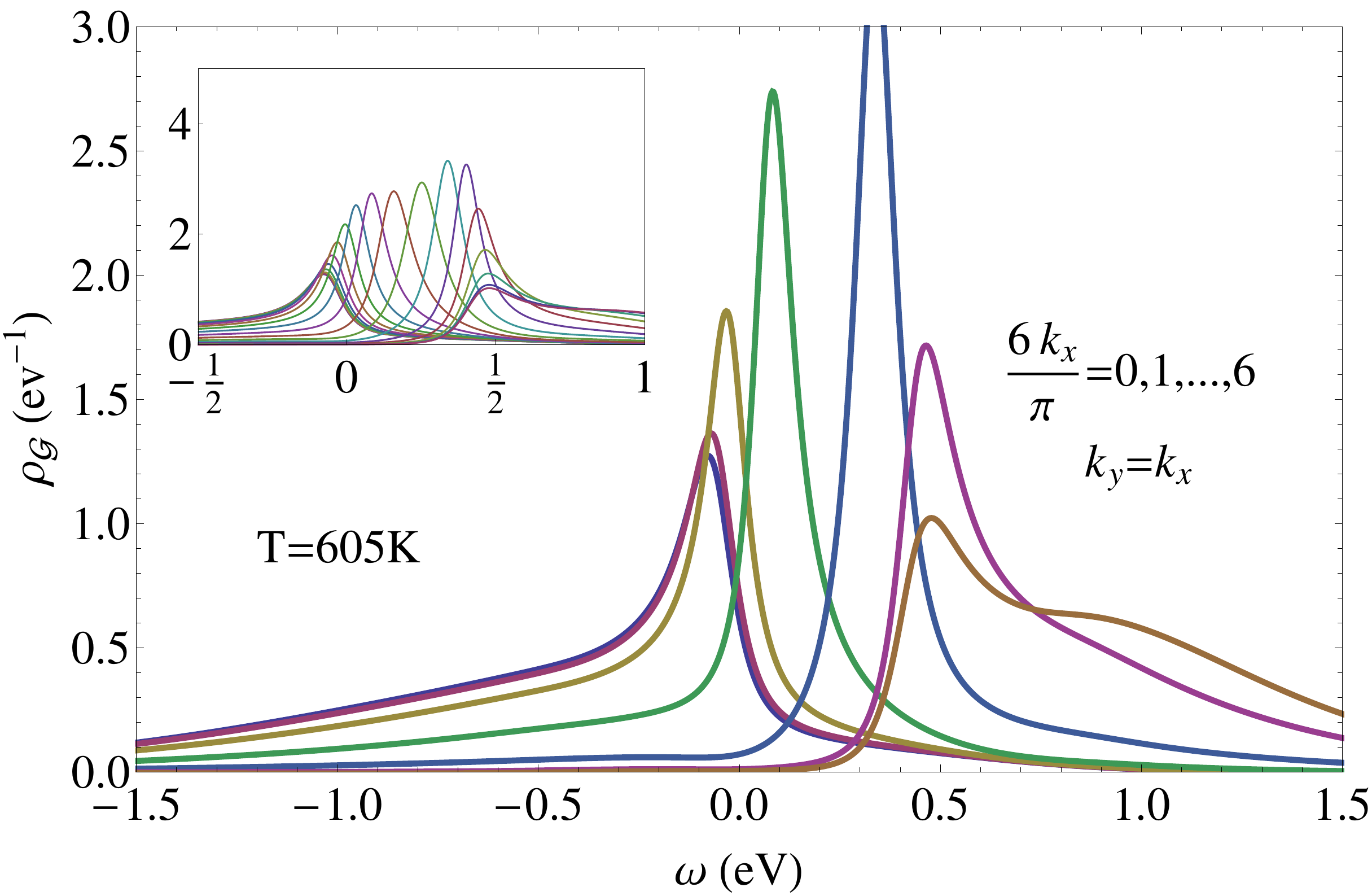}
\caption{$n=.75$. The physical electronic spectral function $\rho_{\G}$ at several selected k points along the $<11>$ direction are plotted, showing the evolution of lineshapes across the FS. {\bf Cases (A)} and {\bf (B)} are on left and right respectively with temperature increasing toward the bottom of the panel. {\bf The inset} in each case zooms out to reveal the height of the lineshapes for all the momenta, enlarging the low frequency range and shrinking the vertical scale.  In this and subsequent figures, we used $L_x=36$; the insets show all positive $k_x$'s and  the main figures display a third of the allowed $k_x$'s.
 While the linewidths near $k_F$ are strongly effected by rising temperature, the incoherent parts particularly at high frequency have very little temperature dependence. Note also that the tails in {\bf Case (A)} demonstrate an interesting structure near $\omega=-.4eV$, a second maximum far from the nominal QP. This secondary peak is responsible for the waterfall  discussed above. A similar structure was previously observed in \cite{Anatomy,Gweon}. Although not seen here for {\bf Case (B)} we note that a similar feature will arise at large positive frequencies as seen in \figdisp{wfpanel}. {\bf Case (B)} (and, to a lesser extent,  {\bf Case (A)}) display a local minimum of the linewidths for some $k > k_F$, as seen more clearly in \figdisp{Gamma}.  }
\label{spectralpanel}
\end{figure}

\begin{figure}[h]
\includegraphics[width=2.75in]{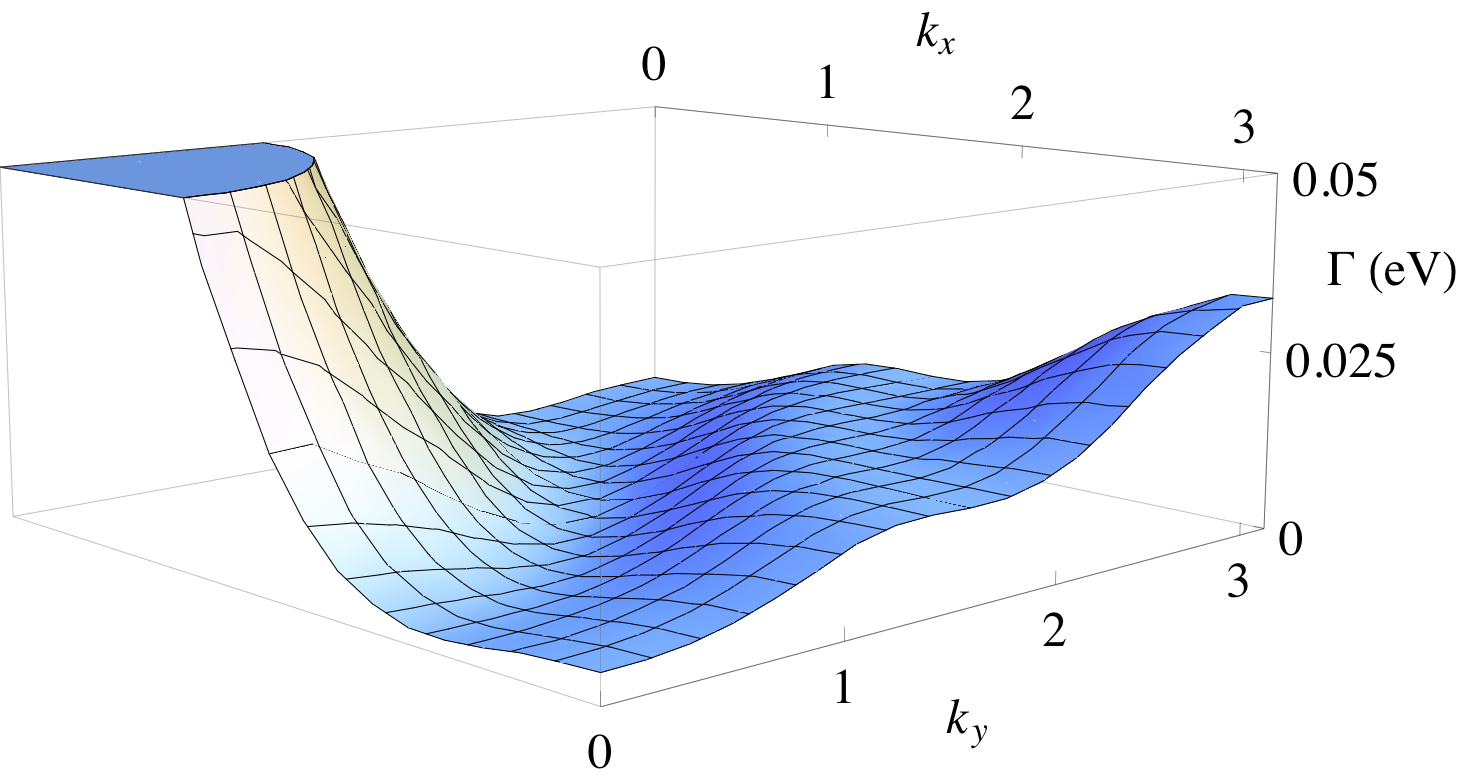}
\includegraphics[width=2.75in]{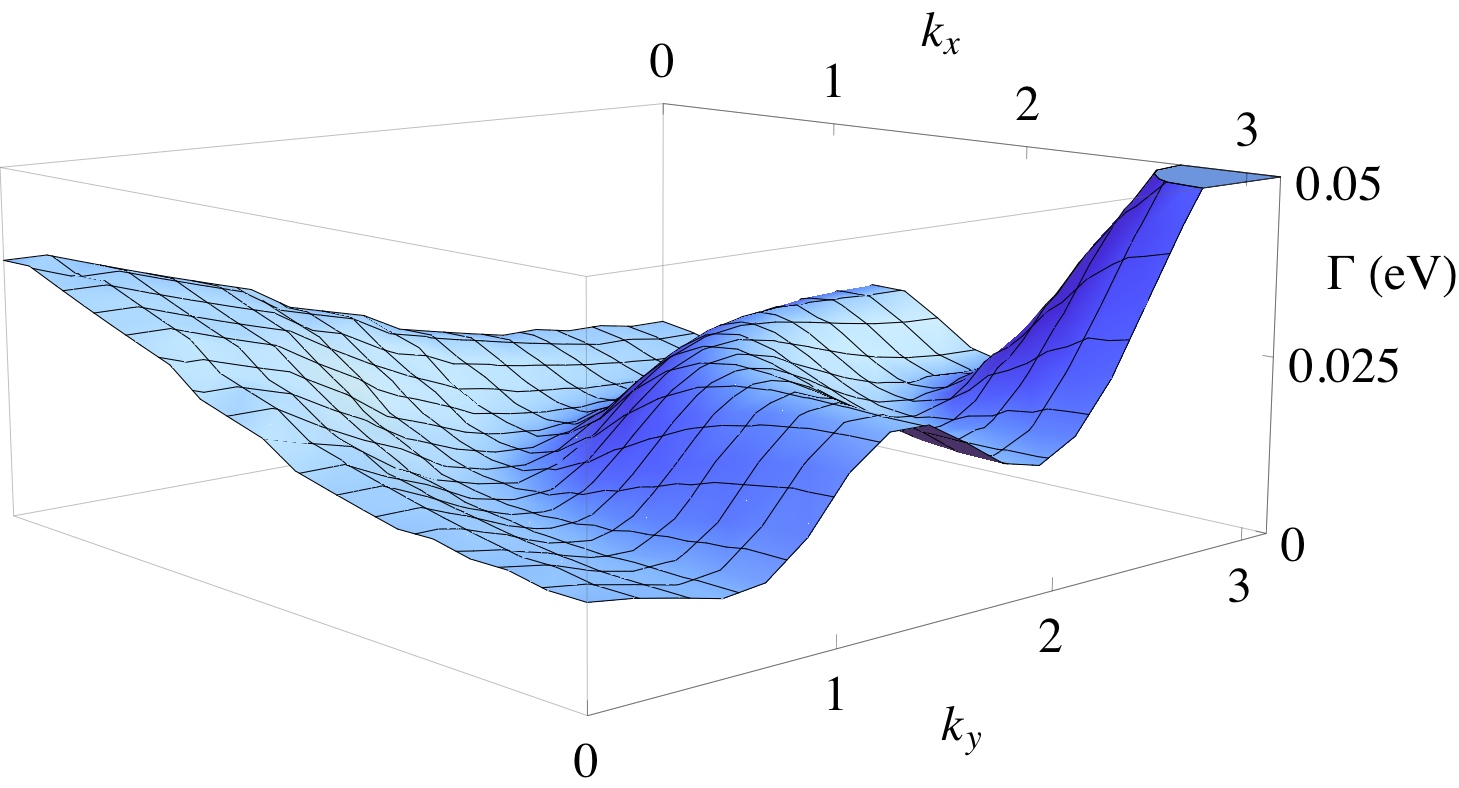}
\caption{For {\bf Cases (A)} and {\bf (B)} at $(n,T)=(.75,130K)$ the decay rate calculated from the imaginary part of the  self energy $\Im \Sigma_{DM}$ : $\Gamma_k=\rho_{\Sigma}(k,E_k)$  is plotted in the first quadrant of the BZ.  Apart from the expected minimum at the  FS,  we also see a  second  set of minima, particularly prominent in {\bf Case (B)}. Note also that the value of $\Gamma$ at the FS is much lower in {\bf Case (A)} than in {\bf Case (B)} where $\chem$ sits near a van Hove singularity  resulting in high scattering.  }
\label{Gamma}
\end{figure}

\begin{figure}[h]
\includegraphics[width=2.75in]{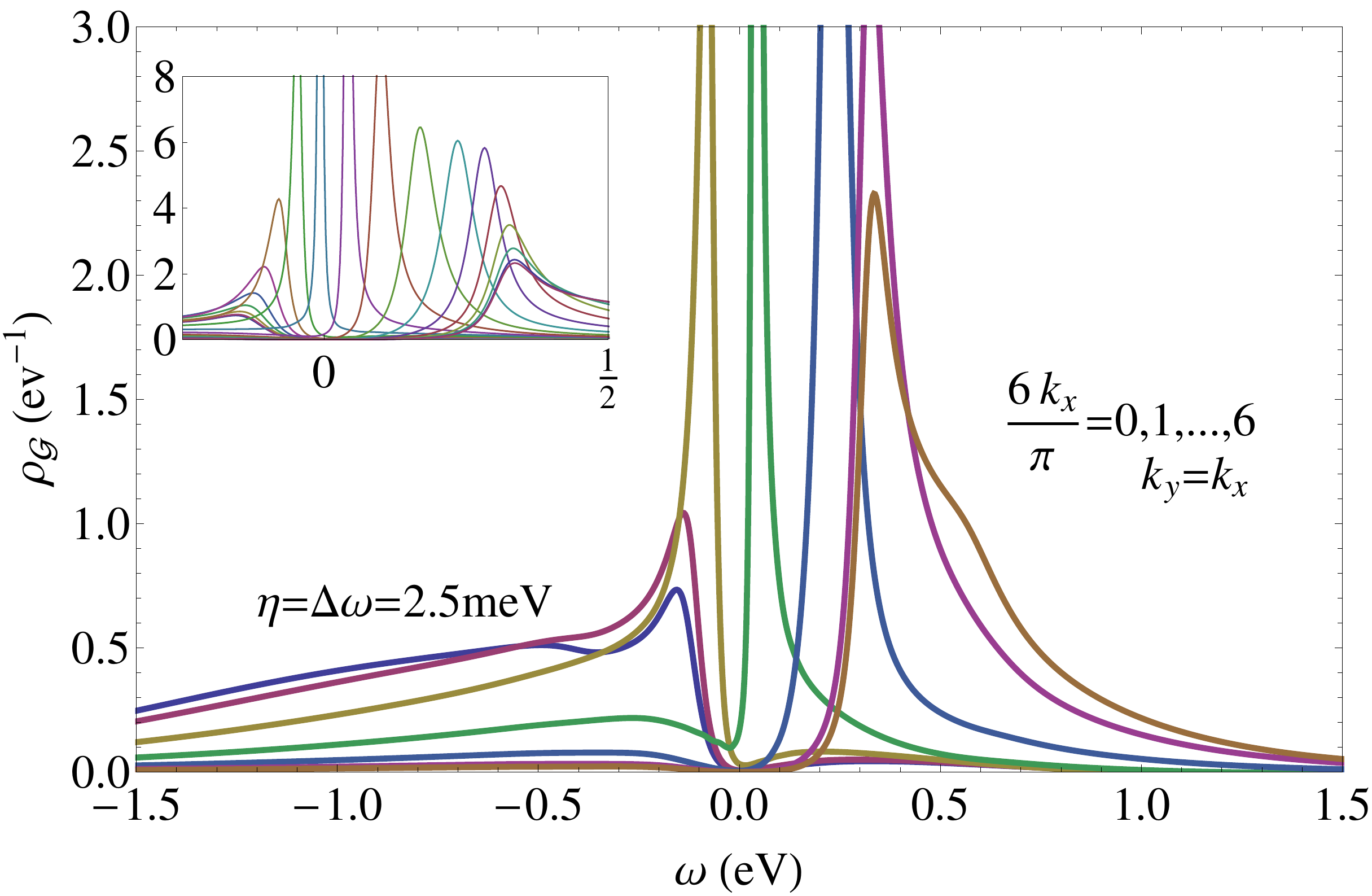}
\includegraphics[width=2.75in]{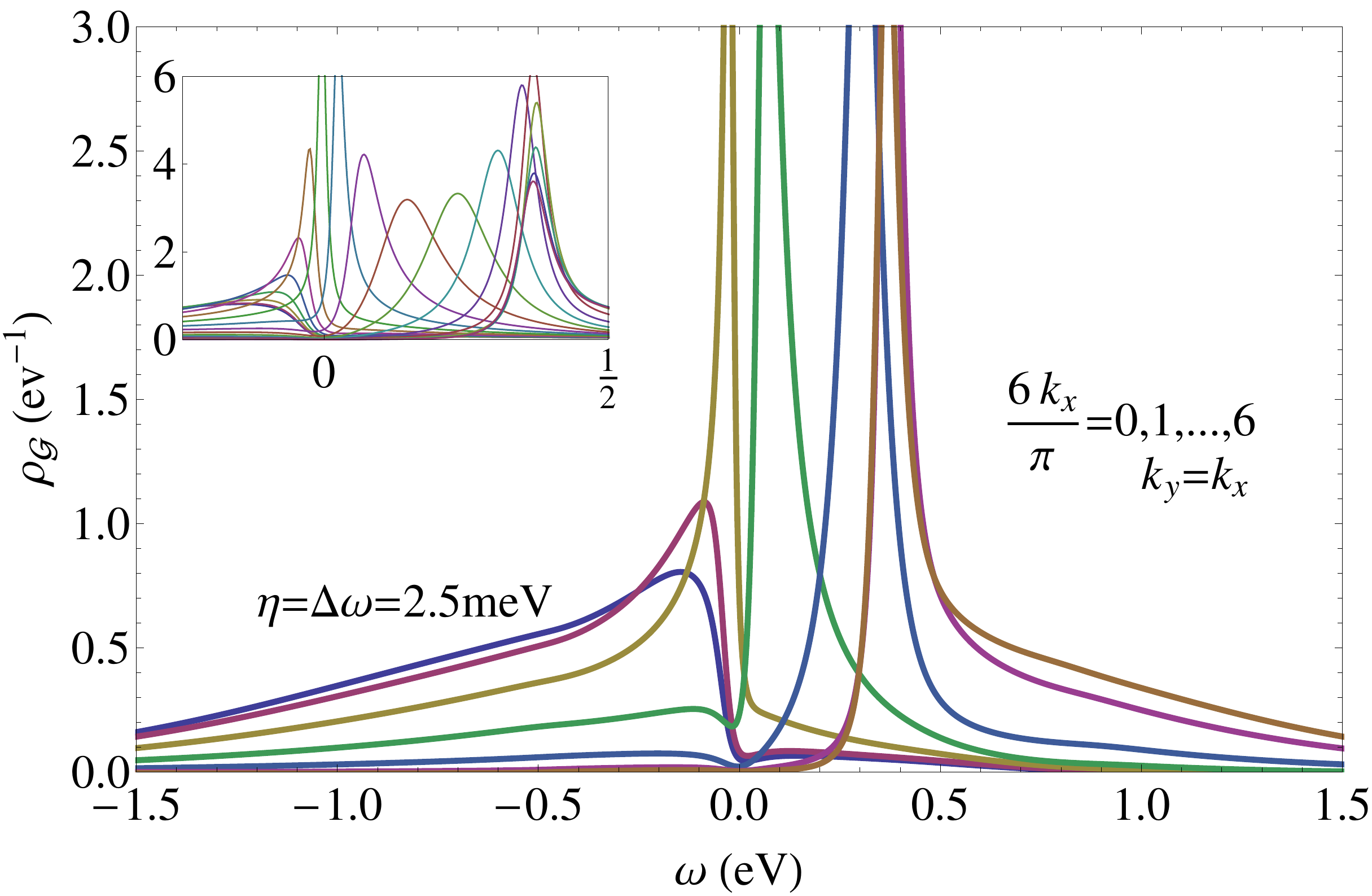}
\includegraphics[width=2.75in]{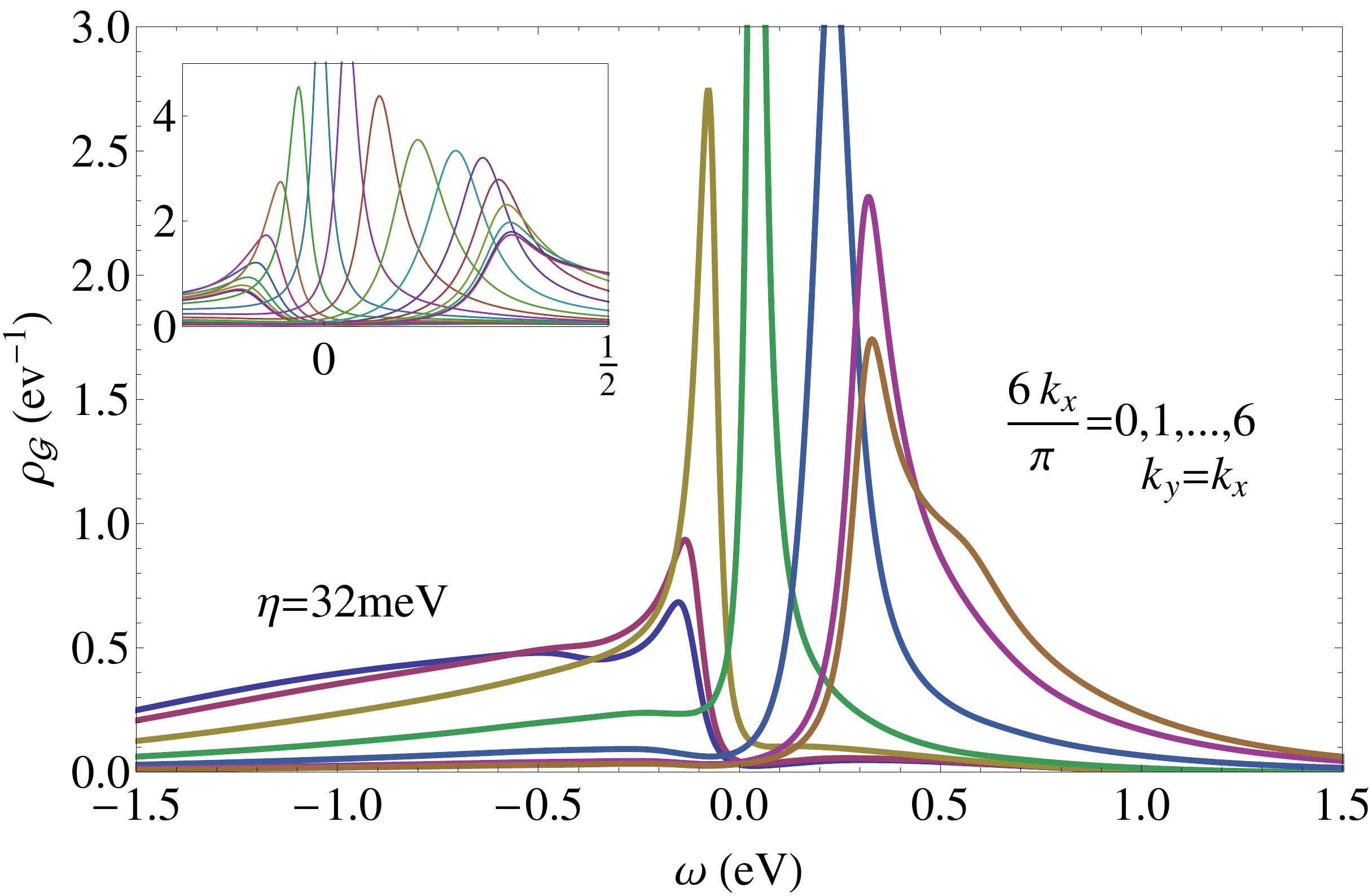}
\includegraphics[width=2.75in]{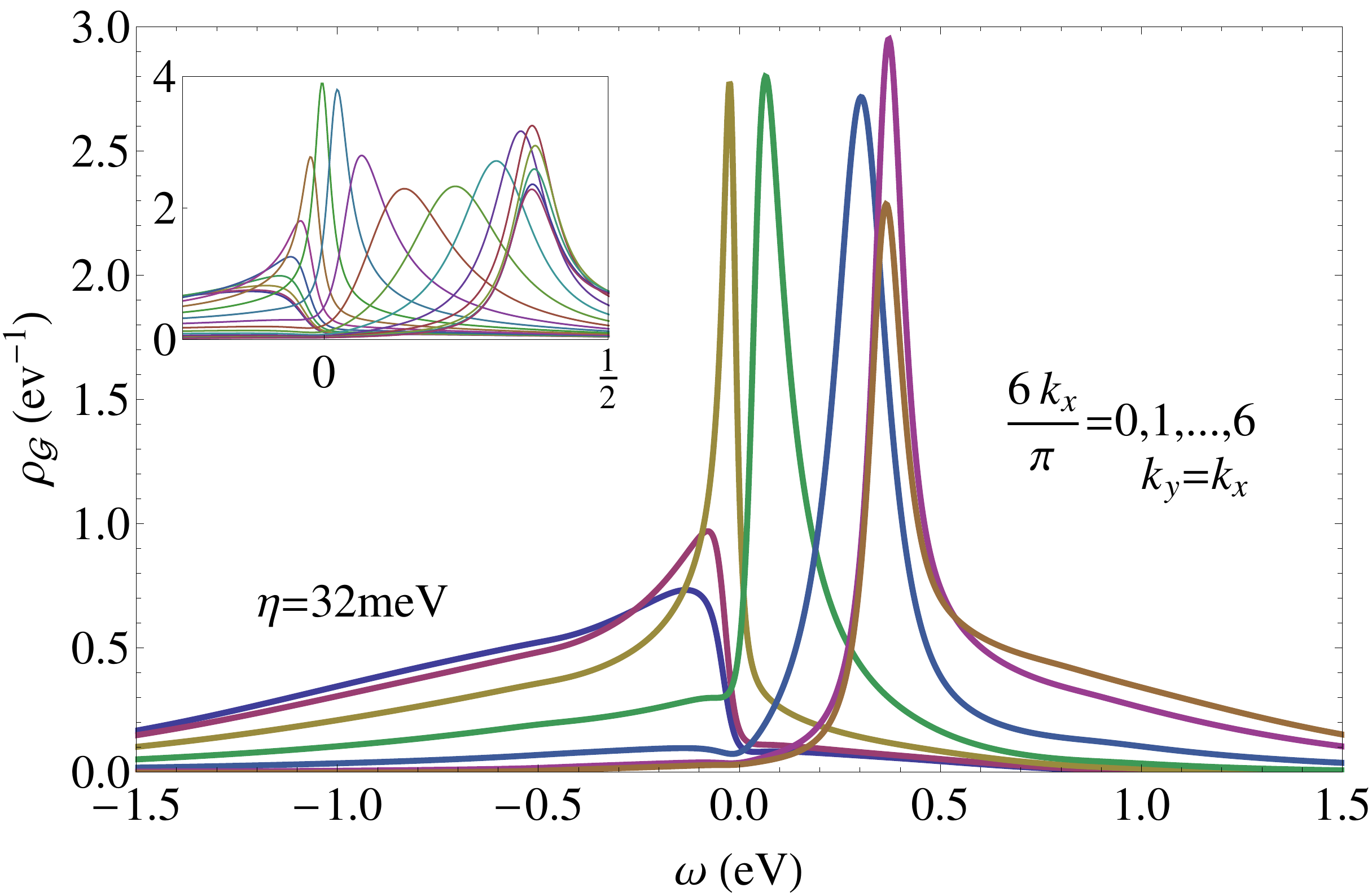}
\includegraphics[width=2.75in]{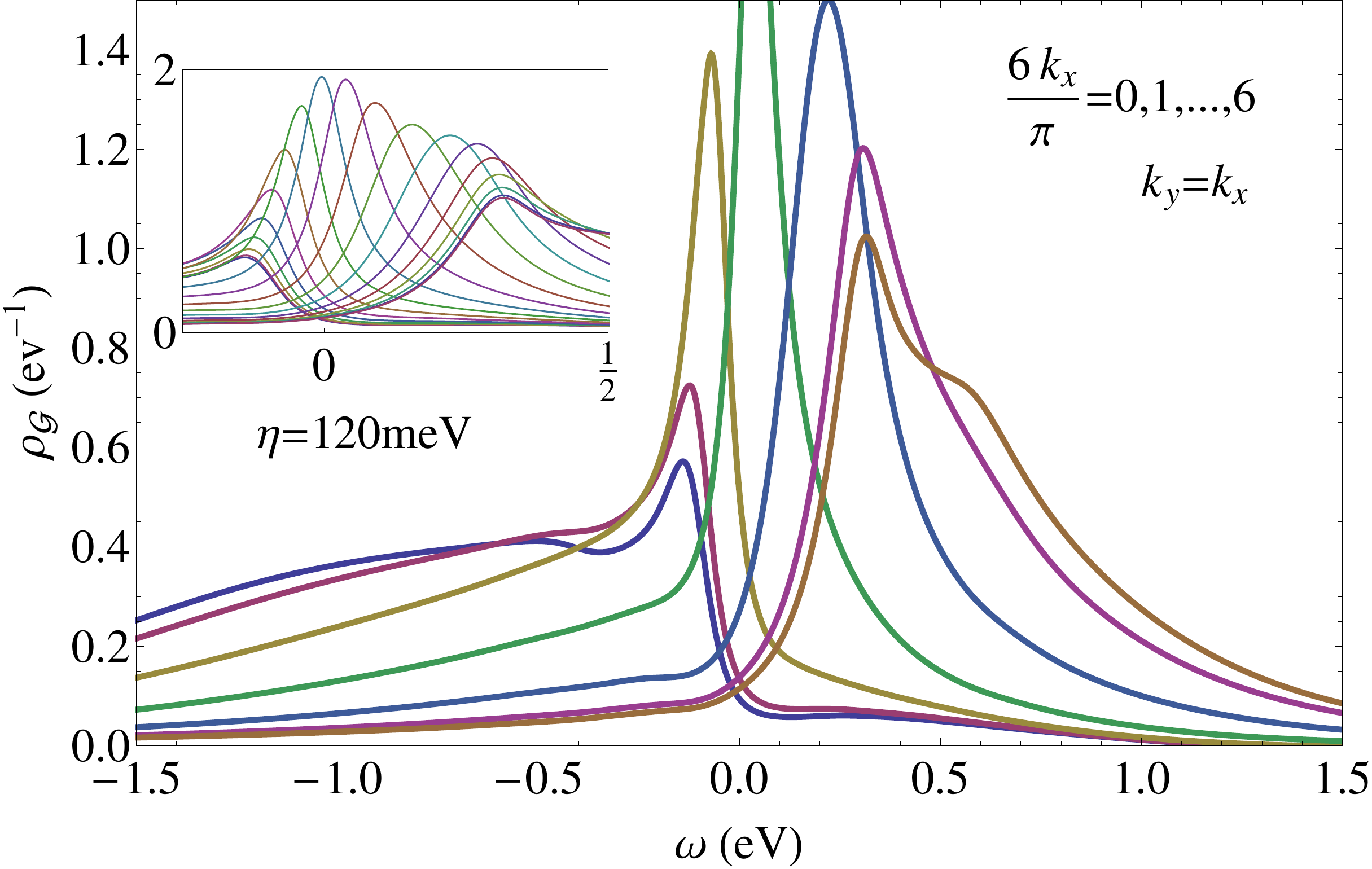}
\includegraphics[width=2.75in]{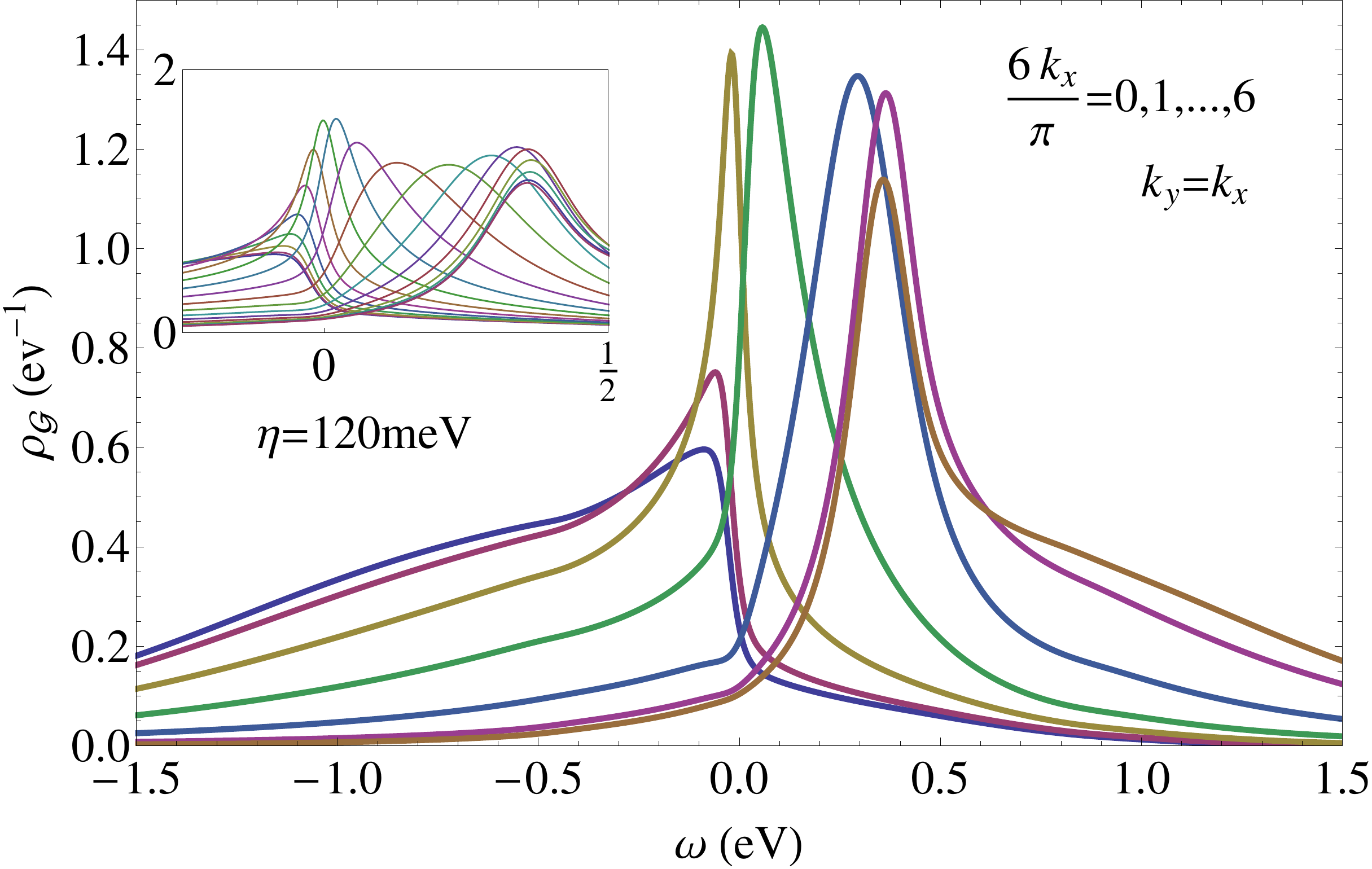}
\caption{$(n,T)=(.75,130K)$ The effect  of the elastic scattering term $\eta$ on the spectral functions $\rho_{\G}$ along $<11>$ in {\bf Cases (A)} and {\bf (B)} (left and right, respectively). The {\bf inset} emphasizes the low energy spectrum, as in \figdisp{spectralpanel}.  For comparison we also include the pristine system with $\eta \sim 0.$ and display the resulting lineshapes for three values  $\eta=0,.032,.12$ eV. The two nonzero values are typical for laser and synchrotron ARPES respectively, as argued in \refdisp{Gweon}. A large (synchrotron ARPES) value of $\eta\sim .12$ eV leads to extended tails in the occupied frequency range,  and a greatly reduced peak value of the spectra, while the smaller (laser ARPES) value of $\eta\sim .032$ has a less drastic effect on the pristine case.  } 
\label{etapanel}
\end{figure}

We next display the spectral lineshapes in a set of figures \figdisp{spectralpanel} to \figdisp{etapanel}.
These explore low and high  temperatures, different amounts of impurity scattering, and different directions along the BZ. The first panel \figdisp{spectralpanel} shows the nodal spectral functions of {\bf Cases (A)} and {\bf (B)} at three different temperatures.  The lines are quite sharp near $k_F$ but broaden out rapidly away from $k_F$. The insets give an idea of the change of spectral density with T. It is interesting that the detailed shapes at higher binding energy differ between the two cases, with {\bf Case (A)} showing a more pronounced tendency for a rise with (binding) energy, giving rise to a local minimum for some $k$. This two-peak structure  causes a waterfall like  feature in the MDC dispersions, as discussed earlier. A feature of this kind has been observed in \refdisp{Anatomy} and \refdisp{Gweon}. For {\bf Case (A)} it is also noteworthy that lines for some $k > k_F$ the lineshapes are  very sharp, and further have tails that spillover into $\omega \leq 0$, thereby contributing to what is commonly relegated as  ``background'' in ARPES.

Both {\bf Case (A)} and {\bf Case (B)} exhibit an interesting behavior in the QP scattering rates at different $k$ values over the first quadrant of the BZ, as  seen in \figdisp{Gamma}. We plot the decay rate from the Dyson-Mori self energy: $\Gamma_k=\rho_{\Sigma}(k,E_k)$ as a function of $\vec{k}$
in \figdisp{Gamma}. In both cases there is a surface of minimum scattering rate which coincides with the FS. However, there is a second surface  with $k>k_F$,  where the scattering rates are a local minimum. This feature is particularly strong in {\bf Case (B)} and can be seen clearly in the lineshapes of \figdisp{spectralpanel} and even at the very low density spectral function seen in \figdisp{lowdensity}.  In {\bf Case (A)} the feature is less pronounced but is nonetheless present. The k-dependence of the self energy will be presented in more detail below.

We next discuss the role of elastic scattering in the lineshapes following \refdisp{Gweon}. 
As discussed in that work the parameter $\eta$ can be used to model the effects of impurity scattering upon ARPES spectra,  and usefully  distinguishes between laser ARPES and synchrotron ARPES.  Using fits from \refdisp{Gweon} we arrive at $\eta=.032eV$ and $\eta=.12eV$ as   typical values  for describing  laser and synchrotron data respectively. It is therefore interesting to see how the lineshapes discussed above in \figdisp{spectralpanel} evolve under the influence of the elastic scattering term. In \figdisp{etapanel} low temperature spectral functions are presented with the additional $\eta$. Its effect  is mainly to increasing the  linewidths.  Increasing temperature also increases linewidths, but predominantly near $k_F$, while $\eta$ broadens the lineshapes uniformly across the BZ. We observe that the main  features of $\rho_{\G}$, namely  the two peak EDCs and long lived QPs at $k>k_F$, remain visible in the spectrum  after the  elastic  broadening.

\figdisp{etapanel} also reveals an interesting connection between impurity scattering contribution  and the spillover weight found for lines at negative energies but  with  $k>k_F$. While a finite spillover is visible for the minimum $\eta$, it becomes quite exaggerated when larger impurity scattering effects are considered. In particular we note the case $\eta=120$meV where even the $k=(\pi,\pi)$ lineshapes have a finite weight at negative $\omega$,  visible at binding energy as high as $1$eV.

\begin{figure}[h]
\includegraphics[width=3.5in]{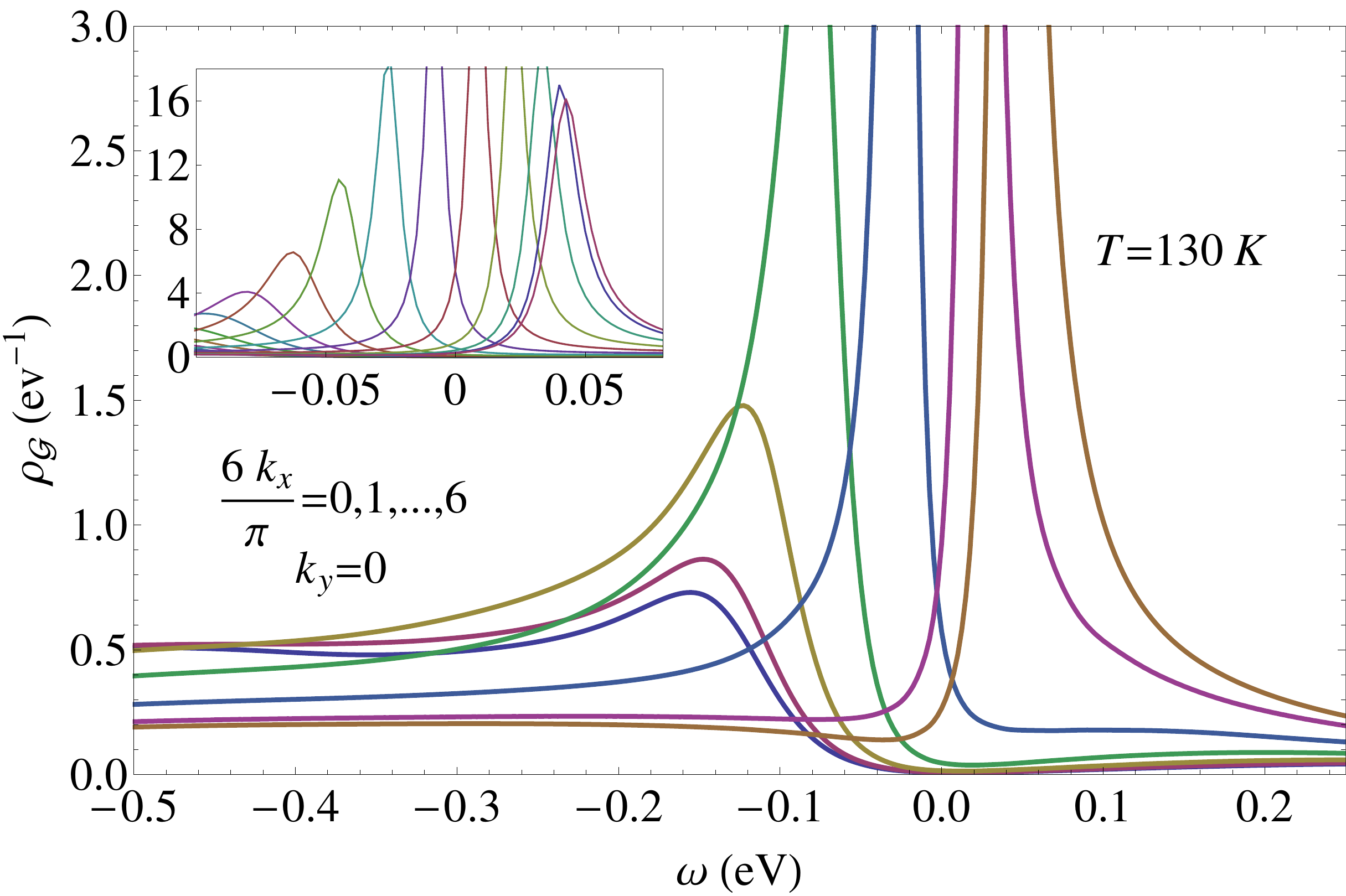}
\includegraphics[width=3.5in]{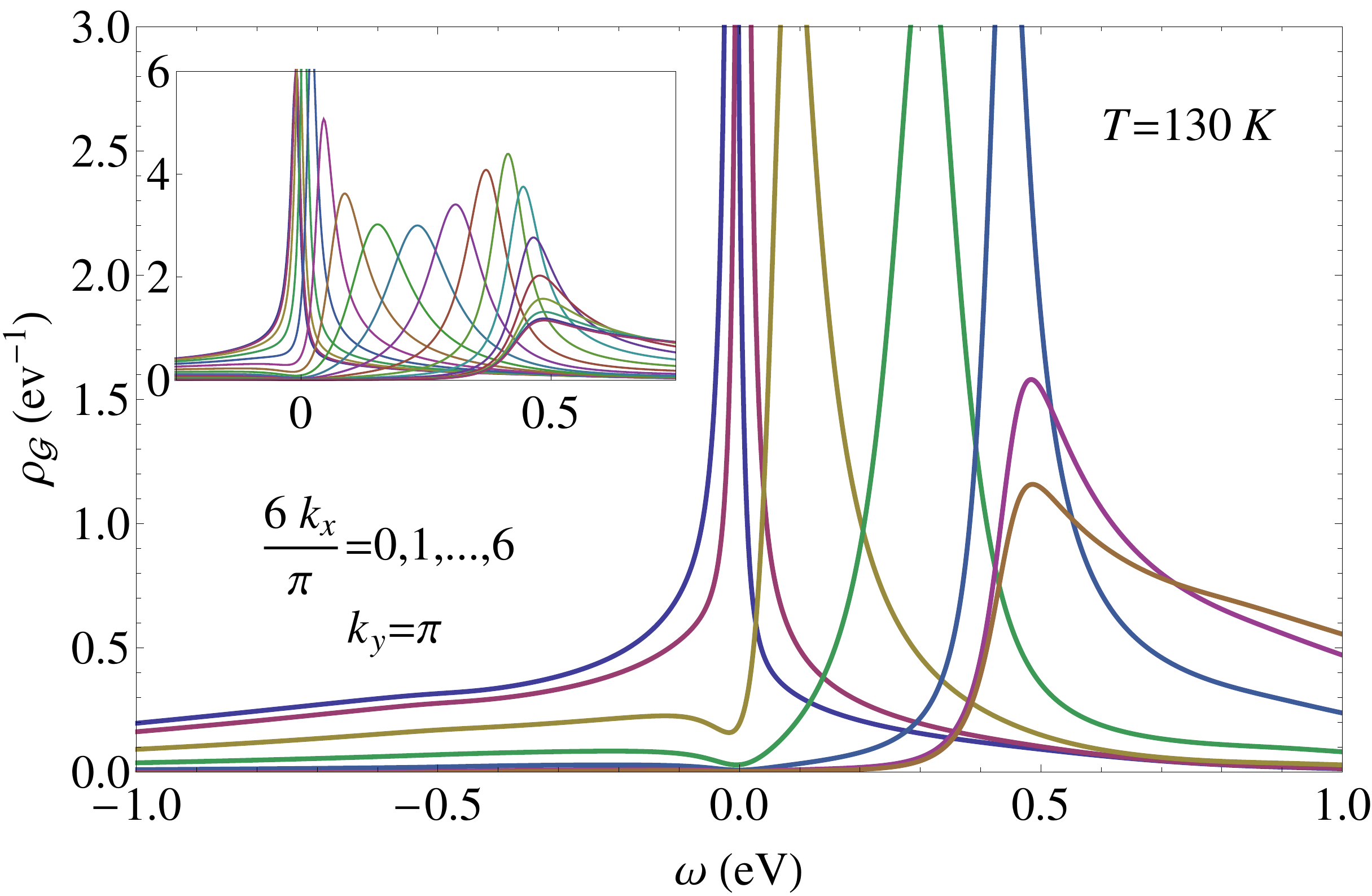}
\caption{For $(n,T)=(.75,130K)$  the spectral function $\rho_{\G}$  is plotted for k's along the line perpendicular to the anti-nodal FS. For {\bf Case (A)} on the left, this corresponds to the $ <01>$ direction,  and  for {\bf Case (B)} on the right the line from $(0,\pi)\rightarrow(\pi,\pi)$, respectively. The bandwidths seen along these directions are much smaller than those seen along the nodal line. } 
\label{line6}
\end{figure}

Finally, in \figdisp{line6}, we show lineshapes for {\bf Case (A)} and {\bf Case (B)} for BZ cuts perpendicular to the anti-nodal FS. Depending on whether the FS is electron or hole like, this can be either the (01) direction or the line from $(0,\pi)\rightarrow(\pi,\pi)$, respectively. The anti-nodal FS in each case has linewidths which are similar to those seen in the nodal direction. The appearance of a small total bandwidth here is due to the fact that these cuts do not connect the minimum and maximum of the bare dispersion $\varepsilon_k$.  This $O(\lambda^2)$  theory at the densities reported does not significantly distinguish between nodal and anti-nodal directions. Note that the two peak EDCs of {\bf Case (A)} responsible for the waterfall are visible along nodal and anti-nodal directions. Likewise, the anomalously sharp QPs at $k>k_F$ in {\bf Case (B)} are also present along the line from $(0,\pi)\rightarrow(\pi,\pi)$. 

\subsection{Optical conductivity}
\begin{figure}
\includegraphics[width=5.5in]{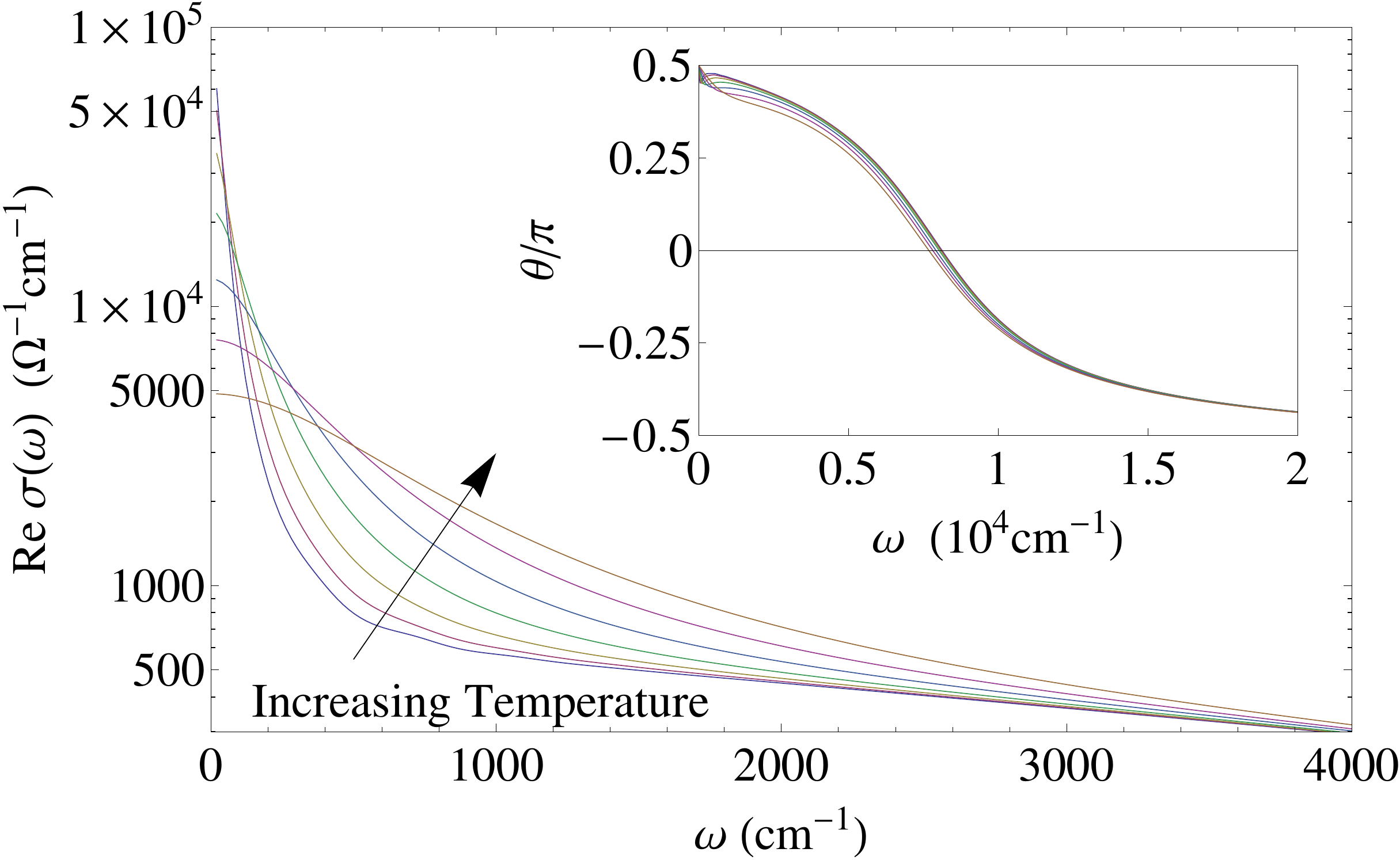}
\caption{{\bf Case (A)} $n=.75$,   $T=60,90,130,190,280,410,605K$.  The optical conductivity is calculated on an absolute scale, and illustrates how  increasing   T rapidly  fills up the regime $0 \leq \omega \leq 1000$ cm$^{-1}$. The rise of conductivity at very low $\omega$ is also inferred from the  DC resistivity displayed in   \figdisp{ScatteringRatesN75}. The phase of the complex  $\sigma$ falls off rapidly beyond 4000 cm$^{-1}$.  Results for {\bf Case (B)} are similar and shown below in \figdisp{puchkov}.  } 
\label{ConductivityFit}
\end{figure}

\begin{figure}
\includegraphics[width=4.5in]{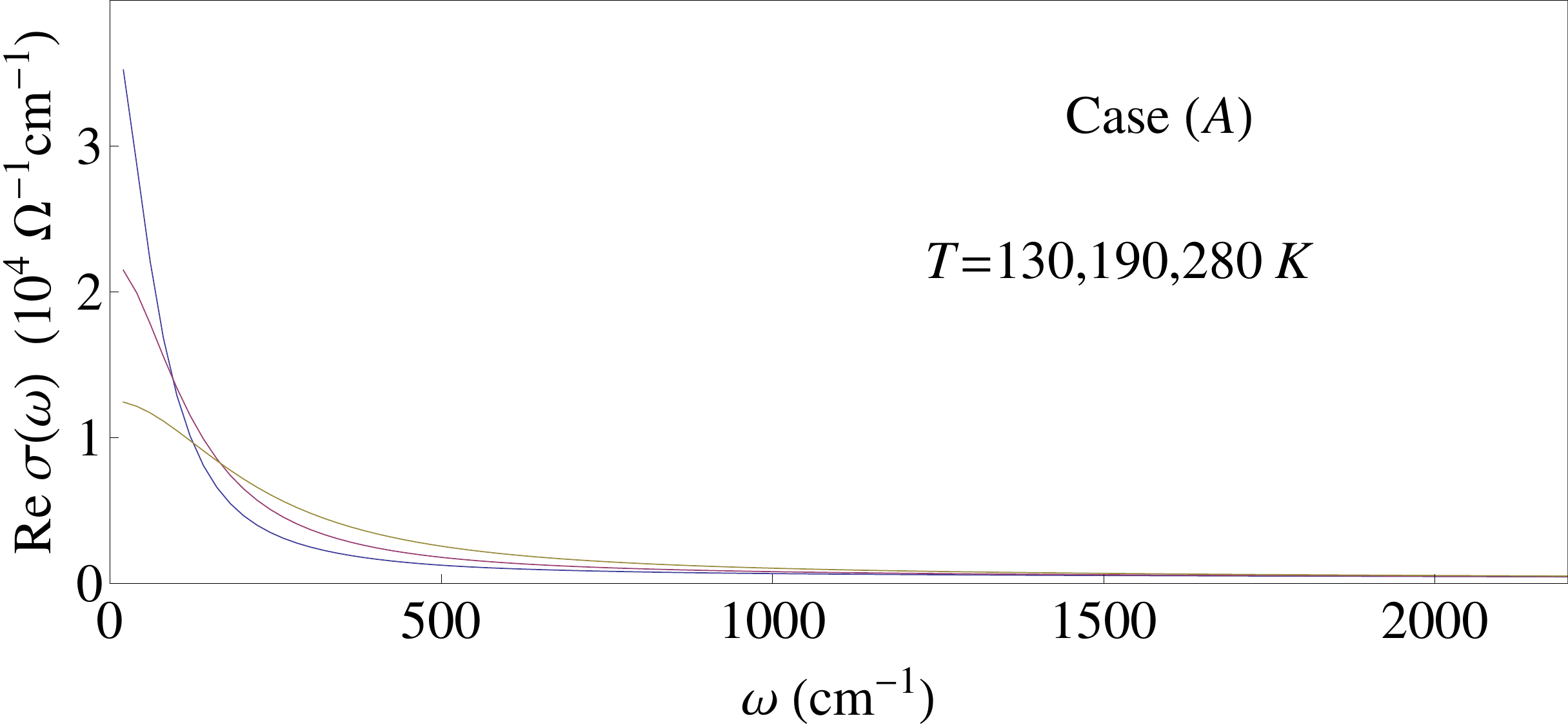}
\includegraphics[width=4.5in]{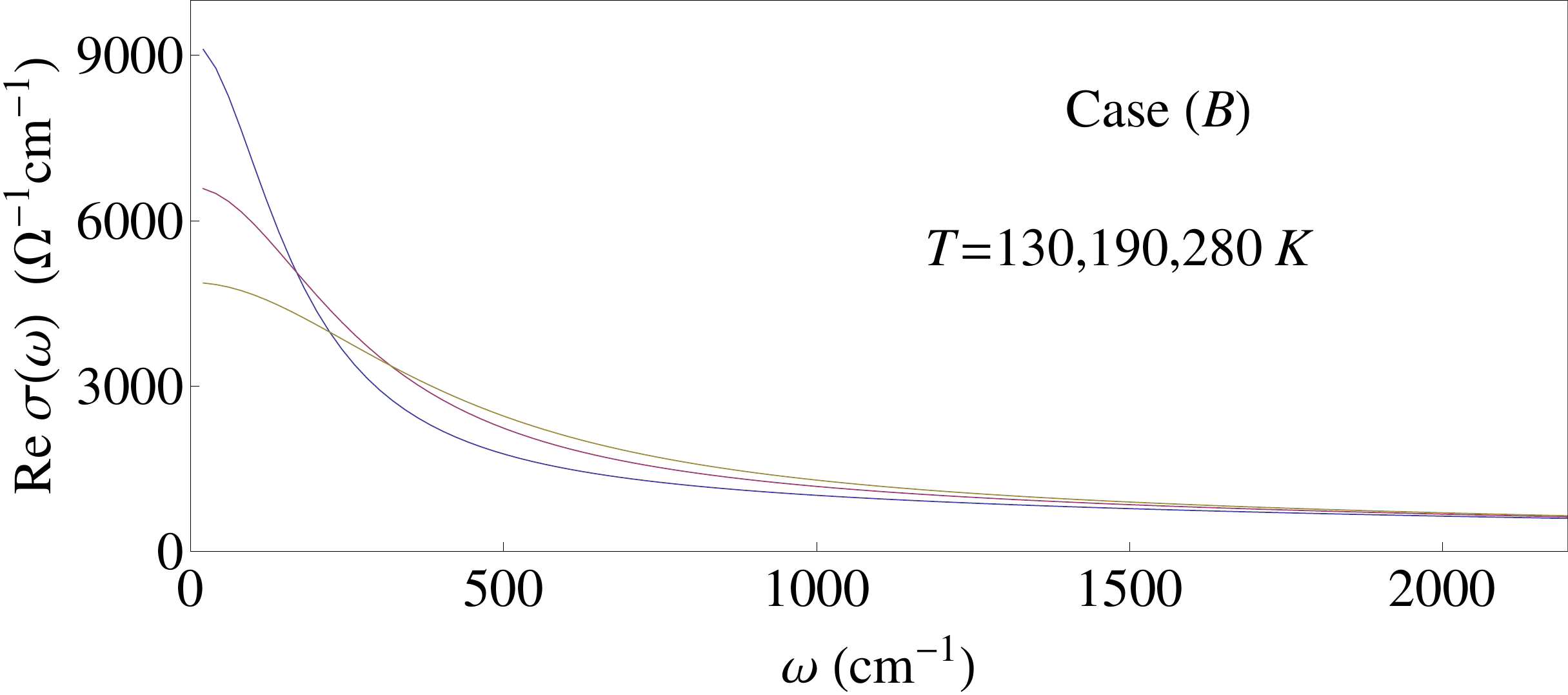}
\includegraphics[width=4.5in]{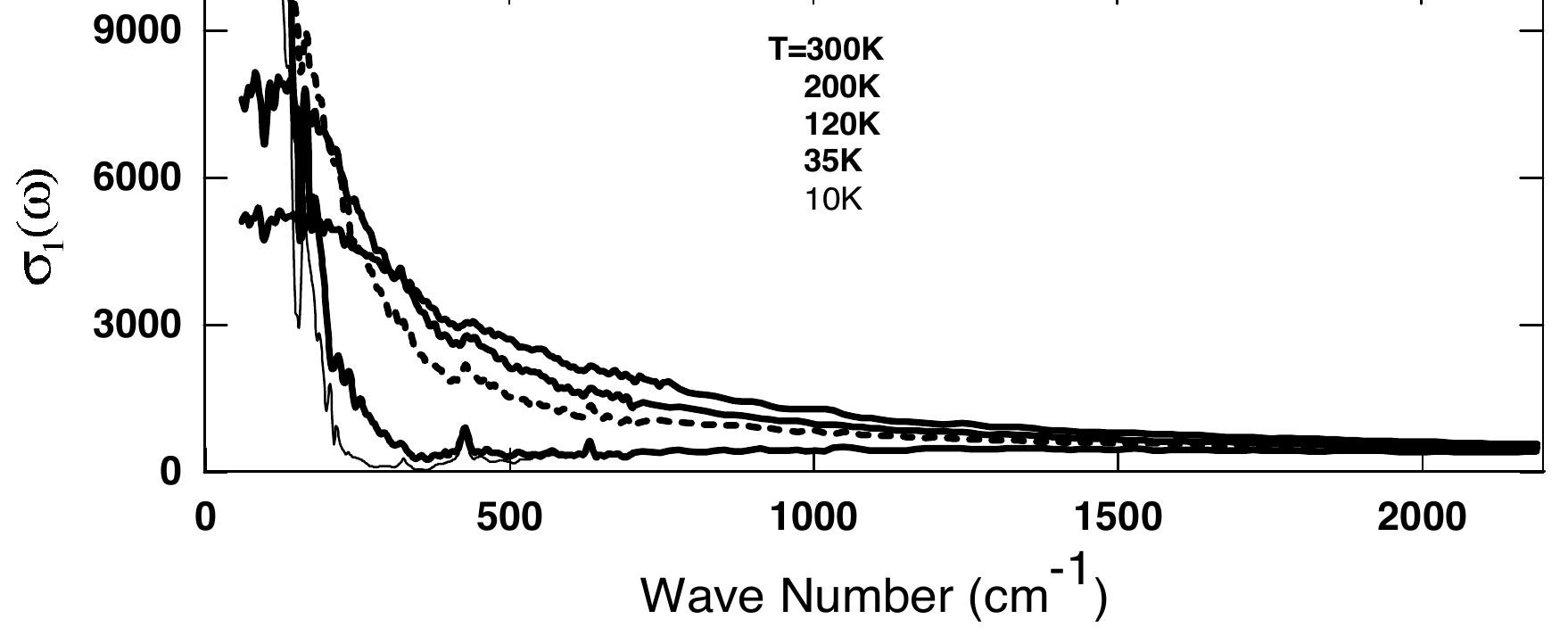}
\caption{We make an explicit comparison here with optical conductivity measurement in Puchkov et al  from \refdisp{basov} with the authors kind permission. It pertains to an overdoped Thallium based compound with $T_c=23K$ with a density that is expected to be close to the value assumed in the calculation $n\approx.75$.  We note the similarity of magnitude and variation with $\omega$ and $T$  with our {\bf Case (B)}, our {\bf Case (A)} leads to a bigger value of $ \Re \  \sigma(\omega)$.} 
\label{puchkov}
\end{figure}

\begin{figure}[h]
\includegraphics[width=3.5in]{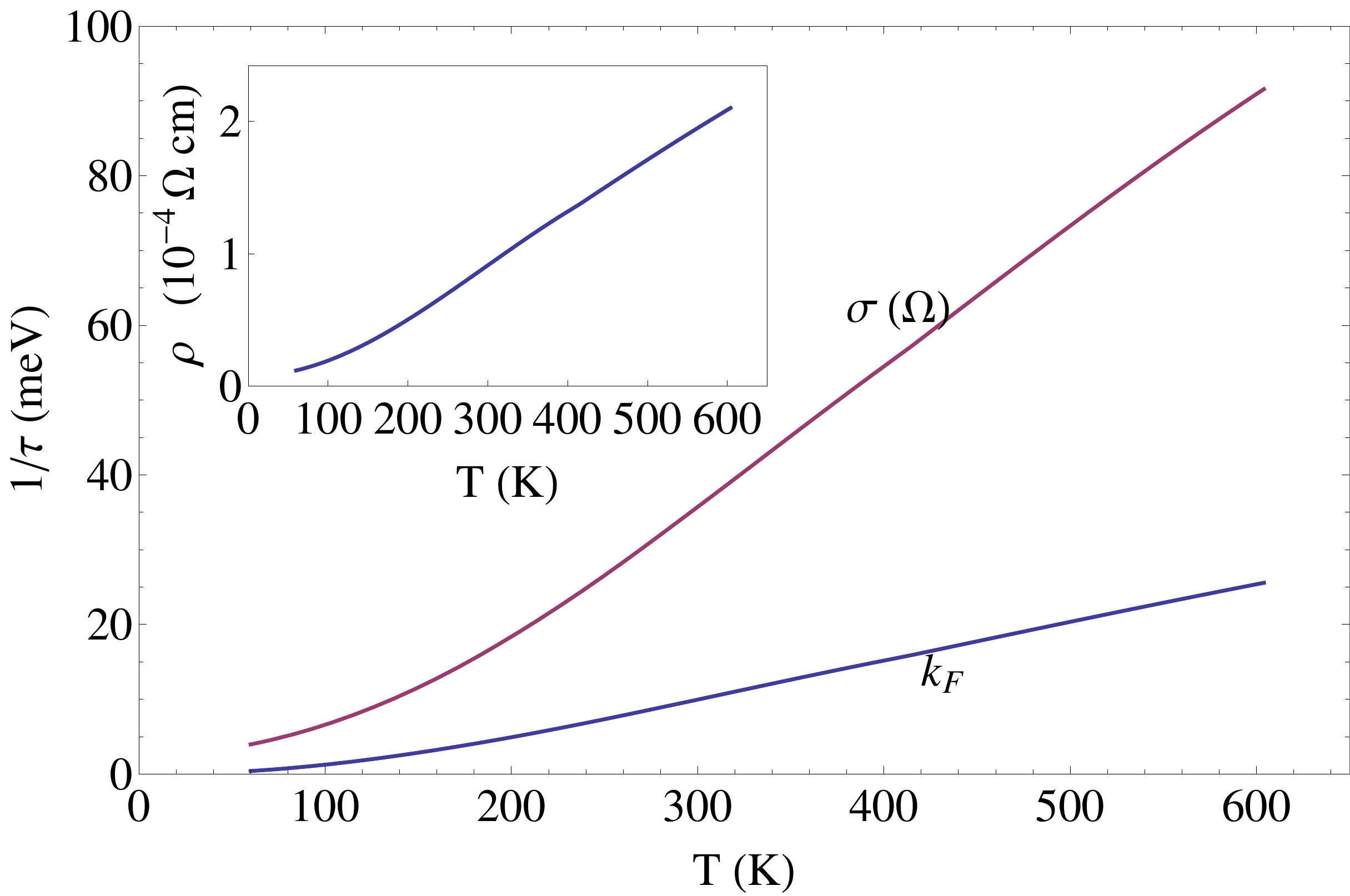}
\includegraphics[width=3.5in]{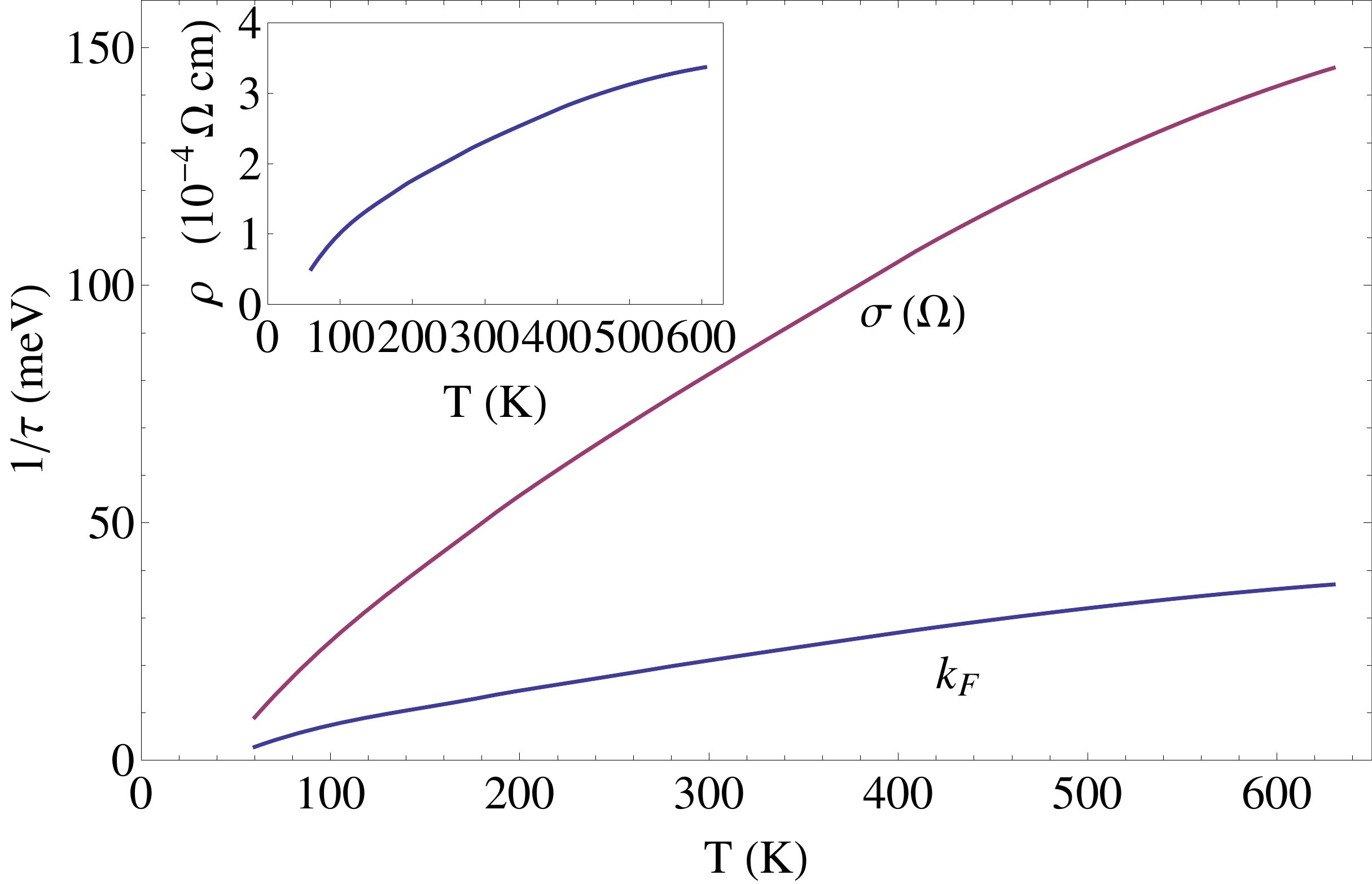}
\caption{$n=.75$: The inverse lifetimes for  QP's  at the FS along $<11>$ obtained from the imaginary part of the Dyson-Mori self energy and the  inverse lifetime obtained from the optical conductivity  as in in \disp{opttau}. The $T^2$ scaling typical of a FL is visible at low temperature for {\bf Case (A)} (on the left). The quadratic dependence ends at a modest crossover temperature ($\sim150K$) partly due to the shrinking band width as seen directly in \figdisp{DispersionPanel}.  {\bf Case (B)} (on the right) appears less FL-like due to the presence of a vHs near the Fermi energy which raises the low energy scattering rates considerably. The {\bf inset} shows the  DC resistivity  obtained from the inverse of \disp{sigEq}   for both cases.  In {\bf Case (B)}, the resistivity  appears to be on track to saturate in the vicinity of the Mott Ioffe Regel (MIR) scale of a 2D metal i.e. $\rho_{2d}^{MIR}\sim0.7 \ m\Omega$cm \cite{saturate}, but actually it  rises indefinitely at higher  temperature as we will discuss later.} 
\label{ScatteringRatesN75}
\end{figure}

The optical conductivity, $\sigma(\Omega)$, is computed within the lowest  approximation of (I) here by discarding the  vertex corrections and working with the auxiliary $\GH$: 
\beq
\real\sigma(\Omega)=\frac{1}{\Omega}\sum_k v_k^2 \int\rho_{\GH}(k,\omega)\rho_{\GH}(k,\Omega+\omega)(f(\omega)-f(\Omega+\omega))d\omega \label{sigEq}
\eeq
The imaginary part of the conductivity can be obtained by a Hilbert transform of the real part.  In this purely $\tJ$ calculation we must be careful how we interpret the  imaginary part of $\sigma$.   A more realistic calculation  should include contributions from the upper Hubbard band and from charge transfer processes which can be appreciable at high frequencies, these are discarded in the \tJ model. For our current purposes we will  discuss  two kinds of relaxation rates. First we compute $1/\tau_{\sigma}$,  describing  a characteristic time which can be extracted from the low frequency behavior  $\sigma(\omega)$ by using
\beq
\tau_{\sigma}\int_0^{1/\tau_{\sigma}} \Re \ \sigma(\omega)/\sigma(0)  d \omega  = \pi/4. \label{opttau}
\eeq
This  momentum averaged quantity  is insensitive to the shape of $\sigma(\omega)$. Secondly we  look at the momentum resolved scattering lifetimes, defined as the inverse width of the ARPES lineshape at the Fermi momentum. These scattering rates are displayed in \figdisp{ConductivityFit}. The ARPES derived $1/\tau$ turns out to be essentially isotropic at the densities reported,   therefore only the nodal $\tau$ is displayed. We find that the $1/\tau$ curves from ARPES and the conductivity have essentially the same temperature dependence, apart from a factor of $O(1)$,    these are shown in \figdisp{ScatteringRatesN75}. 
In {\bf Case (A)} the $1/\tau$  rises quadratically at low temperature in accordance with the standard FL picture.  We  note  that in the overdoped regime, the computed conductivity  matches up, {\em on an absolute scale}   with experiments.  In \figdisp{puchkov} we display the $\Re \sigma(\omega)$ curves for {\bf Case (B)} along with optical conductivity measurements published by Puchkov et al in \cite{basov} for an overdoped Thallium compound.    

A further interesting aspect of the resistivity obtained from this ECFL formalism lies in the high temperature limit. A lack of resistivity saturation has been observed in numerical treatments of strongly coupled models, as in a notable recent DMFT work\cite{georges}. These results are in qualitative agreement with resistivity measurements taken in the cuprates and other strongly correlated compounds.  The ECFL theory leads to a similar result, and provides a simple picture for its origin in terms of the second  chemical potential $u_0$. As noted above, both chemical potentials rise linearly with T at high temperature. Due to the explicit appearance of $u_0$ in the expressions for $\Phi$ and $\Psi$, {\em the magnitude of the self energies}   also grows continuously  with temperature via  $u_0$,    resulting in a  monotonic  broadening of the spectral function. This broadening is insensitive to the Mott Ioffe Regel (MIR) saturation expected in weakly correlated metals, and leads to a non saturating resistivity  at high T, as seen in the inset of \figdisp{ScatteringRatesN75} .   

\subsection{Self energies}
 
\begin{figure}[h]
\includegraphics[width=3.5in]{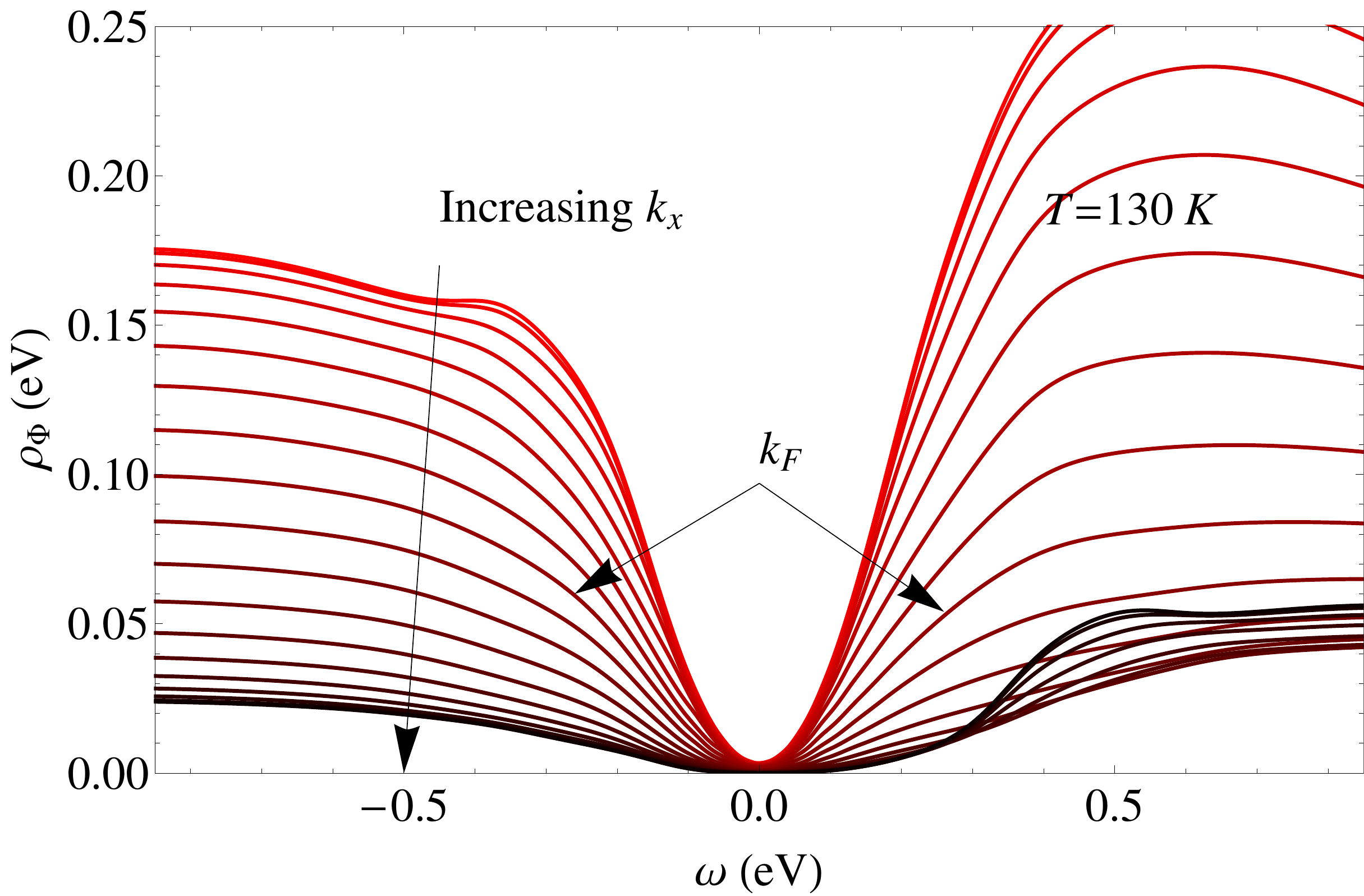}
\includegraphics[width=3.5in]{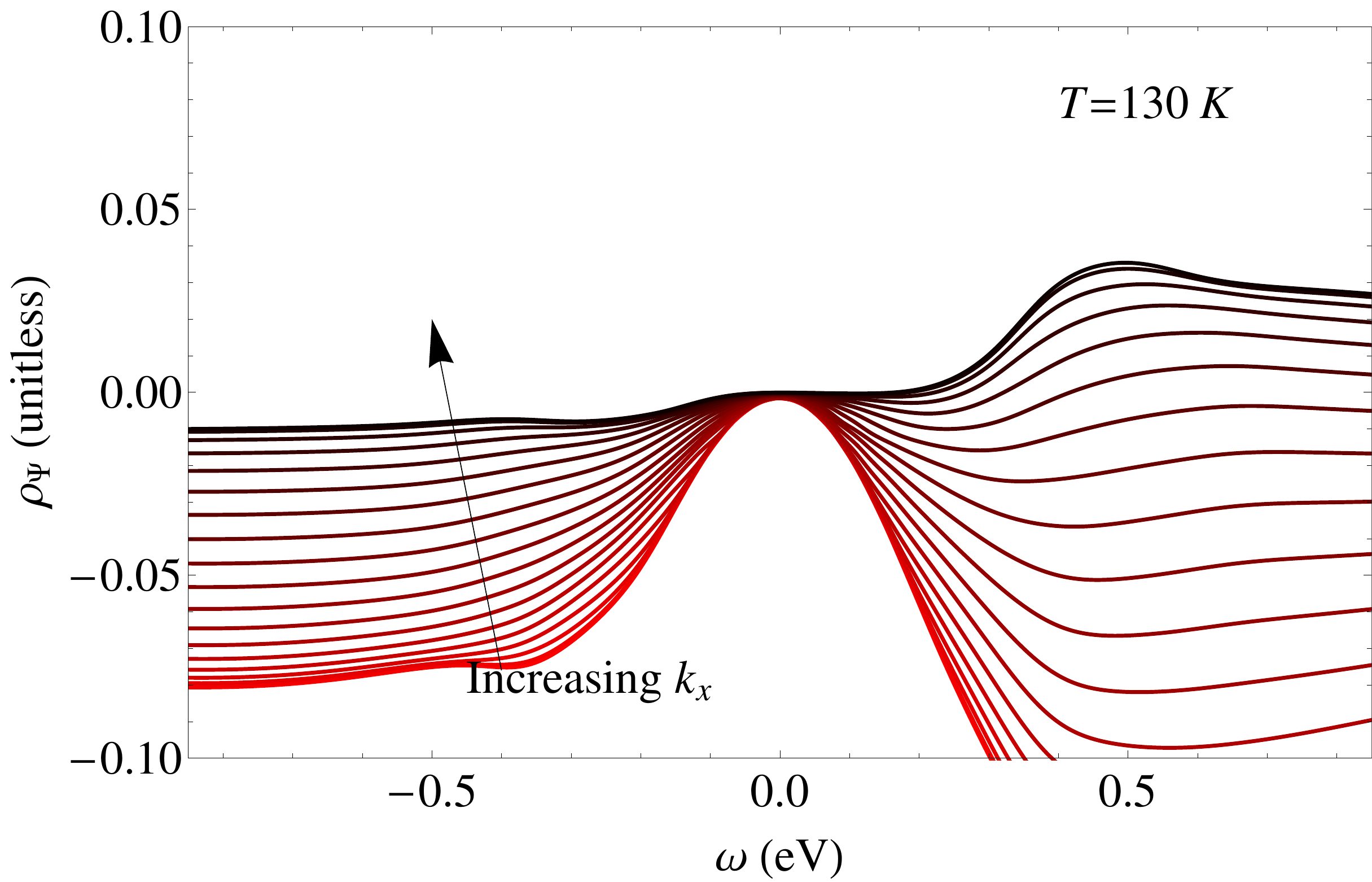}
\caption{{\bf Case (A)} with $(n,T)=(.75,130K)$. The spectral functions (i.e. imaginary parts) for the two self energies $\Phi$   and $\Psi$,  i.e. $\rho_{\overline{\Phi}}$ (left) and $\rho_{{\Psi}}$ (right) in the two panels  at several k points along the $<11>$ direction  vs. frequency.  Both are  roughly  quadratic and symmetric at low frequency,  but have    a strongly k-dependent curvature. In the plot of $\rophi$, the minimum width $\eta$ chops off the bottom of the low frequency minimum. } 
\label{twoselfenergies}
\end{figure}

\begin{figure}[h]
\includegraphics[width=3.5in]{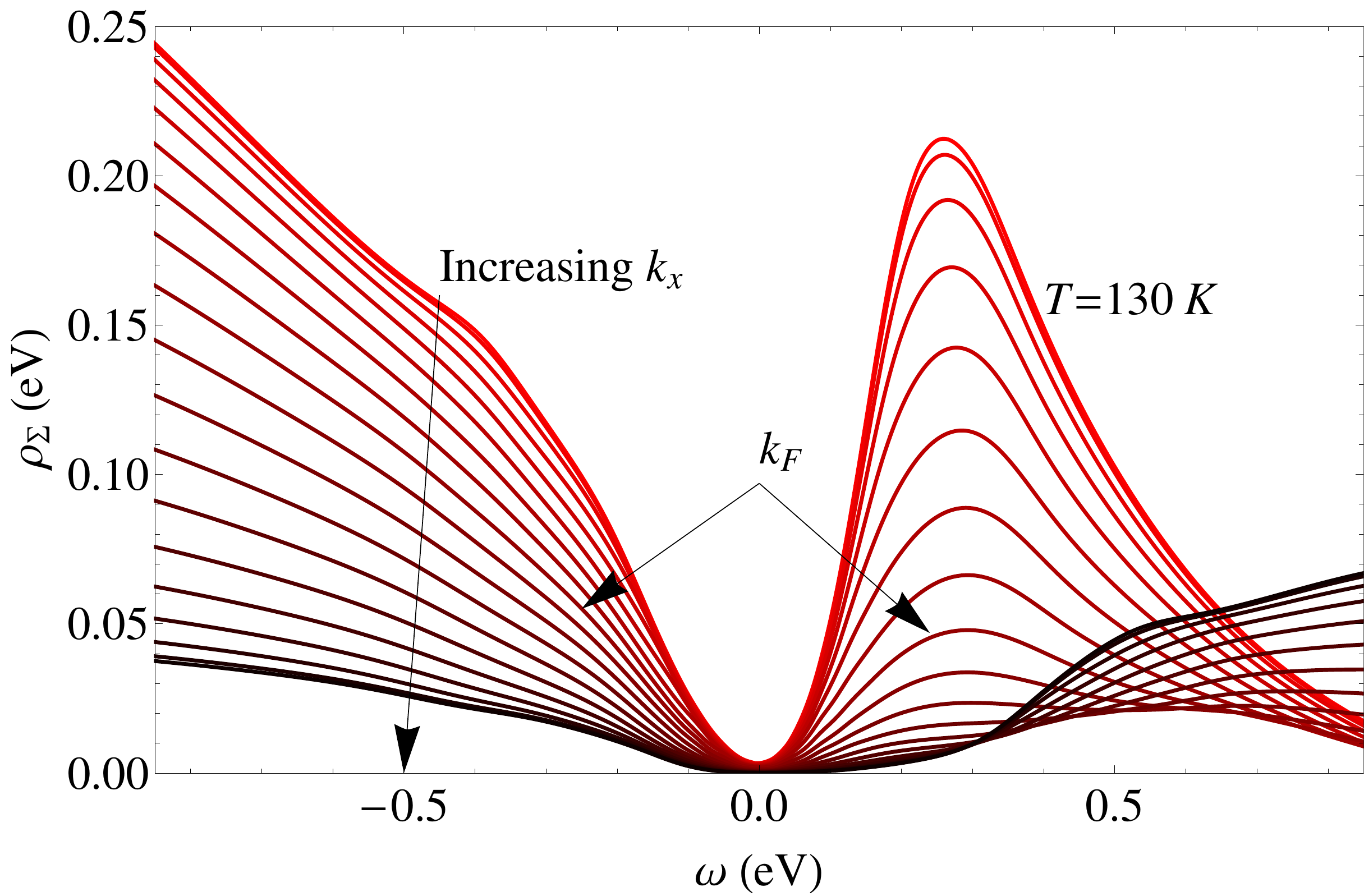}
\includegraphics[width=3.5in]{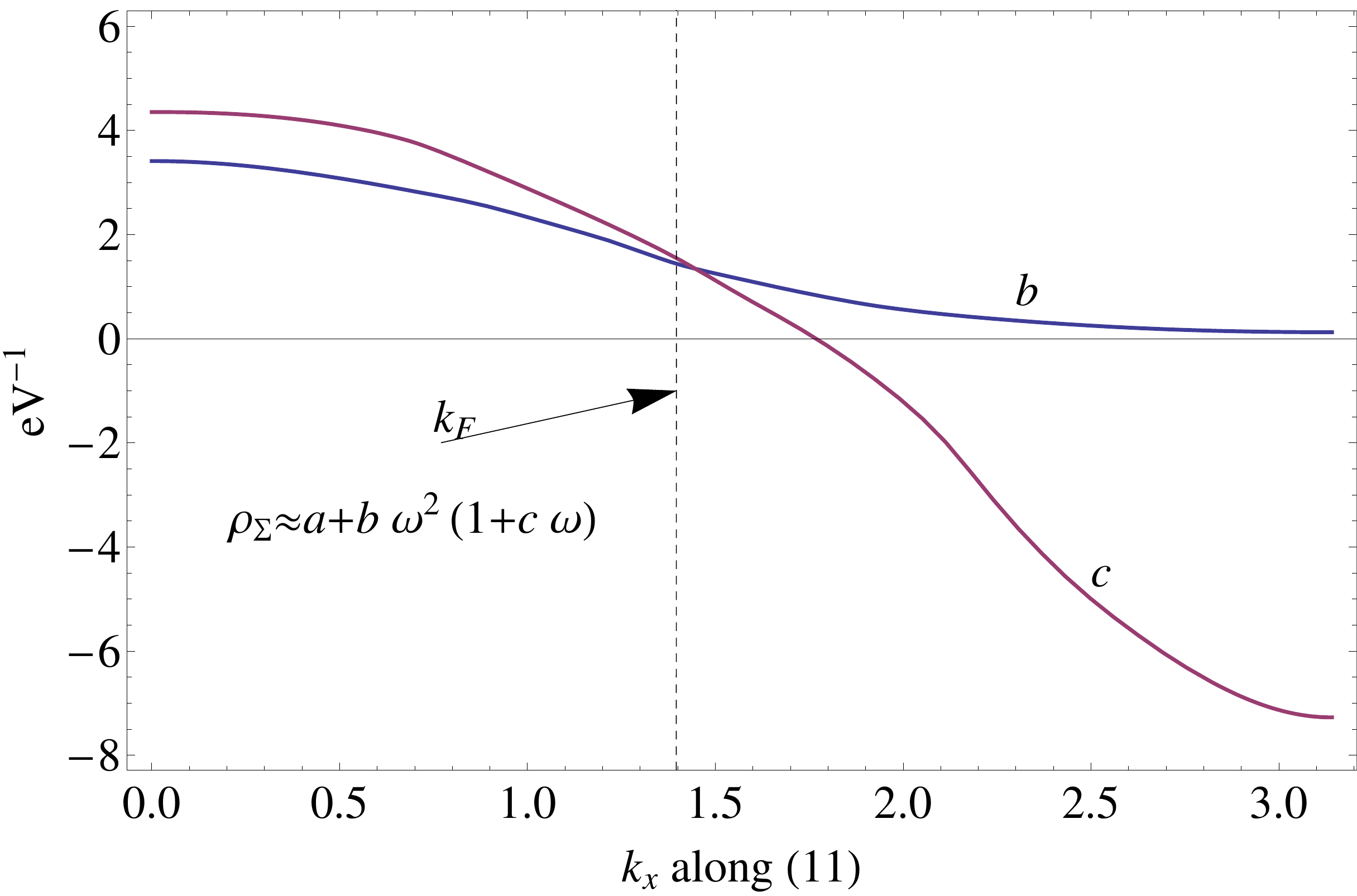}
\caption{{\bf Case (A)} with $(n,T)=(.75,130K)$.  The Dyson-Mori self energy $\Sigma$ can be inferred from the Greens function as in \disp{sigma_def}, and its spectral function $\rho_{\Sigma}$ displayed at several k points along the $<11>$ direction vs. frequency. As with $\rophi$, $\rho_{\Sigma}$ has inherited a strong k dependence. We note here that the primary k-dependence at low frequency is seen in the softening of the quadratic coefficient with increased k. At low frequencies $|\omega | \leq 75 \ meV$, the $\rho_{\Sigma}$ can be fit well with an  expansion up to and including a cubic term, as one sees from the panel at right.
The    cubic term  $\rho_{\Sigma} \propto \omega^3 $  produces particle hole asymmetry as argued in \refdisp{Asymmetry}, and  grows in magnitude with increasing  $k$ beyond $k_F$, where we also see that in  the occupied portion of the BZ  $k < k_F$ the curvature and cubic term have relatively little k-dependence.} 
\label{sigma}
\end{figure}

We now display  the self energies that are involved in calculating the spectral functions. In \figdisp{twoselfenergies} we display $\rho_{\overline{\Phi}}$ and $\rho_{\Psi}$. Both functions exhibit the $\omega^2$ behavior  close to zero, as one finds  for a weakly interacting   FL self energy. Unlike conventional FL's, the magnitude of of the quadratic term is strongly k-dependent.  From these functions and the associated real parts we can construct a Dyson self energy as defined earlier in \disp{sigma_def}. In \figdisp{sigma} we plot the computed imaginary part of the  Dyson self energy,  $\rho_{\Sigma}$. It exhibits a similar magnitude and k-dependence at low frequency to that in $\rho_{\overline{\Phi}}$. However, asymmetries begin to appear at intermediate frequencies.  It is interesting  that at  positive frequency the function is considerably smaller than at negative frequencies, a feature that has already been noted for simplified versions of the ECFL\cite{Asymmetry,Anatomy} and also in recent DMFT studied of the Hubbard model\cite{georges}. In this calculation however, we see an intricate interplay between the momentum and frequency dependences. In particular we see that at positive frequency $0<\omega \lsim  200 \ meV$, $\rho_{\Sigma}$ is extremely k-dependent going from a high maximum suggesting that particle-like excitations near $k=(\pi,\pi)$ are very long-lived while those inside the FS suffer a large scattering rate. This is very different from weakly coupled or local theories where scattering rates are usually determined by {\em frequency alone}.

The right panel of \figdisp{sigma} shows low frequency ($|\omega| \leq 75$ meV) fit parameters of $\rho_{\Sigma}$ as a function of k. An appreciable low frequency asymmetry develops for unoccupied k. At higher frequency much larger asymmetries arise. Specifically, we see that for negative frequencies, a large scattering lobe extends outward to $-\omega\approx  3 \ eV$with considerable weight. This feature is larger for occupied wavevectors. No analogous feature appears for positive frequency. The low frequency asymmetry is usefully described as a FL like quadratic dependence  modified by  a cubic term. Both the quadratic and cubic terms have a strong k-dependence. Fits along the nodal direction are shown in the right panel of \figdisp{sigma}. Note that the sign of the cubic term shifts from the bottom to the top of the band and the quadratic term softens markedly. It is the combination of these two dependences which results in the sharp QPs at $k>k_F$.

Finally, in \figdisp{ScatteringRatesN75} we show the T dependence of the single particle relaxation, obtained from the imaginary part  of the Dyson-Mori self energy at the QP pole $\Im \ \Sigma$.
The  {\bf Case (B)} is clearly  unusual,  but becomes understandable when we recall the proximity of the van Hove point near the Fermi energy for the densities being explored. {\bf Case (A)} shows a  crossover at about $\sim 150$ K  to linear in T behavior, as compared to $T_{\chem'} \sim 400$ K discussed above in \figdisp{Kelvin}. These scales are lower than would be found in  low  order treatments of the weak coupling  Hubbard model,  by roughly a factor of two (at this density). This difference arises primarily due to the bandwidth reduction, thereby reducing the effective Fermi temperature scale. Thus the rather early crossover to linear $T$ behavior is less due to the incoherent contributions to the spectral function, but rather to the shrinking Fermi scale.

\newpage
\section{Concluding remarks}

{\bf In summary}, the present work  is the {\em first microscopic analytical and  controlled calculation  for the \tJ model dynamics}.   We have presented the results of a systematic low density expansion for the \tJ model, using the newly developed formalism of {\em extremely correlated Fermi liquids}, discussed in \refdisp{ECFL} and  \refdisp{ecfl-form}. This calculation goes in a different direction  from the successful  phenomenological simplified ECFL theory in \refdisp{Gweon}. The latter  modeled the optimal doping system lineshapes, and  required the choice of a very small number of parameters, and also assumed  the functional form of $\Im \Phi$ in analogy to FL's.  Here we assume nothing, and our only input parameter is the value of J and the band hopping parameters on the lattice.

The expansion in a parameter $\lambda$  yields a set of  self consistent
equations. The  second order equations  studied here   are argued to be the first step in a systematic microscopic study of the \tJ model,  finally aiming to describe  particle densities closer to optimal doping where the bulk of the experimental effort in the cuprates is focussed. This first step also  helps by 
  validating the forms of the various key elements of the theory- such as the behavior of the two self energies, that are assumed in the  phenomenological implementations of the ECFL ideas in \refdisp{Gweon}.

The range of validity of the present  calculation is somewhat removed from the most interesting regime of optimal doping, fit excellently in \refdisp{Gweon}. It is self consistently determined to be $n \lsim 0.75$ as argued above. The computed forms of the twin self energies found here  indeed have the character assumed in the phenomenological ECFL studies.  The resulting spectral functions  have  skewed lineshapes, that are reminiscent  of the  experimental data,  albeit with reduced magnitude of asymmetry.

In Section (III) we  have presented detailed results for the two  chemical potentials, as well as the momentum distribution function $m_k$, for two standard sets of band parameters of current interest.  We have then detailed the various energy dispersions 
that emerge in our calculation- these include the bare band dispersion, the MDC dispersion and the EDC dispersion. 
We also evaluated  some measures of the line asymmetry. The so called kink feature is displayed and its dependence on the band parameters is explored.  The wave function renormalization $Z_k$ is evaluated at various densities and the Fermi velocity computed along the nodal direction. We  observed  a second   surface, outside the usual Fermi surface,  representing a   shallow  minimum in the QP relaxation rate. It is  different from the usual  Fermi surface in location, size and also in the non zero (but small) dissipation at the surface. It is more  pronounced for {\bf Case (B)} than {\bf Case (A)},  but  not very visible in the  momentum space distribution function $m_k$ in either case.

The relaxation rates of QPs at finite $T$ and resistivity as well as the local DOS are computed. We displayed the spectral function at various densities and temperatures along different directions in detail. We also  explored the effect of elastic impurity scattering on the lineshapes in detail. Finally we displayed the optical conductivity and resistivity from this theory,  after throwing out vertex corrections, and compared with  data on overdoped systems  \cite{basov}. As far as possible, all computed objects are presented {\em on an absolute scale}, using lattice parameters typical of the cuprate materials, and a fundamental energy scale set by the nearest neighbor hopping  $t \sim 3000$K. It is clear that one could tune this scale to better fit a given experiment, e.g. the optical conductivity in \figdisp{puchkov} gives a good fit on an absolute scale  to {\bf Case (A)} if we rescale  $t$, whereas it does well in {\bf Case (B)} with the present scale. Finally, we note the violation of the MIR resistivity saturation in the high T limit.

We have   commented about the absence of the kink at $\Gamma$ in {\bf Case (B)}, and its appearance in {\bf Case (B')} in \figdisp{waterfall} and \figdisp{wfpanel}.  {\bf Case (B)} is the minimal  band dispersion model  with an open Fermi surface near optimum doping, and produces a flat dispersion near the $\Gamma$ point.  More complicated band parameters are  required  to describe the full band, e.g. the parametrization in \refdisp{Bansil} has six parameters, and this restores the curvature near the $\Gamma$ point.  Using the full parametrization,  we computed the spectra  in {\bf Case (B')}, and found that  the waterfall reappears with  a scale similar to that of {\bf Case (A)}.    This exercise   suggests a causal link between the (bare)  band curvature and the appearance of the waterfall at special points in the Brillouin zone.

We next summarizing our key findings:
\begin{itemize}    

\item
The spectral functions at various $k$ values, and their features, probed via the MDC and EDC dispersion relations are detailed. 
 The resulting spectral functions  have features such as a pronounced skew  towards occupied energies $\omega <0$, as seen in most experiments.

\item
This calculation provides a  microscopic identification of the features in the spectral functions that lead to the waterfall phenomenon, and their  dependence on band parameters.
 We find a larger scale of the waterfall for parameters representing electron doped cuprates,
 as also seen in experiments. We also note  an inverse waterfall feature near the top of the band 
 in {\bf Case (B)}, where the waterfall near the $\Gamma$ point is missing.
 
 \item The momentum occupation function $m_k$ indicates a large  spillover for
$k$'s  outside the FS.  Our results quantify the spectral weight at $\omega < 0$ for these wavevectors. These  are of potential use in calibrating  ARPES studies, where often  some  intensity at $\omega < 0$ is often discarded assuming it to be unspecified  background. The Fermi step is weak even at low temperature.
 
 \item
  The ECFL results   for   the Fermi velocities and the optical conductivity are  already in quite reasonable proximity of experimental data {\em on an absolute scale}, and can be made even closer by tuning a single parameter- the hopping parameter- by a factor of 2. 
  
  \item 
 The ECFL calculations naturally lead to  non saturating T dependent resistivities, analogous to those  seen in experiments.  The second chemical potential $u_0$ plays a crucial role here.   This result is also of qualitative value since we are able to identify the cause of the non saturation- {\em  the self energies and relaxation rates  grow in magnitude with $T$ via their dependence on the second chemical potential $u_0$}.
 
 \item
 We find that the scattering mechanism outside the FS is diminished and spectral lines are narrower in the unoccupied part of the Brillouin Zone.
Surprisingly,  this calculation also locates outside the usual FS,  a second surface in k-space,  where the  single particle relaxation rate is a local  minimum.  
It is interesting to speculate the fate of this feature at higher particle densities, where it is imaginable that  the relaxation rates become smaller and almost degenerate with the FS states.

\end{itemize}

\section{ Acknowledgements}
  This work was supported by DOE under Grant No. FG02-06ER46319. We thank Gey-Hong Gweon for stimulating discussions and and Ehsan Khatami for helpful comments on the manuscript.
\appendices
 
 \subsection{Flowchart of the iterative process }
 The self consistency  loop proceeds as follows.

\begin{enumerate}
	\item Initialize all quantities to those of the Fermi gas: $\chem'=\chem_0$, $\ropsi=\rophi=0$, $u_0$ is set to a plausible value of 2t.
	\item Build $\rho_{\GH}$ from latest instance of $\chem'$, $\overline{\Phi}$.
	\item Calculate $\rophi$ from latest instance of $\rho_{\GH}$ and $u_0$. Obtain the real part via Hilbert transform.
	\item Calculate new $\chem'$ using a bisection root finder.
	\item Repeat steps 2-4 until $\chem'$ and $\rophi$ have converged to specified tolerance.
	\item Calculate $\ropsi$ from latest $\rho_{\GH}$ and $u_0$. Obtain real part through Hilbert transform.
	\item Calculate $\sum_k \Psi(k)\GH(k)$ and recalculate $u_0$ with a root finder.
	\item Return to Step 2 and repeat loop. Continue to the next step only when $u_0$ has converged to specified tolerance. 
	\item Calculate $\rho_{\G}$. 
\end{enumerate}

The most computationally expensive step in this loop is the double integration for $\rophi$. If computed by a direct summation the computational time required would scale as $N_s^2 N_{\omega}^2$. Furthermore, this slow step is on the inner most loop so it is repeated many times to find self consistent values of $\chem$ and $u_0$. This leads to unacceptably slow convergence for any reasonable system size. Noting that the summation has the form of a convolution we can make use of FFT routines to calculate $\rophi$ with linear scaling in $N_s N_{\omega}$. This allows us to reach significantly larger systems and lower temperatures than would be possible by a direct approach. The next bottleneck in this flowchart is the calculation of the Hilbert transforms. These can also be made fast through a judicious use of FFT routines. Thus, by using this approach we obtain a scheme which can calculate the full frequency and momentum dependence of $\rho_{\G}$ for lattices of substantial size, $N_s\sim2000$ at temperatures as low as 30K.

Is it useful to discuss the tolerances set on the Lagrange multipliers. $\chem$ is obtained to a relative precision of $10^{-5}$. This is significantly more accurate than is required to satisfy the particle sum rule to within a tenth of a percent. However, we find empirically that this strict convergence criterion for $\chem$ can not be satisfied until the spectral function $\rophi$ is also well converged. Thus, if $\chem$ has successfully converged to this tolerance we can be sure the $\rophi$ is also well converged. The convergence criterion on $u_0$ requires the sum rule $\sum \Psi(k) g(k)=\frac{n^2}{4}(1-\frac{n}{2})$ to be satisfied to less than $10^{-4}$. Again, this is overkill as it concerns the particle density alone. However, $u_0$ appears explicitly in $\bar{\Phi}$ and $\Psi$. The chosen convergence criterion is such that the final $u_0$ lands within $.01t$ of the exact value. This range is comparable to the smallest scales in our calculation, namely  the frequency resolution $\Delta \omega$ and an implicit level broadening scale  $\eta \to \Delta \omega$.
 
To exactly calculate the spectral functions, it is important to capture the entire range of the relevant frequencies. For non-interacting Fermions in 2 spatial dimensions this requires a frequency window no larger than 8t. However, the spectral functions $\rophi$ and $\ropsi$ have long tails which extend to much higher frequency even though the bare bandwidth is considerably decreased according to \disp{eband}. Thus it is important to determine empirically what range of frequency is sufficient to capture the full support of the spectral function. As mentioned before, we employ a frequency grid which extends over the range $|\omega|<15t$, nearly four times the bare bandwidth, and find that this suffices to capture the support of all functions that arise.

\subsection{Fast Fourier Transforms for evaluating  convolutions}
The use of FFTs vastly reduces the time taken to compute the frequency and momentum sums. Each term of $\rophi$ and $\ropsi$ is a convolution of 3 $\GH$'s and has a form which is very similar to the particle-hole bubble diagrams familiar from a second-order perturbation treatment of the Hubbard model in U/t.
\beq
\Sigma(k)_{2^{nd}}\sim \sum_{pq}G(p)G(q)G(p+q-k); \;\; \; \Sigma(i,j) \propto G(i,j)^2 G(j,i)
\eeq
where $i= \vec{R}_i, \tau_i$ is a space time point. The convolution in Fourier space is a simple product in the space time domain and hence the real space version is advantageous.  This is the well-known core idea of the FFT technique, where the time savings arise since the Fourier transforms are performed in $N \log N$ steps rather than $N^2$ (here $N=N_s  N_{\omega}$). The ECFL ``self energies'', $\Psi$ and $\Phi$ have the same frequency convolution structure of $\Sigma(k)_{2^{nd}}$ which appears only through the frequency arguments of $\GH$. However, the $\rophi$ and $\ropsi$ equations suffer from the presence of momentum-dependent decorations which render them not technically convolutions. Nonetheless we can use FFT routines to solve these summations. The strategy is to break up the integral into elementary pieces that do have the form of a convolution. We then avoid the need to do one large integral with quadratic complexity by doing many ($\approx 70$) small FFT's of linear complexity. 
 
To accomplish this we define several $\GH\GH$ correlation functions which are similar to particle-hole bubbles.
\barray
\chi_0(Q)&=&\sum_q \GH(q)\GH(q+Q);\;\;\;\;\;\;\;\;\chi_1(Q)=\sum_q \varepsilon_q\GH(q)\GH(q+Q)\nonumber\\
\chi_2(Q)&=&\sum_q \varepsilon_{q+Q}\GH(q)\GH(q+Q);\;\;\;\chi_3(Q)=\sum_q \varepsilon_{q+Q}^2\GH(q)\GH(q+Q)\nonumber\\
\earray
each of which is a convolution in both frequency and momentum with a spectral function which can be calculated by FFT in linear time.  

With these correlation functions every $\GH\GH\GH$ term (except one to be discussed later) found in \disp{psi} and \disp{phi} can be written in the form
\beq
B_{\GH\GH\GH}(k)=F_1(k)\sum_p F_2(p)\GH(p)\chi_n(p-k)F_3(p-k).\label{conv_structure}
\eeq
where $F_1$,$F_2$, and $F_3$ are each functions of momentum only and their arguments are carefully matched with the arguments of $B_{\GH\GH\GH}$, $\GH$, and $\chi_n$ as they appear in the integral such that all factors fit the form of a convolution in both momentum and frequency. In this way we can massage every term of $\rophi$ into a convolution of one $\GH$ and a $\chi_n$ rather than three $\GH$'s as originally written. There is one term in this problem which cannot be treated in this way because the argument matching  of \disp{conv_structure} cannot be achieved in such a simple way. This problem term looks like
\beq
\overline{\Phi}_{JJ}(k)=\sum_{pq}J_{q-k}J_{p-k}\GH(p)\GH(q)\GH(p+q-k).\nn\\
\eeq
Nonetheless, this term can be treated by the FFT approach if the factor $J_{q-k}$ is broken up using angle addition identities. This is accomplished without difficulty because the locality of $J_{ij}$ ensures that $J_k$ is composed of a small number of Fourier components.

In defining the Fourier transforms, we need to extend the frequency functions to infinity, since it is only then that the frequency convolutions become products in the time domain. Recall that our frequency integrals have been discretized onto $N_{\omega}$ frequency bins which cover the support of our spectral functions. In extending the discretized frequency summations to infinity, we follow the standard procedure of padding the $N_\omega$ frequency bins with an equal number of frequency bins with value zero. By a simple exercise one can verify that padding finite data in this way allows an application of the periodic FFT in such a way that the result of the infinite transform is reproduced. This procedure is discussed in \refdisp{NumRec}. No such considerations are required for the momentum sums which are by definition periodic and discrete, making them naturally suited to treatment by FFT.  

The Hilbert transform \disp{hilberteq} is formally a convolution and can therefore be solved with the advantages of the FFT routines. Once again, however, we face the problem that this convolution is a non-periodic frequency integral. Furthermore, the Hilbert kernel $\frac{1}{\omega}$, unlike other spectral functions with a compact support, falls off very slowly at large frequencies so the padding trick from the $\GH \chi$ convolutions will not work well in this case. It is found that the use of FFT's to calculate a Hilbert transform will always introduce some error. Fortunately, this error can be controlled by increasing the length of padding used. In our code we use a frequency padding of $8N_{\omega}$ for the Hilbert transforms. This relegates the error of the real parts of the various functions to very high frequency, far beyond the compact support of the spectral functions. The error introduced is therefore negligible.


\begin{thebibliography}{99}
\bibitem{fnh}  Alternatively, if we define the Dyson self energy according to $\G=\frac{1}{x-\Sigma'}$, the quasiparticle weight would once again be given by the more familiar expression  $Z'_k=\frac{1}{1-\frac{\partial \Sigma'}{\partial \omega}}$, and moreover  the computed  $Z'_k$  would be the same as in \disp{zk}. Including $A$ in the definition of $\Sigma$ is a convenience employed for two reasons. First, it separates the effects of high and low energy physics formally. The factor $A$ in \disp{sigma_def} represents an overall  depletion of the lower Hubbard band (due to excitations at scale $U$ in a Hubbard model)  and the term $\frac{\partial \Sigma_{DM}}{\partial \omega}$ in the denominator accounts for the depletion of the QP by low frequency FL effects. Secondly,  introducing $A$ results in a self energy $\Sigma_{DM}$ that  reaches a constant at high frequency, unlike the Dyson form  given above with $Z'$ that grows indefinitely.
\end{thebibliography}
\end{document}